\documentclass[iop]{emulateapj}
\usepackage{apjfonts}
\usepackage{epsf}
\usepackage{natbib}
\usepackage[backref,breaklinks,colorlinks,citecolor=blue]{hyperref} 
\usepackage[all]{hypcap}

\newcommand{\drm}{{\rm d}}
\newcommand{\ntot}{n_{\rm tot}}
\newcommand{\avnm}{\langle N|M, z\rangle}
\newcommand{\avncm}{\langle N_{\rm cen}|M, z\rangle}
\newcommand{\avnsm}{\langle N_{\rm sat}|M, z\rangle}

\begin{document}
\title{The FMOS-COSMOS survey of star-forming galaxies at $z\sim1.6$. V: Properties of dark matter halos containing H$\alpha$ emitting galaxies}
 
\author{Daichi Kashino\altaffilmark{1}, 
Surhud More\altaffilmark{2},
John D.~Silverman\altaffilmark{2},
Emanuele Daddi\altaffilmark{3},
Alvio Renzini\altaffilmark{4},
David B.~Sanders\altaffilmark{5},
Giulia Rodighiero\altaffilmark{6},
Annagrazia Puglisi\altaffilmark{6,7},
Masaru Kajisawa\altaffilmark{8,9},
Francesco Valentino\altaffilmark{10,3},
Jeyhan S.~Kartaltepe\altaffilmark{11},
Olivier Le F{\`e}vre\altaffilmark{12},
Nobuo Arimoto\altaffilmark{13},
Naoshi Sugiyama\altaffilmark{14}}

\email{kashinod@phys.ethz.ch}

\altaffiltext{1}{
Institute for Astronomy, Department of Physics, ETH Z{\"u}rich, Wolfgang-Pauli-Strasse 27, CH-8093 Z{\"u}rich, Switzerland
}
\altaffiltext{2}{
Kavli Institute for the Physics and Mathematics of the Universe (WPI), Todai Institutes for Advanced Study, the University of Tokyo, Kashiwanoha, Kashiwa, Chiba 277-8583, Japan
}
\altaffiltext{3}{
Laboratoire AIM-Paris-Saclay, CEA/DSM-CNRS-Universit{\' e} Paris Diderot, Irfu/Service d'Astrophysique,
CEA-Saclay, Orme des Merisiers, F-91191 Gif-sur-Yvette, France
}
\altaffiltext{4}{
INAF Osservatorio Astronomico di Padova, vicolo dell'Osservatorio 5, I-35122 Padova, Italy
}
\altaffiltext{5}{
Institute for Astronomy, University of Hawaii, 2680 Woodlawn Drive, Honolulu, HI 96822, USA
}
\altaffiltext{6}{
Dipartimento di Fisica e Astronomia, Universit{\` a} di Padova, vicolo dell'Osservatorio, 2, I-35122 Padova, Italy
}
\altaffiltext{7}{
ESO, Karl-Schwarschild-Stra{\ss}e 2, 85748 Garching bei M{\"u}nchen, Germany
}
\altaffiltext{8}{
Research Center for Space and Cosmic Evolution, Ehime University, Bunkyo-cho 2-5, Matsuyama, Ehime 790-8577, Japan
}
\altaffiltext{9}{
Graduate School of Science and Engineering, Ehime University, Bunkyo-cho 2-5, Matsuyama, Ehime 790-8577, Japan
}
\altaffiltext{10}{
Dark Cosmology Centre, Niels Bohr Institute, University of Copenhagen, Juliane Maries Vej 30, DK-2100 Copenhagen, Denmark
}
\altaffiltext{11}{
School of Physics and Astronomy, Rochester Institute of Technology, 84 Lomb Memorial Drive, Rochester, NY 14623, USA
}
\altaffiltext{12}{
Aix-Merseille Universit{\'e}, CNRS, LAM (Laboratoire d'Astrophysique de Marseille), UMR 7326, F-13388 Marseille, France 
}
\altaffiltext{13}{
Department of Astronomical Science, SOKENDAI (The Graduate University for Advanced Studies), 2-21-1 Osawa, Mitaka, Tokyo, Japan
}
\altaffiltext{14}{
Division of Particle and Astrophysical Science, Graduate School of Science, Nagoya University, Nagoya, Aichi 464-8602, Japan
}

\begin{abstract}
We study the properties of dark matter halos that contain star-forming galaxies at $1.43 \le z \le 1.74$ using the FMOS-COSMOS survey.  The sample consists of 516 objects with a detection of the H$\alpha$ emission line, that represent the star-forming population at this epoch having a stellar mass range of $10^{9.57}\le M_\ast/M_\odot \lesssim 10^{11.4}$ and a star formation rate range of $15\lesssim \mathrm{SFR}/(M_\odot \mathrm{yr^{-1}}) \lesssim 600$.  We measure the projected two-point correlation function while carefully taking into account observational biases, and find a significant clustering amplitude at scales of $0.04$--$10~h^{-1}~\mathrm{cMpc}$, with a correlation length $r_0 = 5.21^{+0.70}_{-0.67}~h^{-1}~\mathrm{cMpc}$ and a bias $b=2.59^{+0.41}_{-0.34}$.  We interpret our clustering measurement using a halo occupation distribution model.  The sample galaxies appear to reside in halos with mass $M_\mathrm{h} = 4.6^{+1.1}_{-1.6}\times10^{12}~h^{-1}M_\odot$ on average that will likely become present-day halos of mass $M_\mathrm{h} (z=0) \sim2\times10^{13}~h^{-1}M_\odot$, equivalent to the typical halo mass scale of galaxy groups.  We then confirm the decline of the stellar-to-halo mass ratio at $M_\mathrm{h}<10^{12}~M_\odot$, finding $M_\ast/M_\mathrm{h} \approx 5\times10^{-3}$ at $M_\mathrm{h}=10^{11.86}~M_\odot$, which is lower by a factor of 2\textrm{--}4 than those measured at higher masses.  Finally, we use our results to illustrate the future capabilities of Subaru's Prime-Focus Spectrograph, a next-generation instrument that will provide strong constraints on the galaxy-formation scenario by obtaining precise measurements of galaxy clustering at $z>1$.
\end{abstract}

\section{Introduction}

According to our current cosmological model \citep[e.g.,][]{1978MNRAS.183..341W,2006Natur.440.1137S,2008ARA&A..46..385F}, the formation of structure in the Universe is dictated by cold dark matter and dark energy.  On large scales, matter is organized in cosmic web-like structures consisting of dense nodes, filaments, sheets and voids. Dark matter virializes into extended halos at the density peaks within this structure.  Galaxies form within these halos as a result of complex baryonic processes such as star formation and feedback \citep{1977MNRAS.179..541R,1978MNRAS.183..341W}. Understanding the connection between galaxies and dark matter halos can thus shed light on the physical processes that lead to the formation and evolution of galaxies.  Specifically, the mass of halos sets their global properties such as the abundance and spatial distribution \citep{1974ApJ...187..425P,1989MNRAS.237.1127C,2002MNRAS.336..112M} as well as the evolution of their constituent galaxies.  A number of different methods have been used to infer halo masses and constrain the connection between galaxies and dark matter halo masses.  Direct techniques such as measurements of galaxy rotation curves \citep[e.g.,][]{1980ApJ...238..471R,1982AJ.....87..477R}, X-ray emission from the hot intracluster gas \citep[e.g.,][]{2008ApJ...675.1106R} and statistical techniques such as the kinematics of satellite galaxies \citep[e.g.,][]{2009MNRAS.392..801M,2011MNRAS.410..210M} and weak gravitational lensing \citep[e.g.,][]{2005ApJ...635...73H,2006MNRAS.368..715M,2009MNRAS.394..929C,2011ApJ...738...45L,2015MNRAS.446.1356H,2016MNRAS.457.3200M,2016MNRAS.459.3251V} can be used to evaluate dynamical mass of systems, albeit all restricted to fairly local redshifts. At redshifts greater than unity, one has to resort to indirect techniques such as subhalo abundance matching to infer the galaxy--dark matter connection \citep[e.g.,][]{2004ApJ...609...35K,2006ApJ...647..201C,2006ApJ...650..128K,2010ApJ...710..903M,2010ApJ...717..379B,2013MNRAS.436.2286M}.

Analyses of spatial distribution of galaxies can also help to constrain the galaxy--dark matter connection as galaxies reflect the spatial clustering properties of the halos.  For a given cosmological model, the clustering of dark matter halos depends upon halo mass, such that more massive halos display stronger clustering than less massive halos. Therefore, the halo masses can be inferred from the observed clustering amplitude of the hosted galaxies.  Large galaxy surveys such as the two-degree-Field Galaxy Redshift Survey \citep{2001MNRAS.328.1039C} and Sloan Digital Sky Survey \citep[SDSS;][]{2000AJ....120.1579Y} have successfully enabled studies of galaxy clustering at low redshift ($z\sim 0.1$; e.g., \citealt{1994ApJ...431..569P,2000A&A...355....1G,2001MNRAS.328...64N,2002ApJ...571..172Z}).  These studies have also shown that galaxy clustering depends on galactic properties, such as luminosity, color, and morphology, such that relatively luminous (i.e., massive), redder, and bulge-dominated galaxies cluster more strongly (or live in more massive halos), while less massive, bluer, or disk-dominated galaxies have a weaker clustering signal \citep[e.g.,][]{2002MNRAS.332..827N,2005ApJ...630....1Z,2011ApJ...736...59Z}.  Beyond the local Universe ($z\gtrsim0.5$), most studies of galaxy clustering have measured angular correlation functions based on photometric redshift or color selection \citep[e.g.,][]{2004ApJ...611..685O,2005ApJ...619..697A,2008MNRAS.391.1301H,2010ApJ...708..202M,2011ApJ...728...46W,2012A&A...542A...5C,2015MNRAS.449..901M}, while measurements based on spectroscopic samples have been limited \citep[e.g.,][]{2008A&A...478..299M,2009A&A...505..463M,2010MNRAS.406.1306A,2015MNRAS.449.1352C,2015A&A...583A.128D}.

The halo occupation distribution (HOD) framework, which describes the average number of galaxies hosted by a halo as a function of halo mass, has been routinely used to describe the observed abundance and clustering of galaxies \citep[e.g.,][]{2000MNRAS.318.1144P,2001ApJ...546...20S,2002ApJ...575..587B,2003ApJ...593....1B,2004ApJ...609...35K,2005ApJ...633..791Z}.  These observables constrain key quantities of the HOD, such as the minimum mass of halos that host at least one galaxy for a specific population.  The HOD models have been applied to interpret galaxy clustering in the local Universe \citep[e.g.,][]{2004ApJ...608...16Z,2005ApJ...633..791Z,2007ApJ...667..760Z,2009MNRAS.394..929C,2013MNRAS.430..767C,2011ApJ...736...59Z,2011ApJ...738...45L,2015ApJ...806....2M} and at higher redshifts up to $z\sim7$ \citep[e.g.,][]{2011ApJ...728...46W,2012MNRAS.426..679G,2015A&A...583A.128D,2015A&A...576L...7D,2016ApJ...821..123H}.  However, most analyses of the HOD of high-redshift ($z\gtrsim1$) galaxies have been conducted by using photometric samples \citep[e.g.,][]{2011ApJ...728...46W,2012A&A...542A...5C,2015MNRAS.446..169M,2015MNRAS.449..901M}.  While imaging surveys provide large and deep samples at a much lower cost compared to spectroscopy, it is hard to resolve the redshift evolution due to the contamination of back/foreground objects and the associated large uncertainties in the redshift determination.  For example,  even for the estimates in the COSMOS catalogs based on 30-band photometry, the typical error is about $\Delta z/(1+z)\sim0.03$ \citep{2013A&A...556A..55I,2016ApJS..224...24L}.

On the other hand, spectroscopic surveys at higher redshifts generally tend to be restricted to relatively bright, rare objects.  Such samples may not represent the general galaxy population.  Therefore, spectroscopic studies based on galaxies representative of the epoch's average population are required.  It has been established that the bulk of star-forming galaxies follow a tight correlation between stellar mass ($M_\ast$) and star formation rate (SFR) over a wide range of redshift up to $z\sim4$ \citep[e.g.,][]{2007ApJ...660L..43N,2007ApJ...670..156D,2012ApJ...754L..29W,2013ApJ...777L...8K,2009ApJ...694.1517D,2015ApJ...807..141P} or even higher \citep[e.g.,][]{2015ApJ...799..183S}, the so-called ``main sequence''.  Galaxies along this sequence dominate the cosmic stellar mass density at $z\gtrsim1$ \citep{2013A&A...556A..55I} as well as the cosmic star formation rate density over cosmic history \citep{2011ApJ...739L..40R,2015A&A...575A..74S}.  Therefore, main-sequence galaxies are representative of the average star-forming galaxy population at all redshifts.

In this work, we investigate the properties of halos that host galaxies representative of the main-sequence star-forming population using spectroscopic selection of 516 objects at  $1.43 \le z \le 1.74$ down to a stellar mass $10^{9.57}~M_\odot$.  This redshift range marks the transition epoch of cosmic star formation from its peak to the decline phase that continues until today \citep{2014ARA&A..52..415M}.  Galaxies in our sample are drawn from a near-IR spectroscopic campaign using the Fiber Multi-Object Spectrograph (FMOS) on the Subaru telescope, called the FMOS-COSMOS survey \citep{2015ApJS..220...12S}.  For this sample, we successfully map the small scale structures below $1~h^{-1}~\mathrm{comoving~Mpc~(cMpc)}$, where the contribution from galaxies that reside in the same halo (i.e., one-halo term) becomes important.   Such a high sampling rate per unit area is one of the unique advantages of our survey and makes it complementary to the FastSound survey, which is another wide-field spectroscopic survey carried out with FMOS \citep{2015PASJ...67...81T,2016PASJ...68...38O}.

The paper is organized as follows.  In Section \ref{sec:data}, we provide an overview of the FMOS-COSMOS survey and describe our galaxy sample.  We describe the methods employed to measure clustering in Section \ref{sec:clustering}, and corrections for critical biases in Section \ref{sec:corr}.  We present our clustering measurements in Section \ref{sec:results}.  Results are interpreted using an HOD modeling in Sections \ref{sec:halo}, and we discuss the physical implications of the derived quantities in Section \ref{sec:discussion}.  We finally summarize our results and conclusions in Section \ref{sec:summary}.  Throughout the paper, magnitudes are given in the AB system and a flat $\Lambda$CDM cosmology with $\left(\Omega_m, \Omega_\Lambda \right) = \left(0.3, 0.7\right)$ is assumed.  We express distances in comoving units and halo masses with the Hubble parameter $h$ in units of $100~\mathrm{km~s^{-1}~Mpc^{-1}}$, while stellar masses and, subsequently, the stellar mass-to-halo mass ratios are computed assuming $h=0.7$.  We use a \citet{2003PASP..115..763C} initial mass function (IMF).  The conversion factor to a Chabrier IMF is $1/1.7$ for both stellar masses and SFRs given in our companion papers that use a Salpeter IMF \citep{2013ApJ...777L...8K,2017ApJ...835...88K,2015ApJS..220...12S}.  We use ``$\log$'' to denote a logarithm with a base 10 ($\log_{10}$).  For references, Table \ref{tb:symbols} defines symbols used in this paper.

\capstartfalse
\begin{deluxetable}{cl}
\tablecaption{Symbols used in this paper\label{tb:symbols}}
\tablehead{\colhead{Symbol}&\colhead{Definition}}
\startdata
$M_\mathrm{h}$ & halo mass \\
$M_\ast$ & stellar mass \\ 
$M_\ast^\mathrm{lim}$ & stellar mass limit for our sample $(10^{9.57}~M_\odot)$  \\
$\mathrm{SFR}$ & star formation rate \\
$f^\mathrm{pre}_\mathrm{H\alpha}$ & predicted H$\alpha$ flux (after dust obscuration) \\ 
$f^\mathrm{lim}_\mathrm{H\alpha}$ & lower limit of $f^\mathrm{pre}_\mathrm{H\alpha}$ for our sample $(1\times10^{-16}~\mathrm{erg~cm^{-2}~s^{-1}})$ \\
$f_\mathrm{fake}$ & fraction of misidentified objects in our sample \\
$\xi$, $\xi_\mathrm{dm}$ & real space correlation function of galaxies/dark matter \\
$w_\mathrm{p}$, $w_\mathrm{p,dm}$ & projected correlation function of galaxies/dark matter\\
$r_\mathrm{p}$ & comoving tangential separation\\
$\pi$ & comoving line-of-sight separation\\
$b$ & galaxy bias \\
$\omega (\theta)$ & angular correlation function
\enddata
\end{deluxetable}
\capstarttrue

\section{Data} \label{sec:data}
\subsection{Overview of the FMOS-COSMOS survey} \label{sec:FMOS-COSMOS}

The galaxy sample used in this paper is constructed from the data set of the FMOS-COSMOS survey.  Details of the observations, survey design and data analysis are described elsewhere \citep{2013ApJ...777L...8K,2015ApJS..220...12S}.  FMOS is a near-infrared spectrograph on the Subaru telescope with high multiplex capabilities as it allows the placement of 400 fibers, each with an aperture of $1.2\arcsec$ in diameter, over a circular field of 0.19 square degrees \citep{2010PASJ...62.1135K}.  In cross-beam switching mode, roughly 200 objects can be observed at a given time.  FMOS has an OH airglow suppression system that blocks the strong OH emission lines ($\sim 30\%$ of the entire spectral window).  

The survey is designed to detect the H$\alpha$ emission line with the $H$-long grating (1.6--1.8$~\mathrm{\mu m}$, spectral resolution $R \approx 3000$).  The H$\alpha$ and [N{\sc ii}]$\lambda\lambda$6548, 6584 lines are well separated and the accuracy of redshift determination is $\Delta z / (1+z) = 2.2\mathrm{e-4}$, corresponding to $30~\mathrm{km~s^{-1}}$ at $z\sim 1.6$ \citep{2015ApJS..220...12S}.  We carried out additional observations with the $J$-long grating (1.11--1.35~$\mu$m) to detect the H$\beta$ and [O{\sc iii}]$\lambda \lambda $5007, 4959 emission lines to establish the ionization conditions of $z\sim1.6$ star-forming galaxies \citep{2014ApJ...792...75Z,2015ApJ...806L..35K,2017ApJ...835...88K} and confirm redshift measurements for about half of the sample.  Our survey covers the entire COSMOS field with multiple FMOS footprints having some overlap regions.  All data are reduced with the FMOS Image-Based Reduction Package (FIBRE-pac; \citealt{2012PASJ...64...59I}).  

\subsection{Sample selection} \label{sec:selection}

\begin{figure}[tbph] 
   \centering
   \includegraphics[width=3.2in]{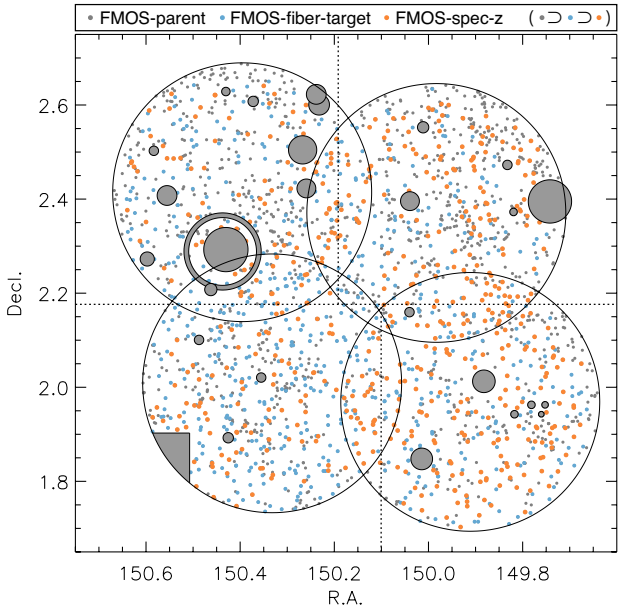} 
   \caption{Four FMOS footprints (large circles) and galaxies in our FMOS samples (gray dots -- the FMOS-parent sample; blue dots -- FMOS-fiber-target sample; orange dots -- FMOS-spec-$z$ sample).  Dotted lines indicate partitions for jackknife resampling (see Section \ref{sec:error}).}
   \label{fig:field}
\end{figure}

In this study, we use a catalog of galaxies with spectroscopic redshifts based on observations carried out from 2012 March to 2014 February, including 11 and 6 pointings in the central region of the COSMOS field with the $H$- and $J$-long gratings, respectively.  The survey field is shown in Figure \ref{fig:field} and has a area of $0.810~\mathrm{deg}^2$.  We remove the sky regions impacted by bright stars (shaded areas), which account for 4.9\% of the survey area.  

Our sample is selected from the COSMOS photometric catalog that includes the Ultra-VISTA/VIRCAM photometry \citep{2012A&A...544A.156M,2013A&A...556A..55I}.  Photometric redshifts and stellar masses are derived by spectral energy distribution (SED) fitting with the Le Phare photometric code \citep{2011ascl.soft08009A}, assuming \citet{2003MNRAS.344.1000B} stellar population synthesis models and a Chabrier IMF.  The photometric catalog reaches a magnitude limit of $K_\mathrm{S}\approx24~\mathrm{mag}$, thus being complete for more than $90\%$ of the galaxies down to a stellar mass $10^{9.57} M_\odot$ \citep{2013A&A...556A..55I}.  We also derive star formation rates (SFRs) from SED fitting assuming a constant star formation history, as extensively described in \citet{2015ApJS..220...12S}, and compute the predicted H$\alpha$ flux ($f^\mathrm{pre}_\mathrm{H\alpha}$) for each galaxy using a relation from \citet{1998ARA&A..36..189K}, converted for use with a Chabrier IMF:
\begin{equation}
f_\mathrm{H\alpha}^\mathrm{pre} = \frac{1}{4\pi d_\mathrm{L}^2}\frac{\mathrm{SFR} (M_\odot~\mathrm{yr^{-1}})}{4.65\times10^{-42}} 10^{-0.4A_\mathrm{H\alpha}},
\label{eq:sfr}
\end{equation}
where $d_\mathrm{L}$ is luminosity distance.  The values of $f^\mathrm{pre}_\mathrm{H\alpha}$ represent the total light coming from galaxies, as opposed to only that seen by a single fiber.  Dust extinction of the stellar component $E_\mathrm{star} (B-V)$ is also derived from our SED fitting.  This value has been converted to a nebular extinction to the H$\alpha$ emission as $A_\mathrm{H\alpha} = 3.325 E_\mathrm{star} (B-V) / 0.66$ (see e.g., \citealt{2013ApJ...777L...8K}), assuming a \citet{2000ApJ...533..682C} extinction curve.  For the remainder of the paper, these SED-based quantities are used for selection, while the spectroscopic redshifts and observed H$\alpha$ fluxes are incorporated into the final determination of their SFRs (Section \ref{sec:bar_conv}).  We describe in detail the steps of our sample selection below.  Each subsample will be labeled at various stages of the selection to make the treatment of biases clear.

Over an effective survey area ($0.77~\mathrm{deg}^2$), we find 7006 galaxies in the COSMOS catalog that have stellar mass above a threshold mass $M_\ast^\mathrm{lim} \equiv 10^{9.57}~M_\odot$ and photometric redshift ($z_\mathrm{phot}$) between 1.46 and 1.72.  We refer to these mass (and photo-$z$) selected galaxies as {\it $M_\ast$-selected} sample.  Our targets are restricted to those with the Ultra-VISTA/VIRCAM photometry $K_\mathrm{S}\le23.5$.  We designate the subset of 6453 galaxies with both $M_\ast \ge M_\ast^\mathrm{lim}$ and $K_\mathrm{S}\le23.5$ as {\it $M_\ast+K_\mathrm{S}$-selected} sample.  In addition, we impose a limit on the predicted H$\alpha$ flux computed from the SED-based SFR and dust extinction (Equation \ref{eq:sfr}), as $f^\mathrm{pre}_\mathrm{H\alpha} \ge f^\mathrm{lim}_\mathrm{H\alpha} \equiv 1\times 10^{-16}~\mathrm{erg~cm^{-2}~s^{-1}}$, to achieve an acceptable success rate of detecting the H$\alpha$ emission line with FMOS.  This flux limit is equivalent to $\mathrm{SFR}\approx20~M_\odot ~\mathrm{yr^{-1}}$ for galaxies with typical extinction $A_\mathrm{H\alpha} = 1$ mag at $z=1.6$.  Of the $M_\ast+K_\mathrm{S}$-selected sample, we find 2139 galaxies having $f^\mathrm{pre}_\mathrm{H\alpha} \ge f^\mathrm{lim}_\mathrm{H\alpha}$, which are referred to as {\it FMOS-parent} sample.  This sample is used as the input catalog for the process of allocating fibers to galaxies.

From the FMOS-parent sample, we observed 1182 galaxies with the $H$-long grating, which are referred to as {\it FMOS-fiber-target} sample.  Of these, we select 516 galaxies that have a positive H$\alpha$ detection ($S/N\ge1.5$) in the $H$-long window ($1.43\le z \le 1.74$; median $z=1.588$) to measure their spatial clustering.  This final sample, refereed to as {\it FMOS-spec-$z$} sample, is restricted to galaxies having one or more emission lines with $S/N\ge3$.  In most cases (503/516), H$\alpha$ is detected at $S/N\ge3$.  For H$\alpha$ detections of low significance ($1.5\le S/N<3$), we require the detection of at least one other line (e.g., [O\,{\sc iii}]$\lambda$5007) at $S/N\ge3$.  Table \ref{tb:samples} summarizes the size, number density, selection criteria and mean/median stellar masses of each subsample.  The total survey volume over the range of spectroscopic redshift ($1.43\le z\le1.74$) is $8.98\times10^5~(h^{-1}~\mathrm{cMpc})^3$.

\capstartfalse
\begin{deluxetable*}{lcclcc}
\tablecaption{Galaxy samples\label{tb:samples}}
\tablehead{\colhead{Sample name}&\colhead{$N$\tablenotemark{a}}&\colhead{$n~\left( 10^{-3} h^3\mathrm{cMpc}^{-3} \right)$}&\colhead{Selection criteria}&\colhead{Mean $M_\ast$\tablenotemark{b}}&\colhead{Median $M_\ast$\tablenotemark{b}}}
\startdata
$M_\ast$-selected & 7006 & 9.29\tablenotemark{c} & $1.46\le z_\mathrm{phot}\le 1.72$, $M_\ast \ge 10^{9.57}~M_\odot$ & 10.42& 10.04 \\
$M_\ast+K_\mathrm{S}$-selected & 6453 & 8.55\tablenotemark{c} & $+$ $K_\mathrm{S}\le 23.5$ & 10.45 & 10.10\\
FMOS-parent & 2319 & 3.07\tablenotemark{c} & $+$ Predicted $f(\mathrm{H\alpha}) \ge 10^{-16}~\mathrm{erg~s^{-1}~cm^{-2}}$ & 10.58 & 10.20 \\
FMOS-fiber-target & 1182 & -- & $+$ Observed & 10.59 & 10.23 \\
FMOS-spec-$z$ & 516 & 0.575\tablenotemark{d} & $+$ H$\alpha$-detected at $1.43\le z\le 1.74$ & 10.54 & 10.22
\enddata
\tablenotetext{a}{Number of galaxies in each sample.}
\tablenotetext{b}{Mean and median stellar masses of each sample, expressed in $\log M_\ast / M_\odot$.}
\tablenotetext{c}{Number density of photometrically-selected sample, computed in the volume over $1.46\le z\le 1.72$ ($7.54\times10^5~h^{-3}\mathrm{cMpc}^{3}$).}
\tablenotetext{d}{Number density of spectroscopic sample, computed in the volume over $1.43\le z\le 1.74$ ($8.98\times10^5~h^{-3}\mathrm{cMpc}^{3}$).}
\end{deluxetable*}
\capstarttrue

\subsection{Sample characteristics} \label{sec:samples}

\begin{figure*}[tbph] 
   \centering
   \includegraphics[width=5in]{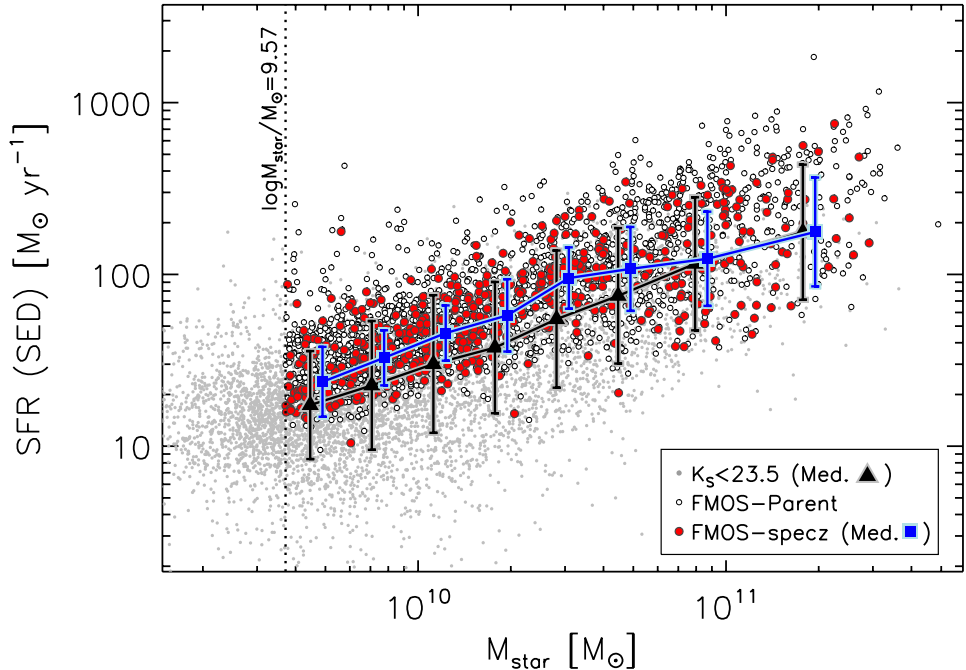} 
   \caption{SFR (based on SED) versus $M_\ast$ for galaxies in our sample.  Gray dots show all galaxies with $K_\mathrm{S} \le 23.5$ and $1.46\le z_\mathrm{phot}\le1.72$ in the FMOS survey area.  The FMOS-parent and FMOS-spec-$z$ samples are shown as open and red filled circles, respectively.  The median SFRs in eight $M_\ast$ bins are indicated for the $M_\ast$+$K_\mathrm{S}$-selected sample (filled triangles) and the FMOS-spec-$z$ sample (filled blue squares).  The error bars indicate the central 68th percentiles in SFR for each bin.  A vertical dotted line indicates the threshold stellar mass ($M_\ast^\mathrm{lim}=10^{9.57}~M_\odot$) for our sample selection.}
   \label{fig:sample}
\end{figure*}

In Figure \ref{fig:sample}, we show the SFR derived from SED as a function of $M_\ast$ for galaxies in our sample.  We note that stellar masses and SFRs of the FMOS-spec-$z$ sample shown here are not revised by incorporating their spectroscopic redshifts, but rather the original values based on photometric redshifts used for the sample selection.  In agreement with many studies, there is a clear correlation between SFR and $M_\ast$, i.e., the star-forming main sequence.  It is evident that the our sample traces the underlying distribution of star-forming galaxies.  We find no significant difference in SFR between the FMOS-parent and FMOS-spec-$z$ samples.  However, it is shown that galaxies having relatively low SFRs (64\%) are missed in the selection due to the self-imposed limitation on the predicted H$\alpha$ flux.  The median SFRs and the 68th percentiles are shown in eight bins of stellar mass for both the $M_\ast+K_\mathrm{S}$-selected (triangles) and FMOS-spec-$z$ (blue squares) samples.  The FMOS-spec-$z$ sample is on average biased toward higher SFRs by $0.1\textrm{--}0.2~\mathrm{dex}$ at $M_\ast \lesssim 10^{10.8}~M_\odot$.  However, such a bias is only about half the scatter in SFR of the $M_\ast+K_\mathrm{S}$-selected sample.  Therefore, we contend that the FMOS-spec-$z$ sample is well representative of the normal star-forming population at these redshifts.

\begin{figure}[tbph] 
   \centering
   \includegraphics[width=3.5in]{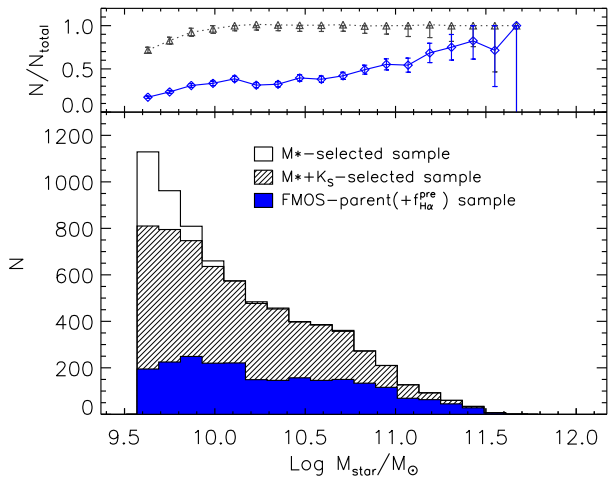} 
   \caption{Stellar mass distributions of our pre-observation samples.  Histograms show the numbers of galaxies in each $0.12~\mathrm{dex}$ bin (empty histogram -- $M_\ast$-selected; hatched histogram -- $M_\ast+K_\mathrm{S}$-selected; filled histogram -- FMOS-parent sample).  The upper panel shows the binned ratios of the number of galaxies in the $M_\ast+K_\mathrm{S}$-selected sample (triangle) and FMOS-parent sample (diamond) to the number of galaxies in the $M_\ast$-selected sample with error bars indicating the Poisson noise.}
   \label{fig:histm}
\end{figure}

We assess the completeness of our sample at a given stellar mass in Figure \ref{fig:histm}, which compares the stellar mass distributions of the $M_\ast$-selected, $M_\ast+K_\mathrm{S}$-selected and FMOS-parent samples.   In the upper panel, the fractions of the number of galaxies in the $M_\ast+K_\mathrm{S}$-selected or in the FMOS-parent samples to the number of galaxies in the $M_\ast$-selected sample are shown as a function of stellar mass.  As evident, approximately 8\% of galaxies in the $M_\ast$-selected sample are missed near the lower mass limit ($M_\ast \lesssim 10^{10}~M_\odot$) by the inclusion of the $K_\mathrm{S}\le 23.5$ limit, while the stellar mass completeness reaches almost unity at $M_\ast > 10^{10.1}~M_\odot$.  In the FMOS-parent sample, a large fraction of galaxies in the $M_\ast$-selected sample are missed due to the limit on the predicted H$\alpha $ flux.  The sampling rate increases slowly with increasing stellar mass.  As a consequence, the FMOS-parent sample has slightly higher mean and median stellar masses as compared to the $M_\ast$-selected or $M_\ast+K_\mathrm{S}$-selected sample (Table \ref{tb:samples}).  Given an expected dependence of the clustering amplitude on stellar mass, such a selection bias may affect the observed correlation function.  The effects of stellar mass incompleteness of our sample are further discussed using mock samples in Appendix \ref{sec:mock_am}.  We ensure that such effects do not significantly impact our results and conclusions.

\begin{figure*}[tbph]
   \centering
   \includegraphics[width=7in]{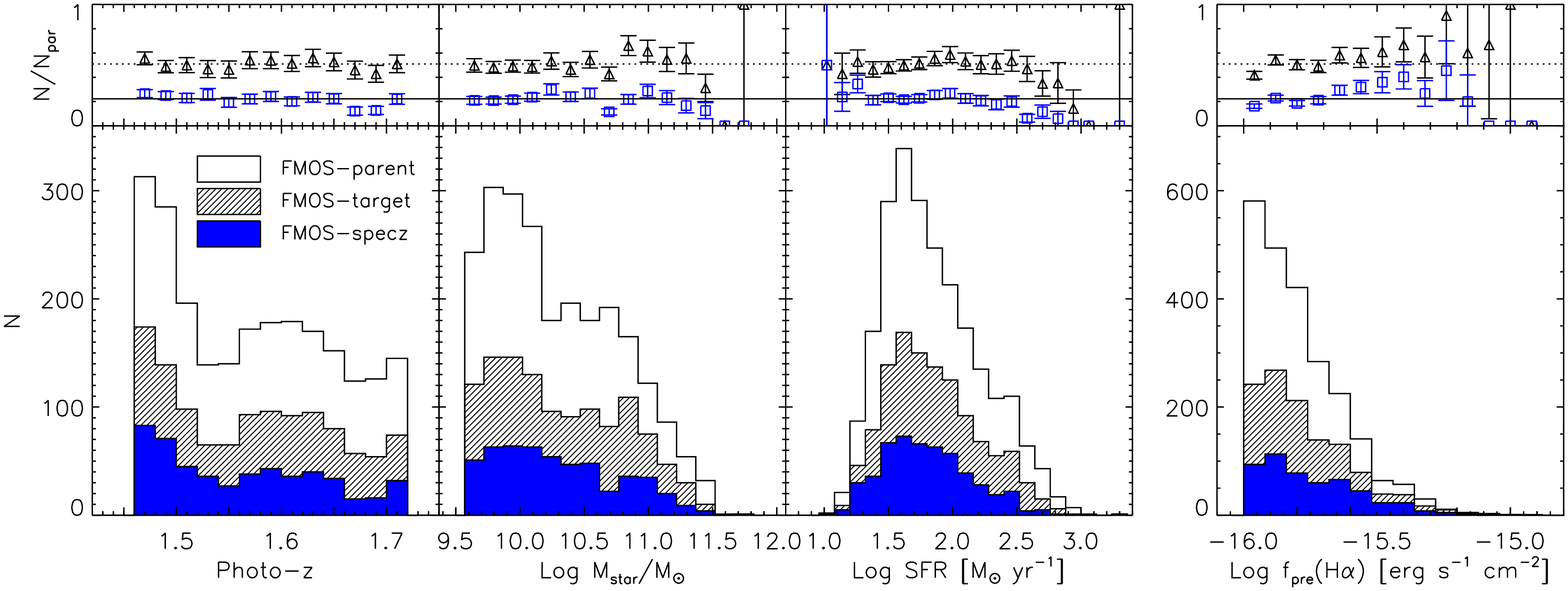} 
   \caption{Distributions of photometric redshift, stellar mass, SFR (based on SED), and predicted H$\alpha$ flux, from the left to right, respectively: empty histograms -- FMOS-parent; hatched histograms -- FMOS-fiber-target; filled blue histograms -- FMOS-spec-$z$ sample.  Upper panels show the binned ratios of the number of galaxies in the FMOS-fiber-target sample (triangles) or FMOS-spec-$z$ sample (squares) to the number of galaxies in the FMOS-parent sample with error bars indicating the Poisson noise.  Note: the left panel shows the distribution of photometric redshifts even for the spec-$z$ sample.}
   \label{fig:3hist}
\end{figure*}

We further evaluate the observational selection effects.  Figure \ref{fig:3hist} shows the distribution of photometric redshift, stellar mass, SED-based SFR, and the predicted H$\alpha$ flux for the FMOS-parent, -fiber-target, and -spec-$z$ samples.  These values are not revised for spectroscopic redshifts.  The fractions of the number of galaxies in the FMOS-fiber-target sample or in the FMOS-spec-$z$ sample to the number of galaxies in the FMOS-parent sample are presented in the upper panels.  We highlight that the sampling rate is almost uniform as a function of photometric redshift, stellar mass and SFR.  As a consequence, the median values of these quantities of the FMOS-spec-$z$ sample are in good agreement with those of the FMOS-parent sample (see Table \ref{tb:samples} for average masses).  In contrast, the right panel of Figure \ref{fig:3hist} indicates that the FMOS-spec-$z$ sample is slightly biased towards having higher $f^\mathrm{pre}_\mathrm{H\alpha}$ as compared to the FMOS-parent sample.  Such a bias is expected from the fact that stronger H$\alpha$ lines are easier to detect.  Even though, the median value of the predicted H$\alpha$ flux of the FMOS-spec-$z$ sample ($\log f^\mathrm{pre}_\mathrm{H\alpha}/(\mathrm{erg~s^{-1}~cm^{-2}})=-15.79$) is close to that of the FMOS-parent sample ($\log f^\mathrm{pre}_\mathrm{H\alpha}/(\mathrm{erg~s^{-1}~cm^{-2}})=-15.83$).  Therefore, we conclude that the observational sampling biases do not significantly affect the characteristics of the spec-$z$ sample, relative to the parent sample.  

\begin{figure}[tbph] 
   \centering
   \includegraphics[width=3.5in]{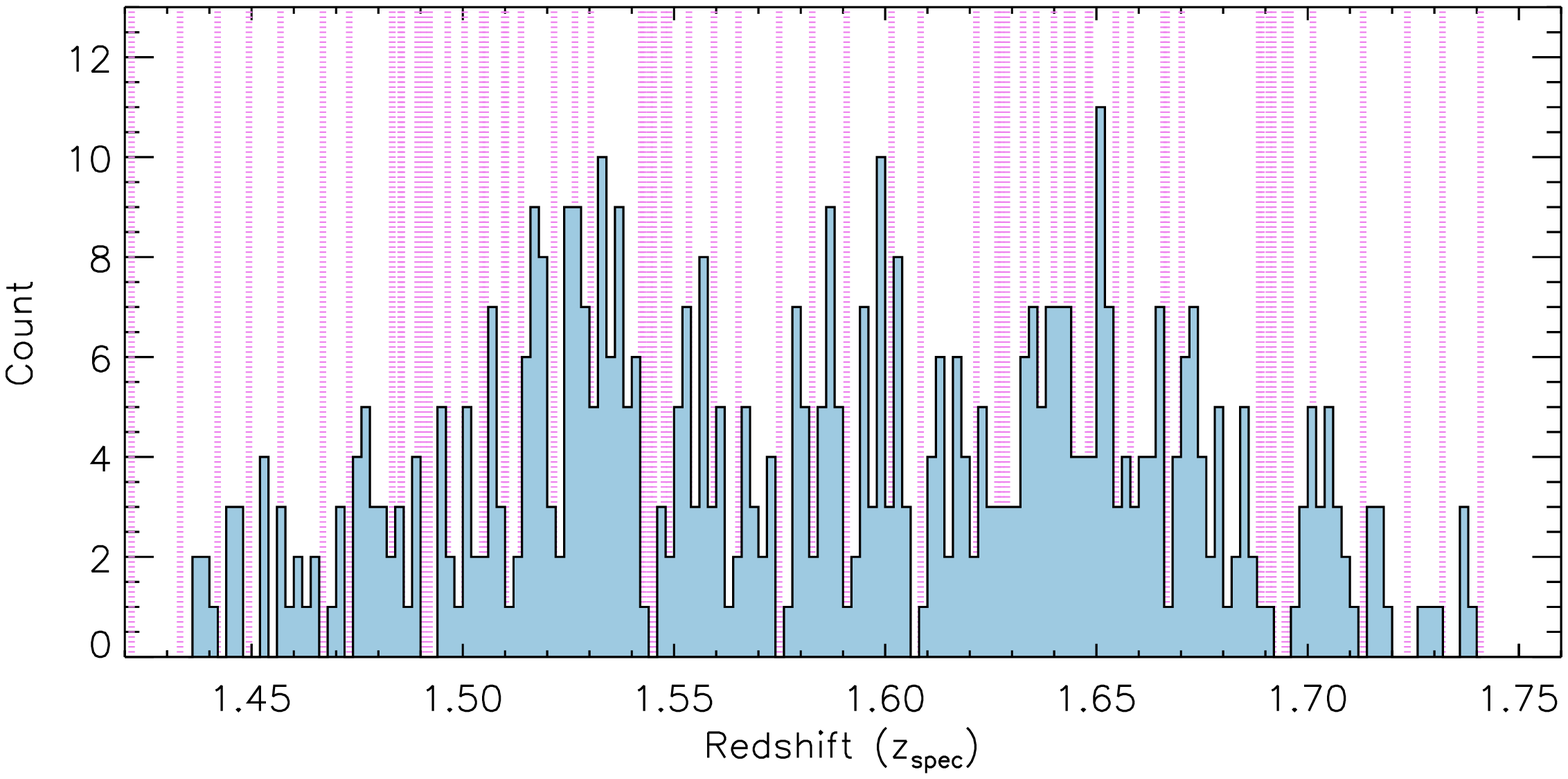}
   \caption{Distribution of the 516 spectroscopic redshifts from the FMOS-spec-$z$ sample.  Magenta stripes indicate positions of the OH airglow lines, which are shifted to the redshift of the H$\alpha$ emission line.}
   \label{fig:histz}
\end{figure}

In Figure \ref{fig:histz}, we show the distribution of spectroscopic redshifts for 516 galaxies in the FMOS-spec-$z$ sample with positions of OH lines highlighted.  Wavelengths of OH lines are converted into redshifts based on the wavelength of the H$\alpha$ emission line as $z=\lambda_\mathrm{OH}/\lambda_\mathrm{H\alpha}-1$.  It is evident that the number of successful detections is suppressed near OH contamination.  We take into account these effects on the radial distribution of the target galaxies, as described in Section \ref{sec:random}.  

\subsection{Mock catalog} \label{sec:mock}
We validate our correction schemes (described in Section \ref{sec:corr}) for observational biases by using a set of mock samples constructed from cosmological $N$-body simulations.  We utilize the new numerical galaxy catalog ($\nu^2$GC, \citealt{2015PASJ...67...61I}) to take advantage of its large simulation volume.  We use the medium-volume simulation ($\nu^2$GC-M) that was conducted by using $4096^3$ particles with a mass resolution of $2.2\times 10^8~h^{-1}~M_\odot$ in a comoving box with a side length of $560~h^{-1}~\mathrm{cMpc}$.  We employ the halo catalog at a scale factor $a=0.384871$ ($z=1.598$), close to our median redshift ($z=1.588$), in which halos and subhalos are identified with the Rockstar algorithm \citep{2013ApJ...762..109B}.  For our purpose, the simulation box is divided into 64 sub-volumes of $70\times70\times560~(h^{-3}~\mathrm{cMpc})^3$, each of which can encloses the entire FMOS survey volume.  In each sub-box, we mimic our observations to construct realistic mock target catalogs.  By doing so, the spatial distribution of mock galaxies reflects the artificial biases due to the fiber allocation and inhomogeneous detection, similarly to the real data.  We use them to establish and examine the correction schemes for these biases.  Details of the construction of the mock catalogs and the correction schemes are described in Appendix \ref{sec:mock_corr}.

Furthermore, we assess the effect of stellar mass incompleteness.  For this purpose, we use the Bolshoi simulation \citep{2011ApJ...740..102K} to take the advantage of its better mass resolution than $\nu^2$GC, and the availability of merger histories of halos.  This simulation traces $2048^3$ particles in a cubic box with a side length of $250~h^{-1}~\mathrm{cMpc}$.  We use the public catalog at $a=0.38435$ ($z=1.602$) to construct mock samples.  We describe further the construction of the mock samples and the assessment of selection effects in Appendix \ref{sec:mock_am}.

\section{Clustering measurement} \label{sec:clustering}

\subsection{Two-point correlation function}
The two-point auto-correlation function is a powerful and commonly-used tool to quantify the spatial distribution of galaxies.  The real-space correlation function $\xi \left( r \right)$ measures the excess of the probability of finding pairs of galaxies as a function of their separation $r$.  As a measure of the correlation function, we use the \cite{1993ApJ...412...64L} estimator:
\begin{equation}
\xi \left(r \right) = \frac{N_R(N_R-1)}{N_D(N_D-1)}\frac{DD\left( r \right)}{RR\left( r \right)} - 2 \frac{N_R-1}{N_D}\frac{DR\left( r \right)}{RR\left( r \right)} +1,
\label{eq:LS}
\end{equation}
where $N_D$ and $N_R$ are the numbers of galaxies and random objects, respectively, $DD \left( r \right) $, $DR \left( r \right) $, $RR \left( r \right) $ are the numbers of data--data, data--random, random--random pairs with a comoving separation within the interval $\left[r, r+dr\right]$, respectively.

The radial distance computed from the redshift is different from the actual distance due to the peculiar motion of the galaxy, which results in the distortion of the correlation function measured in redshift space.  To minimize these effects, we measure the correlation function on a two-dimensional grid, parallel and perpendicular to the line of sight, and integrate along the line-of-sight direction as standard practice for clustering analysis.  We define the separation of objects following \cite{1994MNRAS.267..927F}.  Given a pair of objects at positions $\vec{r}_1$ and $\vec{r}_2$ in the redshift comoving space, the separation $\vec{s}$ and the line-of-sight vector $\vec{l}$ are defined as $\vec{s}=\vec{r}_1-\vec{r}_2$ and $\vec{l}=\left(\vec{r}_1+\vec{r}_2\right)/2$, respectively.  The parallel ($\pi$) and perpendicular ($r_\mathrm{p}$) separations are given, respectively, as
\begin{equation}
\pi \equiv | \vec{s} \cdot \vec{l} |/| \vec{l} |, \qquad r_\mathrm{p} \equiv \sqrt{ |\vec{s}|^2 - \pi^2}.
\end{equation}
The {\it projected} correlation function $w_\mathrm{p} \left( r_\mathrm{p} \right)$ is related to the real-space correlation function as follows:
\begin{equation}
w_\mathrm{p}\left(r_\mathrm{p}\right) = 2 \int_{0}^{\pi_\mathrm{max}} \xi \left( r_\mathrm{p}, \pi\right) d\pi.
\label{eq:wp}
\end{equation}
If $\pi_\mathrm{max}$ is infinity, the redshift space distortions have no effect on $w_\mathrm{p} (r_\mathrm{p})$.  In practice, it should be large enough to eliminate the effect of peculiar motions, but finite to avoid adding noise to the measurements.  We adopt $\pi_\mathrm{max} = 30~h^{-1} ~\mathrm{cMpc}$, which corresponds to $\Delta z\approx \pm0.024$ or $\Delta v\approx \pm2800~\mathrm{km~s^{-1}}$.  We count pairs of galaxies and random objects on the two-dimensional grid that is binned logarithmically in the $r_\mathrm{p}$ direction and linearly in $\pi$.  Considering the small number of galaxy pairs with a small separation ($r_\mathrm{p}\lesssim 1~h^{-1}~\mathrm{cMpc}$), we employ a set of variable-size bins that have larger widths at small scales.

\subsection{Construction of random samples} \label{sec:random}

To measure galaxy clustering with the estimator given in Equation \ref{eq:LS}, we need to construct a reference random sample that follows the same geometrical properties as the real data.  To avoid introducing shot noise, the random sample contains a large number of objects ($N_R=45000$ for our case), which is about 90 times larger than the spec-$z$ sample.  Random objects are distributed uniformly across the effective survey area, while we consider a non-uniform radial distribution to reflect the realistic redshift distribution of our sample.  

Determination of the radial distribution of observed galaxies to create the random sample represents a significant challenge.  The simplest way is to randomly assign objects the same redshifts as the sample galaxies.  However, this method is useful for only wide-field surveys that are not impacted strongly by cosmic variance.  Given the survey area, our spectroscopic redshifts may suffer from this effect as the measured redshift distribution may reflect specific structures in the galaxy distribution.  It can lead to an artificial line-of-sight clustering in the random catalog, and, subsequently, an underestimate of the correlation strength of real galaxies.  In fact, with a reference catalog constructed in this way, we find a smaller correlation length by $\sim 0.6~h^{-1}~\mathrm{cMpc}$ than our final result obtained with the random catalog constructed as described below.
 
\begin{figure}[tbph] 
   \centering
   \includegraphics[width=3.5in]{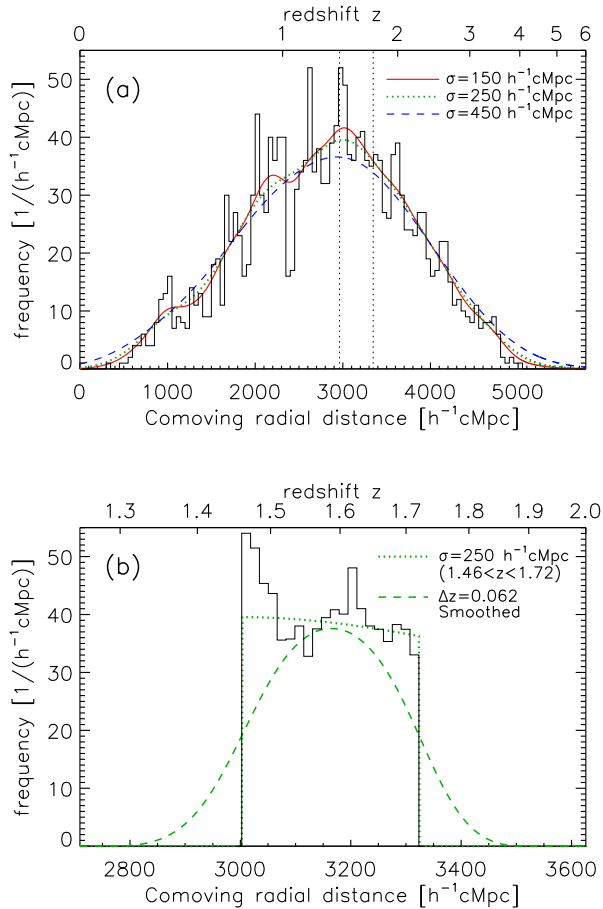} 
   \caption{Distribution of comoving radial distance (and corresponding redshift) for galaxies with $K_\mathrm{S} \le 23.5$ and $M_\ast \ge M_\ast^\mathrm{lim}$ within the COSMOS photometric catalog \citep{2013A&A...556A..55I}.  Panel (a): The solid-line histogram indicates the distribution of all galaxies.  Solid, dotted, and dashed curves show the smoothed distribution with a Gaussian kernel of $\sigma=150$, 250, and 450~$h^{-1}~\mathrm{cMpc}$, respectively.  Vertical dotted lines indicate the photometric redshift range for the FMOS--parent sample ($1.46\le z\le 1.72$).  Panel (b): A solid-line histogram and dotted curve are extracted from those in the top panel within the redshift range of $1.46 \le z \le1.72$.  The dashed curve shows the expected underlying distribution of the true redshift of our parent sample, which is determined by convolving the dotted curve with a Gaussian kernel with $\sigma_z=0.062$.}
   \label{fig:histzp}
\end{figure}

To define the underlying redshift distribution of galaxies, we use $\sim10^4$ galaxies with $K_\mathrm{S}\le 23.5$ and $M_\ast \ge M_\ast^\mathrm{lim}$ in the COSMOS photometric catalog \citep{2013A&A...556A..55I}.  Here, no limitation on the photometric redshift is applied.  Figure \ref{fig:histzp}a shows the distribution of the radial comoving distance calculated from the photometric redshift of those galaxies.  We smooth the binned distribution (histogram) with a Gaussian kernel with a standard deviation of 150, 250, or 450~$h^{-1}~\mathrm{cMpc}$.  The distribution smoothed with the shortest kernel ($150~h^{-1}~\mathrm{cMpc}$) still traces specific structures (red line in Figure \ref{fig:histzp}a).  In contrast, smoothing with the longest kernel ($450~h^{-1}~\mathrm{cMpc}$) may produce a slight artificial enhancement at both ends and suppression around the peak of the distribution (blue dashed line).  Therefore, we decided to use the distribution smoothed with an intermediate scale of $250~h^{-1}~\mathrm{cMpc}$ (green dotted line), which traces well the global shape of the distribution.  We then estimate the intrinsic distribution of true redshifts of the FMOS-parent sample by taking into account the uncertainties on the photometric redshifts.  Figure \ref{fig:histzp}b illustrates our method.  We extract the smoothed distribution between $1.46 \le z_\mathrm{phot} \le1.72$ (dotted line), and convolve it with a Gaussian function having a standard deviation $\sigma_{z_\mathrm{phot}}=0.062$, which is a typical error on the photometric redshift of our sample galaxies.  As a result, we obtain the realistic radial distribution of our parent sample (dashed line).  

\begin{figure*}[tbph] 
   \centering
   \includegraphics[width=6.5in]{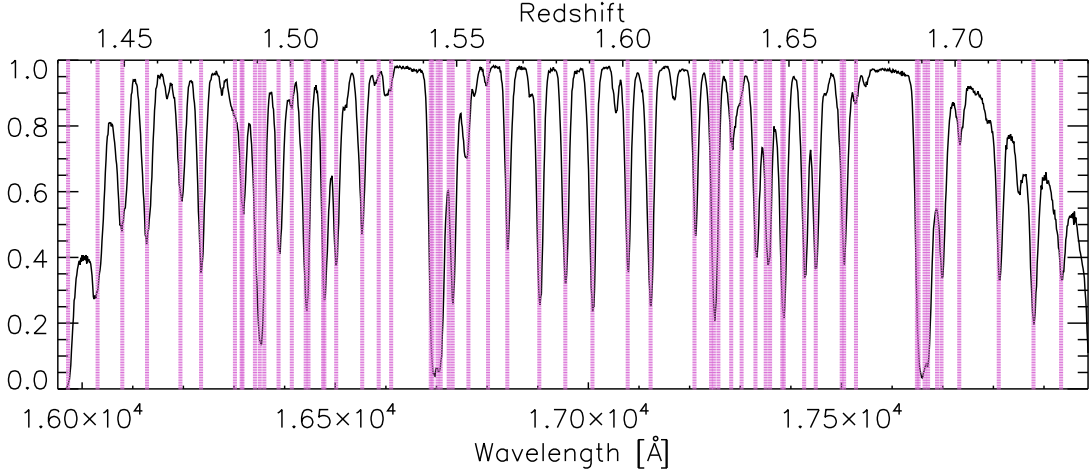} 
   \caption{Average detection rate of the H$\alpha$ emission line at each observed wavelength with corresponding redshift on the upper axis.  Magenta stripes indicate the positions of the OH airglow lines.  The detection rate decreases in the vicinity of the OH lines and towards the ends of the spectral coverage due to relatively higher noise level. }
   \label{fig:weight_z}
\end{figure*}

\begin{figure}[tbph] 
   \centering
   \includegraphics[width=3.2in]{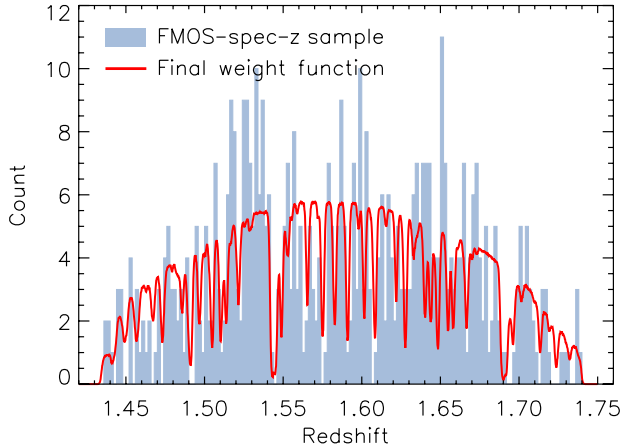} 
   \caption{Comparison between the distribution of observed spectroscopic redshift (filled histogram) and the line-of-sight weight function for the random catalog (red line).  This final weight function is the product of the smooth distribution (dashed line in Figure \ref{fig:histzp}b) and the weight for the inhomogeneous detectability of H$\alpha$ (Figure \ref{fig:weight_z}).}
   \label{fig:specz_pdz}
\end{figure}

We also consider the effects of the OH airglow mask and the inhomogeneous sensitivity.  The noise level of pixels impacted by the sky contamination is much higher than the typical level of $\sim 5\times10^{-19}~\mathrm{erg~cm^{-2}~s^{-1}\textrm{\AA}^{-1}}$  (see Figure 11 in \citealt{2015ApJS..220...12S}) thus these pixels are ignored in the emission-line fitting.  As a consequence, the detection rate decreases at the positions around the OH lines (Figure \ref{fig:histz}).  We assess the detection rate of the H$\alpha$ emission line as a function of redshift over the FMOS $H$-long spectral window by performing a set of Monte-Carlo simulations.  For each of 516 galaxies in the FMOS-spec-$z$ sample, we create artificial spectra with a multi-Gaussian profile that has the measured amplitude and line width for H$\alpha$ and [N{\sc ii}]$\lambda\lambda$6548,6583.  Gaussian noise is added to these artificial spectra based on the noise spectrum of each galaxy.  Here the noise level is intentionally increased by a factor of 1.25 to account for the fact that the noise level tends to be underestimated relative to the actual pixel variance (see \citealt{2015ApJS..220...12S}).  We then perform a fitting procedure for these artificial spectra in the same manner as the data and examine if the H$\alpha$ line is recovered.  We assess the detectability of this artificial signal at all pixels in the $H$-long window by scanning the entire spectral range.  We accept as a success the cases for which the line is detected with $S/N\ge3$ and the difference between the input and measured redshifts $\Delta z/(1+z)<0.001$.  Figure \ref{fig:weight_z} shows the average detection rate with positions of OH lines marked.  As evident, the decrease of the detection rate is seen at the positions of the OH lines.  In addition, the detectability falls down at both ends of the spectral coverage where the noise level becomes relatively high due to the instrumental characteristics.  These features are expected to introduce artificial clustering in the line-of-site direction.  This can be cancelled by using the random catalog that includes the same features.  We take into account these instrumental effects on the radial selection function in the random catalog by multiplying the intrinsic redshift distribution derived above based on the photometric catalog (dashed curve in Figure \ref{fig:histzp}) by this weight function shown in Figure \ref{fig:weight_z}.  The constructed {\it final} weight function is given in Figure \ref{fig:specz_pdz} in comparison with the distribution of observed spectroscopic redshifts.  The global trend of the data is well represented by the weight function, while some spikes in the histogram may reflect the intrinsic clustered structures of galaxies.  We demonstrate the impact of this inhomogeneous selection function and the validity of our correction scheme in Appendix \ref{sec:mock_corr}.

\subsection{Statistical error estimates} \label{sec:error}

We estimate the statistical errors of the observed correlation function using a jackknife resampling method.  We construct twenty jackknife samples of the FMOS-spec-$z$ sample as follows.  We first divide the entire survey area into four contiguous subregions of equal area (see Figure \ref{fig:field}).  The survey volume is then divided into five slices along the line of sight.  Each sub-volume has a typical transverse side length of $20~h^{-1}~\mathrm{cMpc}$ and a depth of $80~h^{-1}~\mathrm{cMpc}$.  For each jackknife sample, a sub-volume (1/20 of the full volume) is omitted in turn.  

Since each spatial structure of galaxies affects the pair counts at different separations, the values of $w_\mathrm{p} ( r_\mathrm{p} )$ at different $r_\mathrm{p}$'s are correlated.  Therefore, we need to use the full covariance matrix to fit a model to the data.  The associated covariance matrix $\bold{C}_{ij}$ is estimated from the jackknife samples as follows:
\begin{equation}
\bold{C}_{ij} =\frac{N-1}{N} \sum_{k=1}^{N} \left[ X^k_i - \left< X_i\right> \right] \left[ X^k_j - \left< X_j\right> \right],
\end{equation}
where $N=20$ is the number of jackknife samples, $X^k_i$ is the projected correlation function at the $i$-th separation measured for the $k$-th jackknife sample, and $\left<X_i \right>$ is the average of $X^k_i$ from $k=1$ to $N$.  Here we use the logarithmic values $\log w_\mathrm{p} (r_\mathrm{p})$ rather than $w_\mathrm{p} (r_\mathrm{p})$ following the suggestion of \citet{2009MNRAS.396...19N}.  The choice of logarithmic or linear value does not change our conclusions.

\begin{figure}[tbph] 
   \centering
   \includegraphics[width=3.5in]{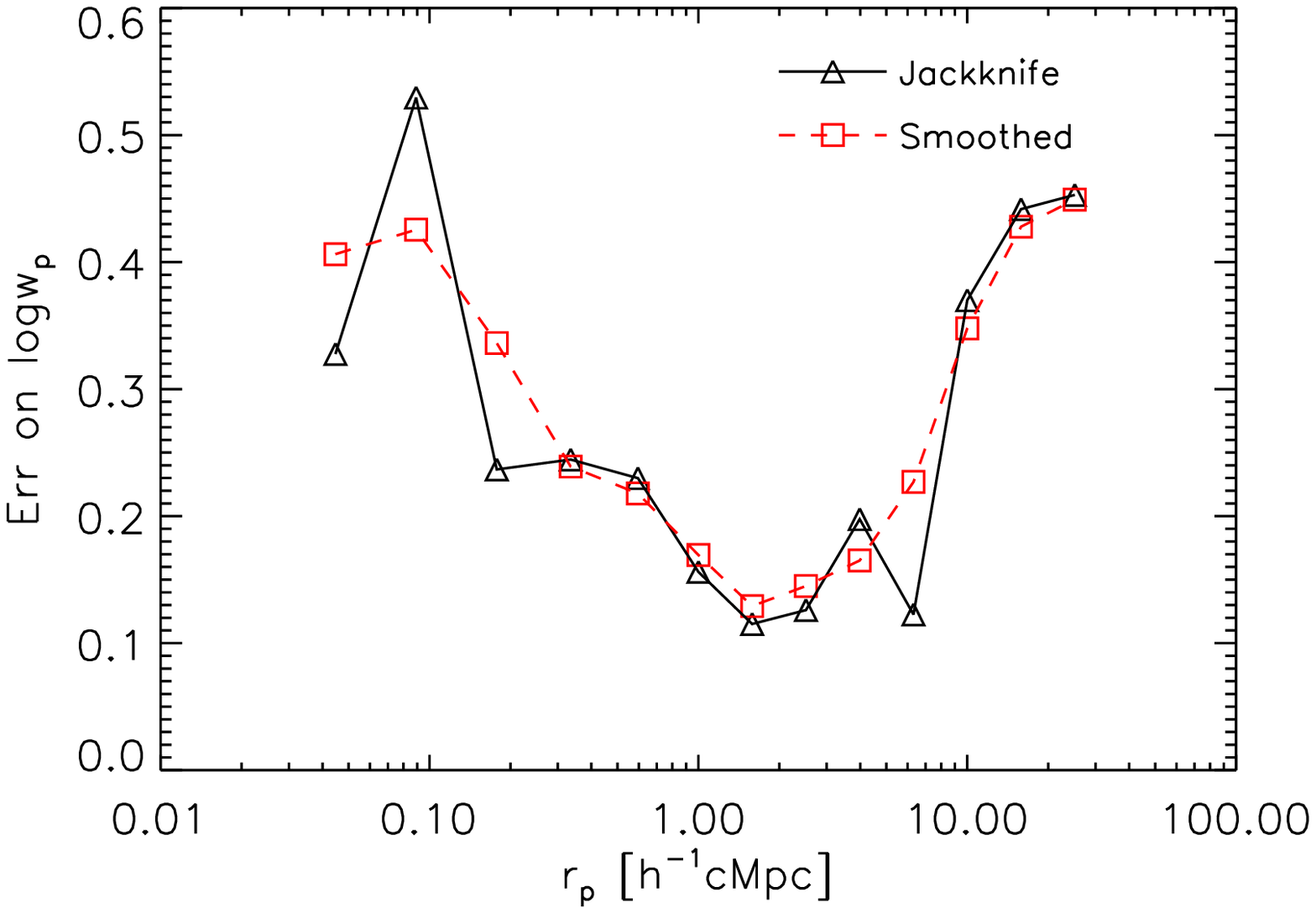} 
   \includegraphics[width=3.5in]{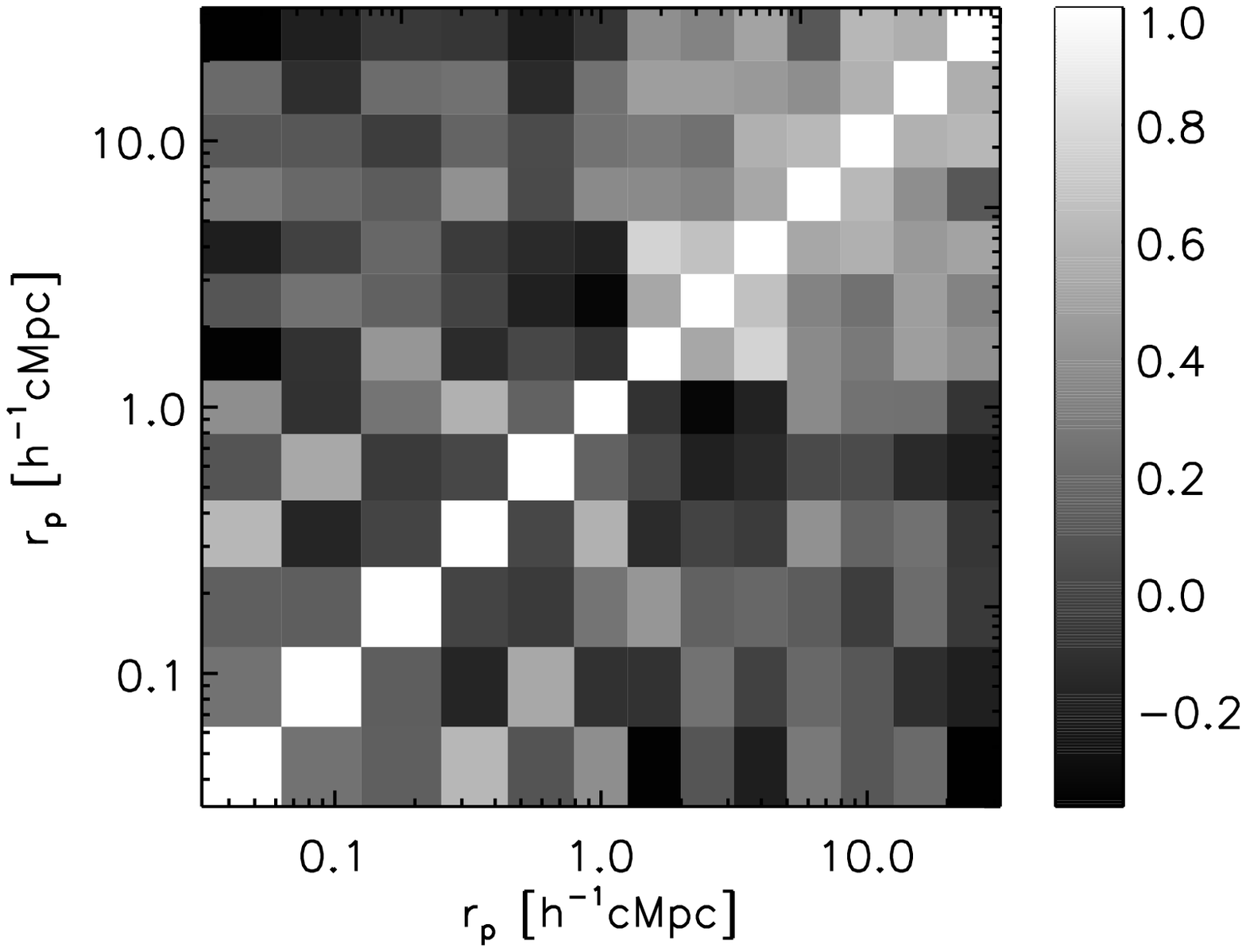} 
   \includegraphics[width=3.5in]{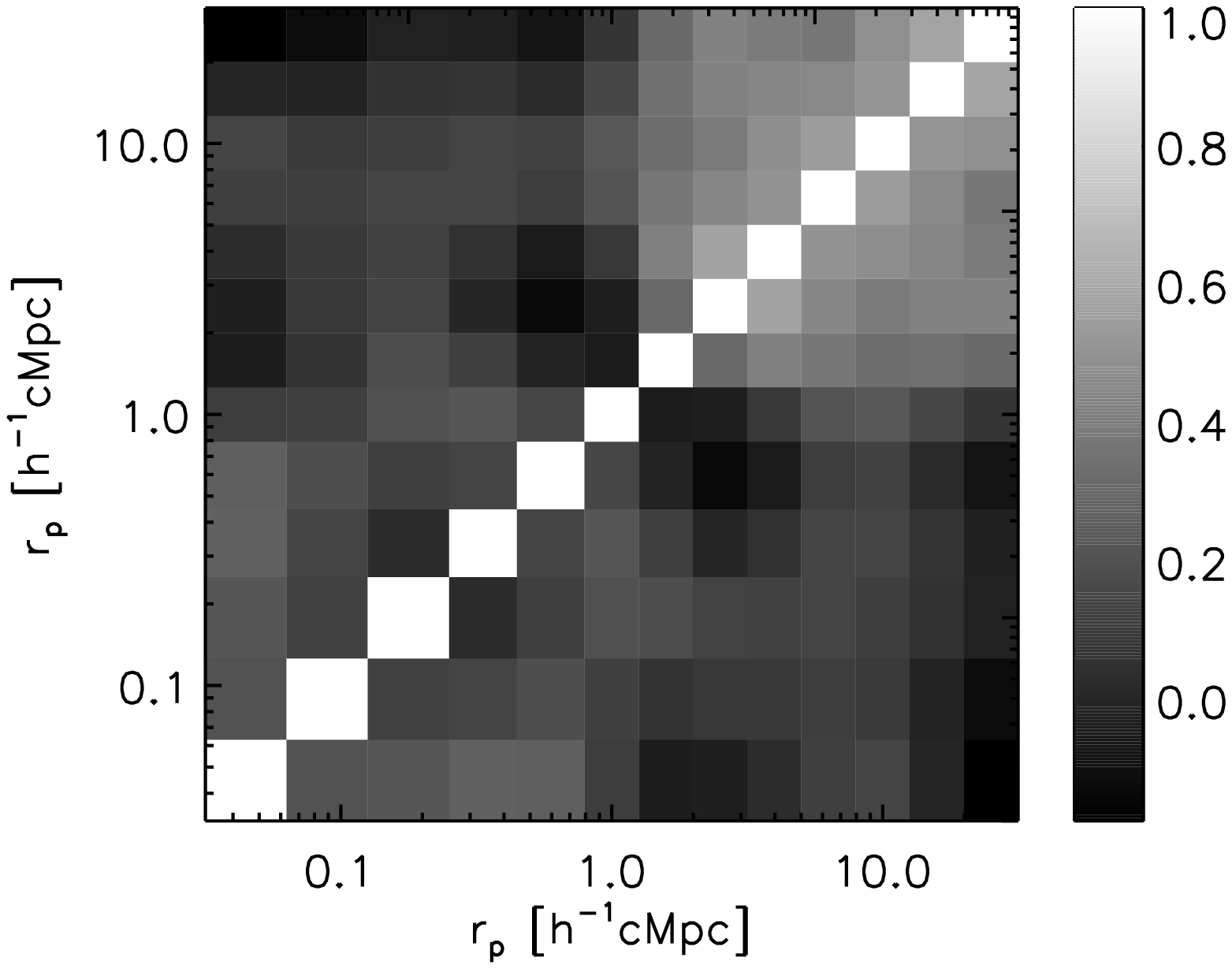} 
   \caption{Top panel: errors on $\log w_\mathrm{p}(r_\mathrm{p})$ (square-root of the diagonal term of the covariance matrix $\bold{C}_{ij}$) derived from the jackknife resampling (triangles and solid line).  The smoothed errors are indicated by squares and a dashed line.  Middle panel: jackknife correlation matrix for $\log w_\mathrm{p}$.  Bottom panel: smoothed correlation matrix.}
   \label{fig:covar}
\end{figure}

Each element of the covariance matrix has large uncertainties due to the small number of jackknife samples.  Therefore, we smooth the covariance matrix separately for the diagonal or off-diagonal elements following \citet{2013MNRAS.432.1544M}.  The diagonal elements (i.e., variance) are smoothed by a center-weighted kernel as follows:
\begin{equation}
\left(\sigma^2_i\right)^\mathrm{smooth} = (\bold{C}_{i-1,i-1}+2\bold{C}_{ii}+\bold{C}_{i+1,i+1})/4.
\end{equation}
Figure \ref{fig:covar} (top panel) shows the jackknife covariance diagonal elements and the smoothed values.  The discontinuities at $r_\mathrm{p}\approx0.15$ and $6~h^{-1}~\mathrm{cMpc}$ are effectively mitigated while the global shape is preserved.  
To smooth the off-diagonal elements, we first define the correlation matrix $\bold{R}_{ij} = \bold{C}_{ij} / \sqrt{\bold{C}_{ii} \bold{C}_{jj}}$, then we smooth it to obtain $\bold{R}^\mathrm{smooth}_{ij}$ with a $3\times3$ kernel as follows:
\begin{equation}
\frac{1}{16}
\left[\begin{array}{ccc}
1 & 2 & 1 \\
2 & 4 & 2 \\
1 & 2 & 1 \\
\end{array}\right]. 
\end{equation} 
The final smoothed covariance matrix $\bold{C}^\mathrm{smooth}_{ij}$ is calculated as follows,
\begin{equation}
\bold{C}^\mathrm{smooth}_{ij} =  \left\{
\begin{array}{ll}
\left( \sigma_i^2 \right)^\mathrm{smooth} & (i=j), \\
\bold{R}^\mathrm{smooth}_{ij} \sqrt{(\sigma^2_i)^\mathrm{smooth} (\sigma^2_j)^\mathrm{smooth}} & (i \neq j).\\
\end{array}\right.
\end{equation}
The original and smoothed correlation matrices are shown in Figure \ref{fig:covar} (middle and bottom panels).  The majority of pixel-to-pixel fluctuations are well eliminated by smoothing while the overall trend is retained.  We finally note that our conclusion does not depend on whether using the full covariance matrix or only the diagonal elements, or details of the smoothing method.

\subsection{Model fit}
To obtain physical insight, we fit a power-law model, a biased dark matter model, and an HOD model to the observed correlation function.  We use 11 data points at $-1.5<\log (r_\mathrm{p}/h^{-1}~\mathrm{cMpc})<1.1$ for the model fitting.  At larger scales ($r_\mathrm{p}\gtrsim15~h^{-1}~\mathrm{cMpc}$), the contribution from the integral constraint becomes greater than 10\% of the measured clustering amplitude (see Section \ref{sec:IC}).  

We define the posterior distribution of our model parameters to be given by
\begin{equation}
\mathcal{P} \propto \exp \left( -\frac{\chi_\mathrm{data}^2 + \chi_\mathrm{prior}^2}{2} \right).
\label{eq:likelihood}
\end{equation}
Here, $\chi_\mathrm{data}^2$ is calculated from the observed correlation function as
\begin{equation}
\chi^2_\mathrm{data} = \sum_{i=0}^N \sum_{j=0}^N \left[ X^\mathrm{mod}_i - X^\mathrm{obs}_i\right] (\bold{C}^\mathrm{smooth}_{ij})^{-1} \left[ X^\mathrm{mod}_j - X^\mathrm{obs}_j\right]
\end{equation}
where $X^\mathrm{mod}_i$ and $X^\mathrm{obs}_i$ are logarithms of the model and observed correlation functions at the $i$-the separation, $N$ is the number of data points, and $(\bold{C}^\mathrm{smooth}_{ij})^{-1}$ is the inverse of the smoothed covariance matrix defined in Section \ref{sec:error}.  To compare with the measurements based on pair counting in bins of $r_\mathrm{p}$, we calculate the average of the model $\log w_\mathrm{p}(r_\mathrm{p})$ in each $r_\mathrm{p}$ bin [$r_\mathrm{min}:r_\mathrm{max}$] as follows,
\begin{equation}
X_i^\mathrm{mod} = \log \left[ \int_{r_\mathrm{min}}^{r_\mathrm{max}} w_\mathrm{p}^\mathrm{mod} (r_\mathrm{p}) dr_\mathrm{p}  / (r_\mathrm{max}-r_\mathrm{min})\right].
\end{equation}
In Equation (\ref{eq:likelihood}), the latter term $\chi_\mathrm{prior}^2$ is the contribution from the prior imposed on the parameters. To sample the posterior distribution of our parameters, we adopt a Markov Chain Monte Carlo (MCMC) technique using the software {\it emcee} \citep{2013PASP..125..306F}. We analyze a chain of 152400 steps that follows 25400 burn-in steps to find the parameter set that provides the maximum posterior probability, and to evaluate the posterior probability distribution of each parameter.  

\section{Corrections for critical biases} \label{sec:corr}

In spectroscopic galaxy surveys, various observational effects can cause non-negligible, artificial biases.  As described in Section \ref{sec:random}, we deal with the effects of the non-uniform detection of the H$\alpha$ emission line along the redshift direction by using a modified random sample (see also Appendix \ref{sec:mock_corr}).  Here we describe our treatment for other systematic effects.

\subsection{Fiber allocation} \label{sec:fiber}
The most important issue is the impact of fiber allocation that artificially distorts the on-sky distribution of objects, if not all galaxies in the input catalog are observed.  For our FMOS observations, galaxies are selected from an input catalog by using the Echidna Spine-to-Object allocation software \citep{2008SPIE.7018E..94A} to maximize the operational efficiency.  The FMOS fibers are uniformly embedded in the field-of-view, and they can move within a limited circular patrol area of $174~\mathrm{arcsec}$ in diameter.  Once a pair of fibers is allocated for one galaxy, the opportunity for its neighboring galaxies to be observed at the same time decreases due to the lack of fibers and/or the avoidance of fiber entanglement, although the patrol areas of adjacent fibers overlap with each other.  As a consequence, the sampling rate of close galaxy pairs is suppressed at scales less or similar to the minimum separation of fibers ($\sim1.6^\prime$). In addition, the sampling rate is different across the survey area due to different number of repeated exposures covering the same footprint and the overlapping regions.  These characteristics of our observations affect the observed correlation function of galaxies and thus need to be properly removed to measure the galaxy clustering.

We correct the observed correlation function for these biases by using a simple weighting scheme, in which each galaxy-galaxy pair is weighted in response to their angular separation \citep{2011MNRAS.412..825D,2015A&A...583A.128D}.  The weight is defined as a ratio of the probability of finding pairs with a given angular separation in the input catalog to the sample of galaxies that were assigned to be observed.  The weight function can be expressed by the angular correlation function of these two samples ($\omega_\mathrm{par} \left( \theta \right)$ and $\omega_\mathrm{tar} \left( \theta \right)$, respectively) as follows:
\begin{equation}
f\left( \theta \right) = \frac{1+\omega_\mathrm{par} \left( \theta \right)}{1+\omega_\mathrm{tar}  \left( \theta \right)}.
\label{eq:weight}
\end{equation}
It may be straightforward to determine the weight function based on the real data (i.e., the FMOS-parent and FMOS-fiber-target samples).  However, the statistical uncertainties in the resulting weight function becomes large because the sample size is not sufficiently large.  Therefore, we decided to use the mock samples to avoid such large statistical uncertainties.  In Appendix \ref{sec:mock_corr}, we fully describe the construction of the mock samples and the determination of the weight function, and demonstrate the validity of our correction scheme.

\subsection{Integral constraint}
\label{sec:IC}
Due to the finite survey area, the observed correlation function is underestimated by a scale-independent constant value $C$, which is known as the integral constraint:
\begin{equation}
w_\mathrm{p} \left( r_\mathrm{p} \right) = w^\mathrm{obs}_\mathrm{p} \left( r_\mathrm{p} \right) + C.
\label{eq:wp_IC}
\end{equation}
First we calculate the integral constraint for the real-space correlation function $\xi ( r)$ following \citet{1999MNRAS.306..538R} as
\begin{equation}
\label{eq:ICxi}
C_\xi=\frac{\sum_i \xi^\mathrm{mod}\left( r_i \right) RR\left( r_i \right)}{\sum_i RR\left( r_i \right)},
\end{equation}
where $RR\left( r_i \right)$ is the number of random-random pairs whose separation is in the linearly-spaced $i$-th bin (bin size $\Delta r=1~h^{-1}~\mathrm{cMpc}$).  The summation is taken over the entire survey volume.  The relation between $C$ and $C_\xi$ is simply given by
\begin{equation}
\label{eq:IC}
C = 2\int_0^{\pi_\mathrm{max}} C_\xi d\pi  = 2C_\xi \times 30 ~h^{-1}~\mathrm{cMpc}.
\end{equation} 
We use a biased non-linear correlation function of dark matter as the model function, i.e, $\xi^\mathrm{mod} = b^2 \xi_\mathrm{dm}$.  We first evaluate $b^2$ by comparing the non-linear projected correlation function of dark matter with the observed $w^\mathrm{obs}_\mathrm{p} (r_\mathrm{p})$, and calculate the integral constraint using Equations \ref{eq:ICxi} and \ref{eq:IC}.  We then revise $b^2$, and recalculate $C$.  We find $C=0.80~h^{-1}~\mathrm{cMpc}$ for our data by repeating this process until convergence.  This value is comparable to the observed correlation amplitude at scales greater than $r_\mathrm{p} \sim 20~h^{-1}~\mathrm{cMpc}$.  Therefore, we use only measurements at scales smaller than this for analyses.

\subsection{Contamination by fake detection}
\label{sec:fake}
We correct the observed correlation function for contamination by fake sources for which a spurious signal or non-H$\alpha$ emission line is misidentified as H$\alpha$.  Such contamination is expected to reduce the amplitude of the observed correlation function since these sources are not correlated with other real galaxies.  Assuming that such sources are randomly distributed over the survey  volume, the correlation function needs to be corrected by increasing its amplitude by a factor $1/(1-f_\mathrm{fake})^2$ where $f_\mathrm{fake}$ is the fraction of fake sources in the sample.  

We assess the reliability of our redshift measurements using an independent spectroscopic survey that includes the same galaxies.  In the FMOS-spec-$z$ sample, we find 28 galaxies that have a robust redshift measurement (confidence class 3 or 4) from the zCOSMOS-deep survey \citep{2007ApJS..172...70L}.  Of these, 24 galaxies have the redshift measurements from the FMOS-COSMOS and zCOSMOS surveys that are mutually consistent ($\Delta z<0.01$), while four other objects have inconsistent measurements.  Assuming that our FMOS measurements are wrong for these objects and all 28 zCOSMOS redshifts are correct, we find a plausible fraction of fake detection to be $f_\mathrm{fake}=0.14$.  This fraction corresponds to an underestimate of the correlation function by 26\%.  We note that the four objects with a possible wrong measurement are not either revised or removed from the sample for analysis, although one of them match to the zCOSMOS measurement if we suppose that the [O{\sc iii}]$\lambda 5007$ is misidentified as H$\alpha$.

In the fitting process, we use the observed correlation function without the correction for the contamination from fake detections.  Instead, we include this effect in our analyses by handling $f_\mathrm{fake}$ as a parameter to be estimated together with other model parameters.  Namely, we compare the model multiplied by $(1-f_\mathrm{fake})^2$ to the data.  Assuming Poisson statistics, 4/28 approximately translates to $0.14\pm0.06$, which is imposed as a prior probability distribution on $f_\mathrm{fake}$.  However, the zCOSMOS-deep survey is targeting rather higher redshift galaxies at $z >2$, while having a less sampling rate at the redshift range of our FMOS sample.  Therefore, the fake fraction may be overestimated.  Considering the large uncertainty in the estimation of $f_\mathrm{fake}$, we report results obtained by using this prior on $f_\mathrm{fake}$ as fiducial, while presenting results with $f_\mathrm{fake}$ fixed to zero for reference.  In all cases, our results for these two cases are consistent at a $1\sigma$ level.  We note that this prior dominates the posterior probability distribution of $f_\mathrm{fake}$ for the power-law and biased dark matter models (see Sections \ref{sec:PL} and \ref{sec:DM}, respectively) because there is no other information that can constrain the amplitude of the correlation function. This prior on $f_\mathrm{fake}$ broadens the posterior distribution of the parameter that determines the correlation amplitude, i.e., correlation length or galaxy bias.  In contrast, in the HOD modeling (Section \ref{sec:hod}), the posterior of $f_\mathrm{fake}$ is modified from the prior as the amplitude and shape of correlation function, and the total galaxy abundance are linked through the model.

\section{Results} \label{sec:results}

\begin{figure}[tbph]
   \centering
   \includegraphics[width=3.5in]{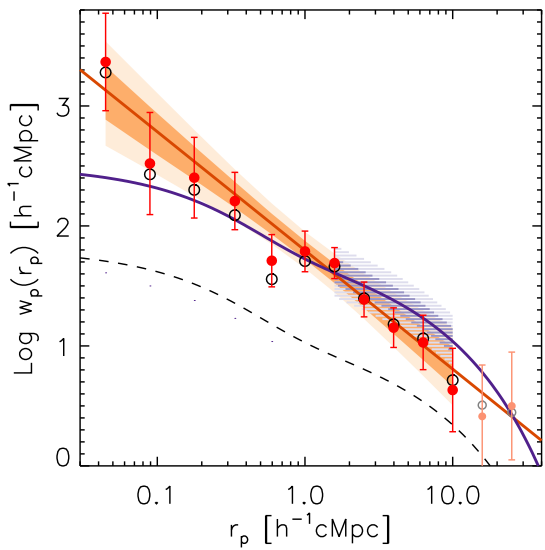} 
   \caption{The projected two-point correlation function $w_\mathrm{p} (r_\mathrm{p})$ of our galaxy sample at $1.43\le z\le 1.74$.  Filled red circles show the fiducial measurements of $w_\mathrm{p} (r_\mathrm{p})$ corrected for the fiber allocation effects, while open circles indicate the uncorrected $w_\mathrm{p} (r_\mathrm{p})$.  The error bars indicate 1-$\sigma$ uncertainties (see Section \ref{sec:error}).    The best-fit power-law and biased dark matter models are shown by orange and purple lines, respectively.  The 68\% and 95\% confidence intervals of each model are shown by dark and light shaded regions.  Note that the observed $w_\mathrm{p} (r_\mathrm{p})$ is not corrected for the suppression of the amplitude due to the fake detections, while the best-fitting models are reduced by multiplying by a factor of $(1-f_\mathrm{fake})^2$ where $f_\mathrm{fake}=0.14$.  The dashed line indicates the correlation function of dark matter ($b=1$).}
   \label{fig:wp}
\end{figure}

The projected correlation function $w_\mathrm{p} (r_\mathrm{p} )$ is computed for a sample of 516 star-forming galaxies at $1.43\le z\le 1.74$ in the central $0.81~\mathrm{deg}^2$ (effectively $0.77~\mathrm{deg}^2$) of the COSMOS field.  Figure \ref{fig:wp} presents the observed $w_\mathrm{p} (r_\mathrm{p} )$ with and without the correction for the scale-dependent effect of fiber allocation.  In both cases, the measurements are corrected for the integral constraint, but not for the effect of fake detections.  Instead, the model functions are reduced by $(1-f_\mathrm{fake})^2$ where $f_\mathrm{fake}=0.14$.  As evident, the amplitude of the correlation function without the correction for the fiber allocation effect is slightly suppressed as compared to the corrected values at small scales below $r_\mathrm{p}\sim1~h^{-1}\mathrm{Mpc}$ (see Section \ref{sec:fiber} and Appendix \ref{sec:mock_corr} for details).  We use the corrected values as the fiducial measurements throughout the paper.  The error bars indicate the standard deviation of $\log w_\mathrm{p}(r_\mathrm{p})$ that is estimated from jackknife resampling (see Section \ref{sec:error}).  

\capstartfalse
\begin{deluxetable*}{ccccc}
\tablecaption{Parameter constraints\label{tb:params}}
\tablehead{\colhead{Model}&\colhead{Params.}&\colhead{Prior\tablenotemark{a}}&\colhead{Best-fit\tablenotemark{b}}&\colhead{Best-fit ($f_\mathrm{fake}=0$)\tablenotemark{c}}}
\startdata
Power-law & $r_0/(h^{-1}~\mathrm{cMpc})$ & -- & $5.21_{-0.67}^{+0.70}$ & $4.46_{-0.56}^{+0.56}$ \\
& $\gamma$ & -- & $1.99_{-0.17}^{+0.13}$ & $1.99_{-0.21}^{+0.15} $ \\
& $f_\mathrm{fake}$ & $\ge0, G(0.14,0.06)$ & $0.14_{-0.06}^{+0.06}$ & -- \\
\hline
Dark matter & $ b$ & -- & $2.59_{-0.34}^{+0.41}$ & $2.23_{-0.34}^{+0.36}$ \\
& $f_\mathrm{fake}$ & $\ge0, G(0.14,0.06)$ & $0.14_{-0.06}^{+0.06}$ & --
\enddata
\tablenotetext{a}{Prior probability distribution for each parameter.  $G(x_\mathrm{0},\sigma)$ denotes a Gaussian function with a mean $x_0$ and standard deviation $\sigma$.}
\tablenotetext{b}{The best-fitting model parameters that gives the minimum $\chi^2$ with the broad prior on $f_\mathrm{fake}$ (see Section \ref{sec:fake}).}
\tablenotetext{c}{The best-fitting parameters for $f_\mathrm{fake}=0$.}
\end{deluxetable*}
\capstarttrue

\subsection{Power-law model} \label{sec:PL}

\begin{figure}[tbph]
   \centering
   \includegraphics[width=3.5in]{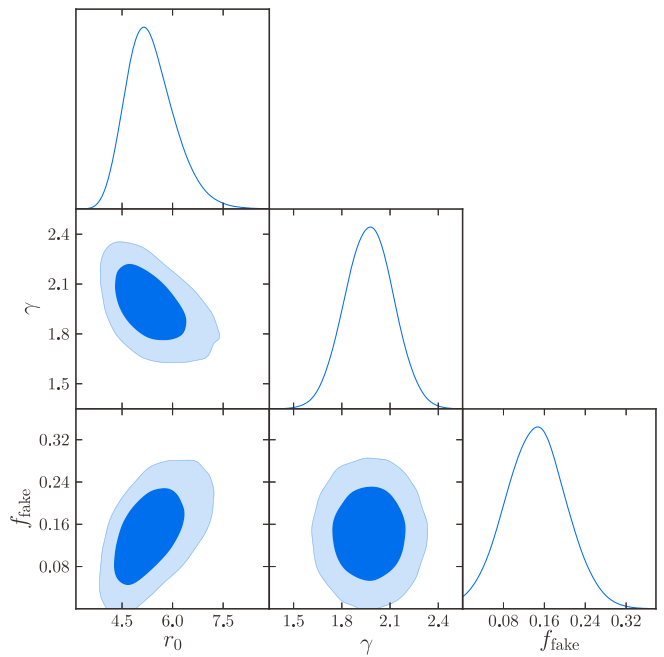} 
   \caption{Constraints on the power-law parameters ($r_0$, $\gamma$) and $f_\mathrm{fake}$.  Contours show the 68 and 95 percent confidence levels.  Solid lines show the posterior probability distribution of each parameter.}
   \label{fig:plpar}
\end{figure}

It is known that a real space galaxy correlation function $\xi \left( r \right)$ can be well described by a power-law function as $\xi \left( r \right) = (r/r_0)^{-\gamma}$\citep[e.g.,][]{1969PASJ...21..221T,2002ApJ...571..172Z}, where $r_0$ and $\gamma$ are a correlation length and power-law slope, respectively.  The correlation length denotes how strongly galaxies are clustered.  With this form of $\xi \left( r \right)$, from Equation (\ref{eq:wp}), $w_\mathrm{p} \left( r_\mathrm{p} \right)$ can be expressed as
\begin{equation}
\label{eq:pl}
w_\mathrm{p} \left( r_\mathrm{p} \right) = r_\mathrm{p} \left(\frac{r_\mathrm{p}}{r_0}\right)^{-\gamma} \frac{\Gamma \left( \frac{1}{2}\right) \Gamma \left(\frac{\gamma -1}{2}\right)}{\Gamma \left(\frac{\gamma}{2}\right)},
\end{equation}
where $\Gamma$ is Euler's Gamma function and the integration limit $\pi_\mathrm{max}$ of Equation (\ref{eq:wp}) is taken as infinity.  We fit the form of Equation (\ref{eq:pl}), multiplied by the contamination factor $(1-f_\mathrm{fake})^2$, to the observed $w_\mathrm{p} \left( r_\mathrm{p} \right)$ (see Figure \ref{fig:wp}), and search the parameter space ($r_0$, $\gamma$, $f_\mathrm{fake}$) with the MCMC procedure.  We find a correlation length to be $r_0=5.21_{-0.67}^{+0.70}~h^{-1}~\mathrm{cMpc}$ with a slope $\gamma = 1.99_{-0.17}^{+0.13}$ for our sample.  The parameter constraints are shown in Figure \ref{fig:plpar}.  The best-fit parameters and the associated uncertainties (68\% confidence intervals) are listed in Table \ref{tb:params}.  

\subsection{Biased dark matter model} \label{sec:DM}

\begin{figure}[tbph]
   \centering
   \includegraphics[width=3.5in]{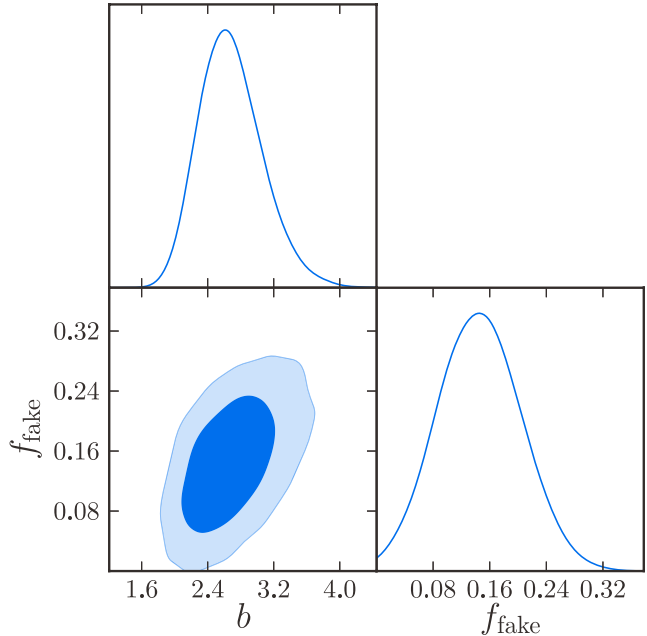} 
   \caption{Constraints on the galaxy bias $b$ and $f_\mathrm{fake}$.  Contours show the 68 and 95 percent confidence levels.  Solid lines show the posterior probability distribution of each parameter.}
   \label{fig:dmpar}
\end{figure}

The galaxy distribution is biased relative to the underlying matter distribution because galaxies form at peaks of the dark matter density fluctuations.  A correlation function of galaxies is related to that of dark matter $\xi_\mathrm{dm} (r)$ by the galaxy bias $b$ as,
\begin{equation}
\xi (r) = b^2 \xi_\mathrm{dm} (r),
\label{eq:xi_bias}
\end{equation}
where we calculate $\xi_\mathrm{dm}$ from the non-linear matter power spectrum derived by \citet{2003MNRAS.341.1311S}.  For a scale-independent bias, Equation (\ref{eq:xi_bias}) can be simply rewritten for the projected form as 
\begin{equation}
w_\mathrm{p} (r_\mathrm{p}) = b^2 w_\mathrm{p, dm} (r_\mathrm{p}),
\label{eq:wp_bias}
\end{equation}
where $w_\mathrm{p,dm}$ is the projection of $\xi_\mathrm{dm}$.  We fit the $b^2 w_\mathrm{p, dm} (r_\mathrm{p})$ to the 7--11th data points at $r_\mathrm{p}>1~\mathrm{h^{-1}cMpc}$ to avoid an enhancement of the resulting galaxy bias due to the significant one-halo term of the data.  We find $b=2.59_{-0.34}^{+0.41}$ (see Table \ref{tb:params}).  The best-fit model and the parameter constraints are shown in Figures \ref{fig:wp} and \ref{fig:dmpar}, respectively.  We note that $b=2.32_{-0.22}^{+0.25}$ is obtained when only the errors on each data point are used to calculate $\chi^2$, rather than the full covariance matrix.

\subsection{Comparisons of clustering strength}

\begin{figure*}[tbph] 
   \centering
   \includegraphics[width=7in]{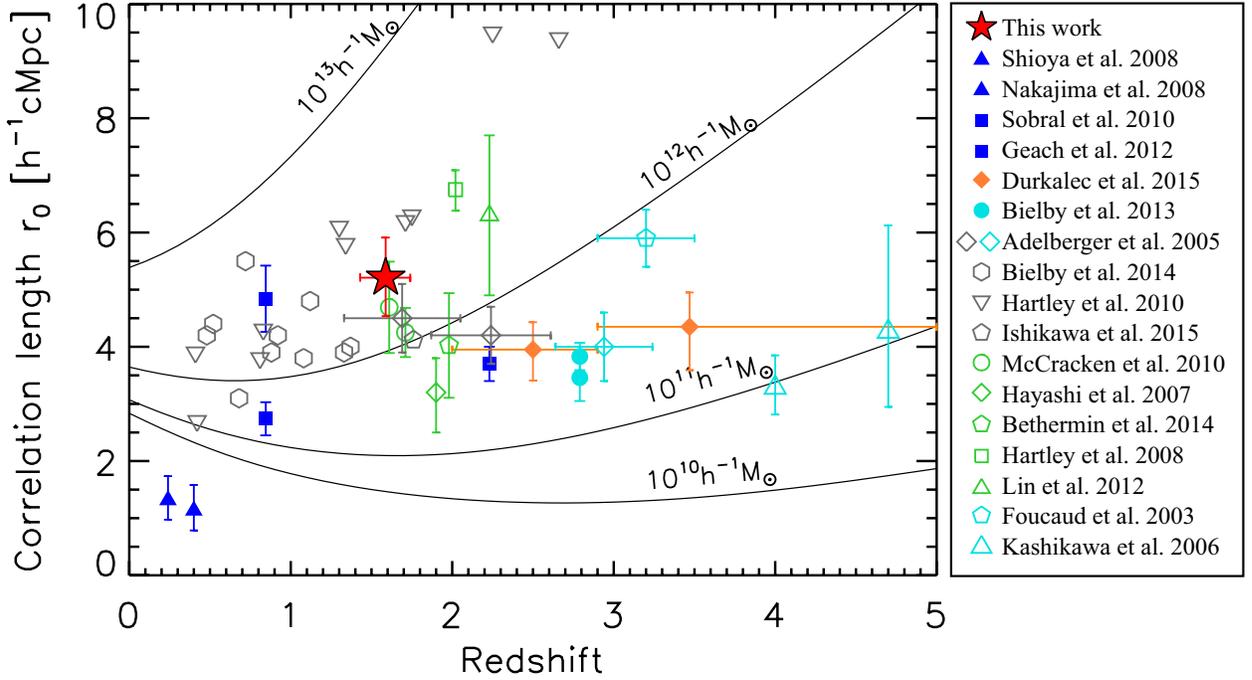} 
   \caption{Correlation length $r_0$ as a function of redshift.  The red star indicates our measurement.  Open and filled symbols show measurements from the literature based on photometric and spectroscopic (or narrow-band imaging) observations, respectively: {\it Blue} -- H$\alpha$ emitters in narrow-band surveys, {\it Orange} -- VIMOS Ultra Deep Survey, {\it Green} -- sBzK galaxies, {\it Gray} -- other color-selected star-forming galaxies, {\it Cyan} -- LBGs.  Four solid curves indicate the correlation length of dark halos of different masses, as labelled.}
   \label{fig:r0}
\end{figure*}

In Figure \ref{fig:r0}, we compare the observed correlation length ($r_0=5.21_{-0.67}^{+0.70}~h^{-1}~\mathrm{cMpc}$) with other previous measurements up to $z\sim 5$ and predictions from the dark halo model.  Most past studies presented here used samples of star-forming galaxies with stellar masses or luminosity similar to our sample (typically $M_\ast\sim10^{10}~M_\odot$ and $L_\mathrm{H\alpha}\sim10^{42}~\mathrm{erg~s^{-1}}$), thus can be straightforwardly compared to each other.

At lower redshifts, \citet{2008ApJS..175..128S} and \citet{2008PASJ...60.1249N} measured clustering of narrow-band-selected H$\alpha$-emitters (HAEs) at $z\sim 0.24$ and $z\sim 0.4$, respectively.  They found a correlation length to be $r_0=1.3~h^{-1}\mathrm{cMpc}$ ($z\sim0.24$) and $r_0=1.1~h^{-1}\mathrm{cMpc}$ ($z\sim0.4$), which are smaller than our results and others (filled triangles in Figure \ref{fig:r0}).  \citet{2008PASJ...60.1249N} mentioned that the low H$\alpha$ luminosity limits of their samples ($L_\mathrm{H\alpha}\ge10^{39.8}~\mathrm{erg~s^{-1}}$), which is lower by roughly two orders of magnitude than that of our sample, and relatively large H$\alpha$ equivalent widths (i.e., their lower stellar masses) are likely responsible for the observed weak clustering strengths.  

The HiZELS team conducted a wide-field, near-infrared narrow-band survey of HAEs.  \citet{2010MNRAS.404.1551S} presented a clustering analysis of HAEs at $z\sim0.84$ and found $r_0\sim2.7~h^{-1}\mathrm{cMpc}$ and $\sim4.8~h^{-1}\mathrm{cMpc}$ for their entire sample and a brighter subsample ($L_\mathrm{H\alpha}\ge10^{42}~\mathrm{erg~s^{-1}}$), respectively.  The latter measurement is in rough agreement with our result, as expected from the similar H$\alpha$ luminosity limit.  In addition, \citet{2012MNRAS.426..679G} measured a correlation function at $z\sim2.23$ with $\mathrm{SFR} \gtrsim 7~M_\odot~\mathrm{yr^{-1}}$, and found $r_0\sim3.7~h^{-1}~\mathrm{cMpc}$.   This slightly smaller correlation length (than ours and the HiZELS result at $z=0.84$) may be reasonable given the lower SFR limit for their sample.

There have been many studies that carry out a clustering analysis for color-selected star-forming galaxies.  \citet{2005ApJ...619..697A} measured the clustering for BM ($z\sim1.7$) and BX ($z\sim2.2$) galaxies \citep{2004ApJ...607..226A} with $r_0=4.5~h^{-1}\mathrm{cMpc}$ and $4.2~h^{-1}\mathrm{cMpc}$.  \citet{2010MNRAS.407.1212H} presented a measurement for rest-frame $U-V$ selected star-forming galaxies and found that the correlation length of subsamples at a fixed $K$-band luminosity (a proxy of stellar mass) decreases since $z\sim3$ to the present day.  Their measurements at $z\sim1.6$ are similar to ours.  \citet{2014A&A...568A..24B} used a sample of $NUV-r$ selected star-forming galaxies at $z\sim0.5$--1.75, showing $r_0\sim 4$--$5~h^{-1}~\mathrm{cMpc}$ for galaxies with $\log M_\ast / M_\odot = 9.57$--$11$.  \citet{2015MNRAS.454..205I} found $r_0=4.12\pm0.07~h^{-1}~\mathrm{cMpc}$ for $gzK$-selected star-forming galaxies with $K<23$.  
The $BzK$ color selection is known to well select star-forming (sBzK) galaxies at $1.4\lesssim z\lesssim 2.5$ \citep{2004ApJ...617..746D}.  \citet{2007ApJ...660...72H} found $r_0=3.2~h^{-1}~\mathrm{cMpc}$ for a sample of sBzK galaxies with $K<23.3$ and $\left<z\right>=1.9$.  \citet{2008MNRAS.391.1301H} measured a stronger clustering strength with $r_0\sim 6.8~h^{-1}~\mathrm{cMpc}$ and argued that contamination from highly clustered galaxies at higher redshifts ($z\gtrsim 2.5$) may be responsible for such a high amplitude.  \citet{2012ApJ...756...71L} found $r_0\sim6.3~h^{-1}~\mathrm{cMpc}$ for sBzK galaxies of $10<\log (M_\ast/M_\odot) < 10.5$.  At $z\sim 1.6$--$1.7$, \citet{2010ApJ...708..202M} measured an angular correlation function of $K_\mathrm{s}<23~(22)$ sBzK galaxies in the COSMOS field with $r_0\sim 4.25~(4.69)~h^{-1}~\mathrm{cMpc}$.  Another study in COSMOS \citep{2014A&A...567A.103B} found $r_0 = 4.0\pm0.9$ for sBzK galaxies of $M_\ast \sim 10^{10} M_\odot$.  While these measurements are slightly different from one study to another, our measurement is in rough agreement and bracketed by measurements from past studies using a sample with a range of stellar mass or luminosity similar to our sample.

At higher redshifts, there have been efforts to measure the clustering of Lyman break galaxies (LBGs) at $z\sim3$--5 \citep[e.g., ][]{2013MNRAS.430..425B,2006ApJ...637..631K,2003A&A...409..835F,2005ApJ...619..697A} or beyond \citep{2016ApJ...821..123H}, finding correlation lengths to be $r_0\sim4$--$5~h^{-1}~\mathrm{cMpc}$.  \citet{2015A&A...583A.128D} measured the projected correlation function of $\sim3000$ galaxies from the VIMOS Ultra Deep Survey, including the COSMOS field, and found $r_0\sim4~h^{-1}\mathrm{cMpc}$ at $z\sim2.5$--$5$.  These measurements are similar to those for star-forming galaxies at $z\sim1\textrm{--}2$, including our own, although the average properties of LBGs may not be identical to lower-redshift color- or H$\alpha$-selected star-forming galaxies.

Given these comparisons, we conclude that our measurement is in general agreement with other previous measurements for star-forming galaxies at similar redshifts.  In Figure \ref{fig:r0}, no clear evolutionary trend can be seen since $z\lesssim4$.  This is consistent with the moderate evolution of $r_0$ for a single population of galaxies predicted by the $\Lambda$CDM framework \citep[e.g.,][]{1999MNRAS.307..529K}.  However, it is generally difficult to compare the clustering strengths at different redshifts because measurements are based on different selection functions.  

\section{Connection between galaxies and dark matter halos}
\label{sec:halo}
Knowledge of the connection between galaxies and dark matter halos is essential to understand how and in what environments galaxy form and evolve. We investigate the properties of halos of ``main-sequence'' star-forming galaxies with $M_\ast\gtrsim10^{9.57}~M_\odot$ at $z\sim 1.6$ by interpreting the observed correlation function with an HOD model.

\subsection{The halo model and occupation distribution}
\label{sec:hod}
In the standard CDM paradigm, dark matter halos form at peaks in the matter density field. The global properties of halos such as their abundance (or the halo mass function) and large scale clustering amplitude (or the halo bias) are primarily determined by halo mass \citep{1974ApJ...187..425P,1989MNRAS.237.1127C,2002MNRAS.336..112M}.  Galaxies reside in dark matter halos. Therefore, the observable abundance and clustering of galaxies can be used to constrain the connection between galaxies and dark matter halos. 

The HOD framework is a convenient parametric way to describe the galaxy--dark matter connection to model the abundance and clustering of galaxies. In its simplest form, the HOD describes the average number of galaxies, $\avnm$, that reside in a halo of mass $M$ at redshift $z$, and assumes that this number does not depend upon the formation history and the environment of halos. For the model implemented here, we subdivide galaxies into being either central or satellite galaxies \citep[e.g.,][]{2005ApJ...633..791Z}, depending upon their location within their dark matter halos, such that
\begin{equation}
    \avnm = \avncm + \avnsm\,.
\end{equation}
The average number density of galaxies is then simply given by
\begin{equation}
    \ntot = \int \avnm n(M, z) \drm M \,,
\label{eq:ntot}
\end{equation}
where the halo mass function, $n(M, z)\drm M$, gives the number density of halos of mass $M\pm \drm M/2$ at redshift $z$.  The average number density of central ($n_\mathrm{cen}$) or satellite galaxies ($n_\mathrm{sat}$) is calculated by replacing $\left< N | M,z\right> $ in Equation (\ref{eq:ntot}) with $\left< N_\mathrm{cen} | M,z\right> $ or $\left< N_\mathrm{sat} | M,z\right> $.  For calculations, we fix the redshift to the median of the sample ($z=1.588$).  Hereafter, the variable $z$ is omitted from equations.

The clustering of galaxies can be quantified as the excess probability over the random case of finding two galaxies separated by a distance and is described by the two-point correlation function $\xi$. The clustering of galaxies arises from a combination of two contributions.  The one-halo term corresponds to the pairs of galaxies that reside within the same halo and the two-halo term to the pairs of galaxies which reside in distinct halos:
\begin{eqnarray}
\xi(r) = \xi^\mathrm{1h} (r)+\xi^\mathrm{2h} (r),
\label{eq:xi}
\end{eqnarray}
where the superscripts ``1h'' and ``2h'' stand for the one-halo and the two-halo terms, respectively. The correlation function $\xi(r)$ and the power spectrum $P(k)$ form a Fourier transform pair such that
\begin{equation}
\xi(r) = \frac{1}{2\pi^2} \int_0^\infty \drm k k^2 P(k) \frac{\sin kr}{kr}\,,
\end{equation}
which implies that Equation (\ref{eq:xi}) can also be written as
\begin{equation}
P(k) = P^{1h} (k) + P^{2h} (k).
\end{equation}

The one-halo term can be further expressed as the sum of contributions from the central-satellite and satellite-satellite galaxy pairs hosted by the same halo:
\begin{eqnarray}
& P^\mathrm{1h}(k) = \frac{1}{\ntot^2} \int  n\left(M\right)\drm M\\
& \left[ 2\left< N_\mathrm{cen} |M\right> \left< N_\mathrm{sat} |M\right> u (k, M)  + \left< N_\mathrm{sat}|M \right>^2 u^2 (k, M) \right],
\label{eq:1hterm}
\end{eqnarray}
where $u(k, M)$ describes the Fourier transform of the density profile of
satellite galaxies within dark matter halos. We assume that central galaxies
reside at the center of halos and that the occupation numbers of centrals and
satellites are independent of each other, so that $\left<N_\mathrm{cen}
N_\mathrm{sat} \right> = \left<N_\mathrm{cen} \right>\left< N_\mathrm{sat}
\right>$. We have further assumed that $N_\mathrm{sat}$ follows Poisson
statistics, so that $\left<N_\mathrm{sat} (N_\mathrm{sat} -1)\right> =
\left<N_\mathrm{sat} \right>^2$.  This is supported both by observations
\citep{2008ApJ...676..248Y} and numerical simulations
\citep{2004ApJ...609...35K}. 

The two-halo term consists of contributions from the central--central, central--satellite, and satellite--satellite galaxy pairs hosted by distinct halos:
\begin{eqnarray}
P^\mathrm{2h}(k) = \frac{1}{\ntot^2} \int \drm M_1 \int \drm M_2 ~ n\left(M_1\right) n\left(M_2\right) \\
\left[ \left< N_\mathrm{cen}|M_2 \right> \left< N_\mathrm{cen}|M_2 \right> + 2 \left< N_\mathrm{cen}|M_1\right> \left< N_\mathrm{sat}|M_2\right> u (k, M_2) \right. \\
\left. + \left< N_\mathrm{sat}|M_1 \right> \left< N_\mathrm{sat}|M_2 \right> u(k, M_1)u(k, M_2) \right] P_\mathrm{hh} (k|M_1, M_2),
\label{eq:2hterm}
\end{eqnarray}
where $P_\mathrm{hh} (k|M_1, M_2)$ describes the cross power spectrum of halos of masses $M_1$ and $M_2$. Following \citet{2013MNRAS.430..725V}, we express it as a product of the large scale bias of halos of masses $M_1$ and $M_2$, and the non-linear matter power spectrum \citep{2003MNRAS.341.1311S}, and account for the radial dependence of the bias as well as for halo exclusion.

We assume that the HOD of central galaxies is given by 
\begin{equation}
\left< N_\mathrm{cen}|M \right> = \frac{1}{2} \left[ 1 + \mathrm{erf} \left( \frac{\log M  - \log M_\mathrm{min}}{\sigma_{\log M}} \right) \right],
\label{eq:cen}
\end{equation}
and that of satellite galaxies is given by
\begin{equation}
\left< N_\mathrm{sat}|M \right> = \frac{1}{2} \left[ 1 + \mathrm{erf} \left( \frac{\log (M/M_\mathrm{min})}{\sigma_{\log M}} \right) \right] \left( \frac{M-M_\mathrm{cut}}{M_1^\prime} \right)^\alpha.
\label{eq:sat}
\end{equation}
For $M<M_\mathrm{cut}$, $\left< N_\mathrm{sat}|M \right>=0$.  The parameter $M_\mathrm{min}$ is a halo mass above which a halo has a central galaxy.  This transition is relaxed with a smoothing scale $\sigma_{\log M}$ \citep{2005ApJ...633..791Z,2007ApJ...667..760Z}.  The average number of satellite galaxies increases with increasing halo mass by a power-law parametrized by a slope $\alpha$ and normalization $M_1^\prime$, which is related to the halo mass ($M_1$) at which a halo is expected to have a single satellite galaxy, as $M_1=M_1^\prime+M_\mathrm{cut}$. 

For computing all observables, we use the halo mass function from \cite{2010ApJ...724..878T}, which defines a halo as a spherically-collapsed region with an average density 200 times greater than the background matter density of the Universe, and a large-scale halo bias proposed by \cite{2010ApJ...724..878T} with an empirical radial scale dependence derived by \citet{2012ApJ...745...16T} with corrections described in \citet{2013MNRAS.430..725V}. In addition, we assume that the radial distribution of satellite galaxies follows the density distribution of dark matter in halos.  We use the Navarro-Frenk-White profile \citep{1997ApJ...490..493N} with the mass-concentration relation calibrated by \cite{2007MNRAS.378...55M} for this purpose.

Once a set of HOD parameters is given, the following physical quantities are inferred:
\begin{itemize}
\item Effective large scale bias
\begin{equation}
b_\mathrm{eff} = \frac{1}{\ntot} \int b_h (M) \left< N | M\right> n \left(M\right) \drm M 
\label{eq:beff}
\end{equation}
\item Satellite fraction
\begin{equation}
f_\mathrm{sat} = \frac{1}{\ntot} \int \left< N_\mathrm{sat} | M\right> n \left(M \right) \drm M
\label{eq:fsat}
\end{equation}
\item Effective halo mass
\begin{equation}
M_\mathrm{eff} = \frac{1}{\ntot} \int M \left< N | M\right> n \left(M\right) \drm M 
\label{eq:Meff}
\end{equation}
\end{itemize}
The effective large scale bias is the number-weighted average of the halo bias $b_h (M)$.  The effective halo mass is the number-weighted average mass of halos that host galaxies in the sample.  

\subsection{Limitations for the HOD parameters}
\label{sec:prior}

We sample the posterior distribution of the HOD parameters given the abundance and clustering measurements using a MCMC technique. While the HOD model defined above has five parameters, our data do not have enough statistics to constrain all the parameters simultaneously.  We here describe the prior limitations imposed on some model parameters to resolve degeneracies and to avoid overfitting.

We fix the power-law slope $\alpha$ in Equation (\ref{eq:sat}) to be 1, which is supported observationally \citep{2005ApJ...633..791Z} and theoretically \citep{2004ApJ...609...35K}, and has been commonly applied in past studies \citep[e.g.,][]{2006ApJ...647..201C}.   We further impose a relation between $M_\mathrm{cut}$ and $M_1^\prime$ as
\begin{equation}
\log M_\mathrm{cut}/(h^{-1}M_\odot) = 0.76 \log M_1^\prime/(h^{-1}M_\odot) + 2.3,
\label{eq:Mcut}
\end{equation}
following \citet{2006ApJ...647..201C}.  In addition, we impose priors on $\sigma_{\log M}$ independently by using a stellar-to-halo mass relation derived by \cite{2013ApJ...770...57B} with the uncertainties on the stellar mass estimate taken into account (see Appendix \ref{sec:lim_sigmalogM} for details).  The prior probability distribution is given by
\begin{equation}
P(\sigma_{\log M}) \propto \left\{
\begin{array}{ll}
\exp \left[ -\frac{(\sigma_{\log M}-0.24)^2}{2\times 0.03^2} \right] & \textrm{for }\sigma_{\log M} \ge 0, \\
  0 & \textrm{for }\sigma_{\log M} < 0.
\end{array}\right.
\label{eq:sigma_logM}
\end{equation}
The prior on the contamination fraction $f_\mathrm{fake}$ is given as $f_\mathrm{fake}=0.14\pm-0.06$ (see Section \ref{sec:fake}).  We use uniform non-informative priors on the halo mass parameters $\log M_\mathrm{min} / (h^{-1}M_\odot)$ and $\log M_1^\prime / (h^{-1}M_\odot)$, in the range $[9, 15]$.

With the HOD model parametrized by Equations (\ref{eq:cen}--\ref{eq:sat}), the number density of all galaxies with $M_\ast \ge M_\ast^\mathrm{lim}$, i.e., the $M_\ast$-selected sample, can be predicted using Equation (\ref{eq:ntot}), and compared with the observed number density of $9.29\times 10^{-3} ~h^{-3}\mathrm{cMpc}^{-3}$ (see Table \ref{tb:samples}).  We emphasize that this is not the number density of the FMOS-parent sample with additional selection based on $K_\mathrm{S}$ and predicted H$\alpha$ flux. In doing so, we have implicitly assumed that our FMOS-spec-$z$ sample is representative of the $M_\ast$-selected sample (see Section \ref{sec:samples}).  We estimate the cosmic variance of the number density by using the subhalo mock catalog from $\nu^2$GC simulation and find approximately 10\% fluctuation from one to another sub-box with the same volume as our survey.  Therefore, we use the observed abundance to be $n_\mathrm{tot} = (9.29 \pm 0.93) \times 10^{-3} ~h^{-3}\mathrm{cMpc}^{-3}$ as an additional independent observable to be compared to the predictions in the MCMC procedure.  The prior information is summarized in Table \ref{tb:params}.

\subsection{HOD model fit} \label{sec:halomodelfit}

\begin{figure}[tbph] 
   \centering
   \includegraphics[width=3.5in]{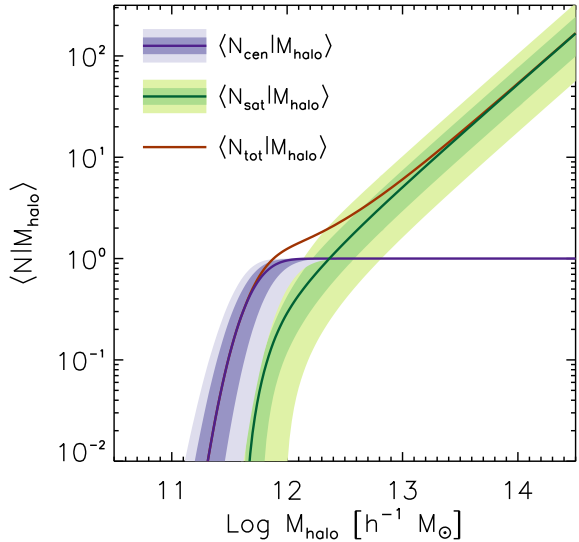} 
   \caption{Halo occupation distribution for our sample.  Purple, green and dark red lines show the average numbers of central, satellite, and all galaxies in a halo as a function of halo mass, respectively.  Dark and light shaded regions show the 68 and 95 percent confidence intervals, respectively.}
   \label{fig:hod_n}
\end{figure}

We compare the projected correlation function computed from the HOD model parameters to the observed $w_\mathrm{p} ( r_\mathrm{p} )$. We have varied three HOD parameters ($\log M_\mathrm{min}$, $\sigma^2_{\log M}$, $\log M^\prime_1$), while using 11 data points at $-1.5<\log (r_\mathrm{p}/h^{-1}~\mathrm{cMpc})<1.1$.  As a result, we have 10 degrees of freedom in this analysis:
\begin{eqnarray}
\mathrm{d.o.f} = 10 &=& 11 \left[ \text{data points of } w_\mathrm{p} (r_\mathrm{p})\right] \\
	&& +1 \left[ \text{observed constraint on }n_\mathrm{tot}\right] \\
	&& +2 \left[ \text{priors on }\sigma_{\log M}, f_\mathrm{fake}\right] \\
	&& -4 \left[ \text{parameters: } M_\mathrm{min}, \sigma^2_{\log M}, M_1^\prime, f_\mathrm{fake} \right].
\end{eqnarray}

\begin{figure*}[tbph] 
   \centering
   \includegraphics[width=5in]{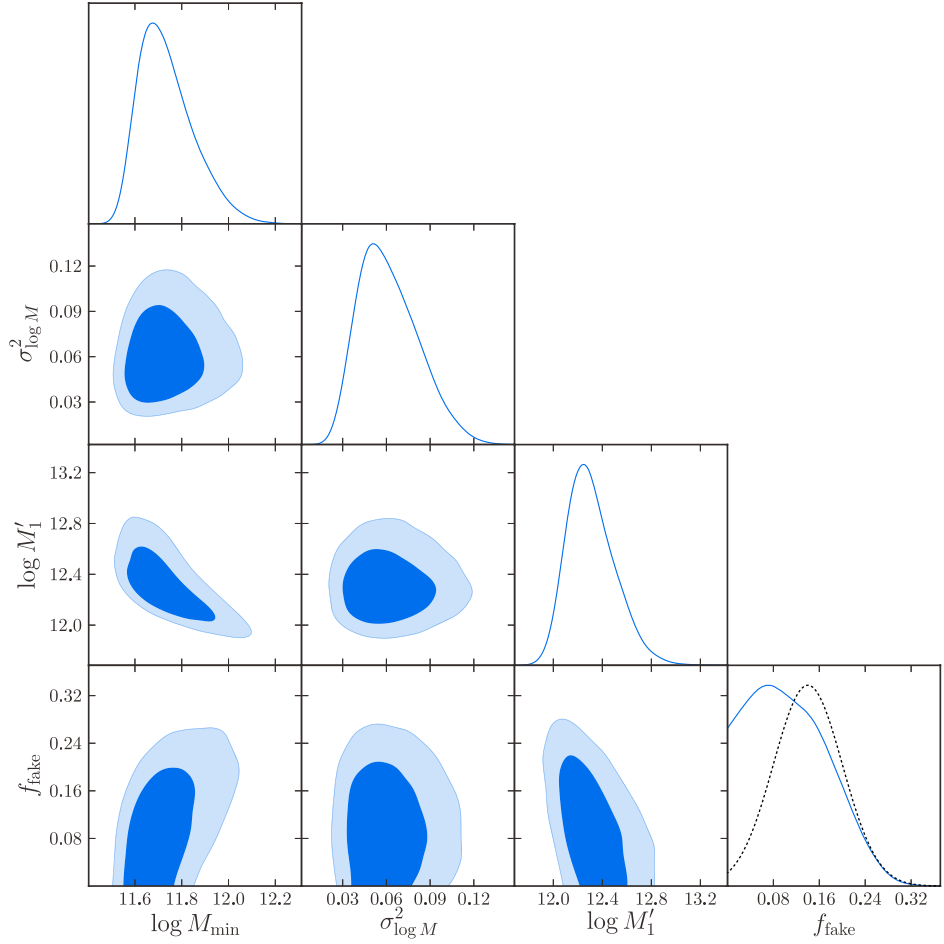} 
   \caption{Constraints of the HOD parameters ($\log M_\mathrm{min}$, $\sigma^2_{\log M}$, $\log M^\prime_1$) and $f_\mathrm{fake}$.   Contours show the 68 and 95 percent confidence levels.  Solid lines show the posterior probability distribution of each parameter.   A dashed line indicates the prior probability distribution of $f_\mathrm{fake}$.}
   \label{fig:hod_params}
\end{figure*}

\begin{figure}[tbph] 
   \centering
   \includegraphics[width=3.3in]{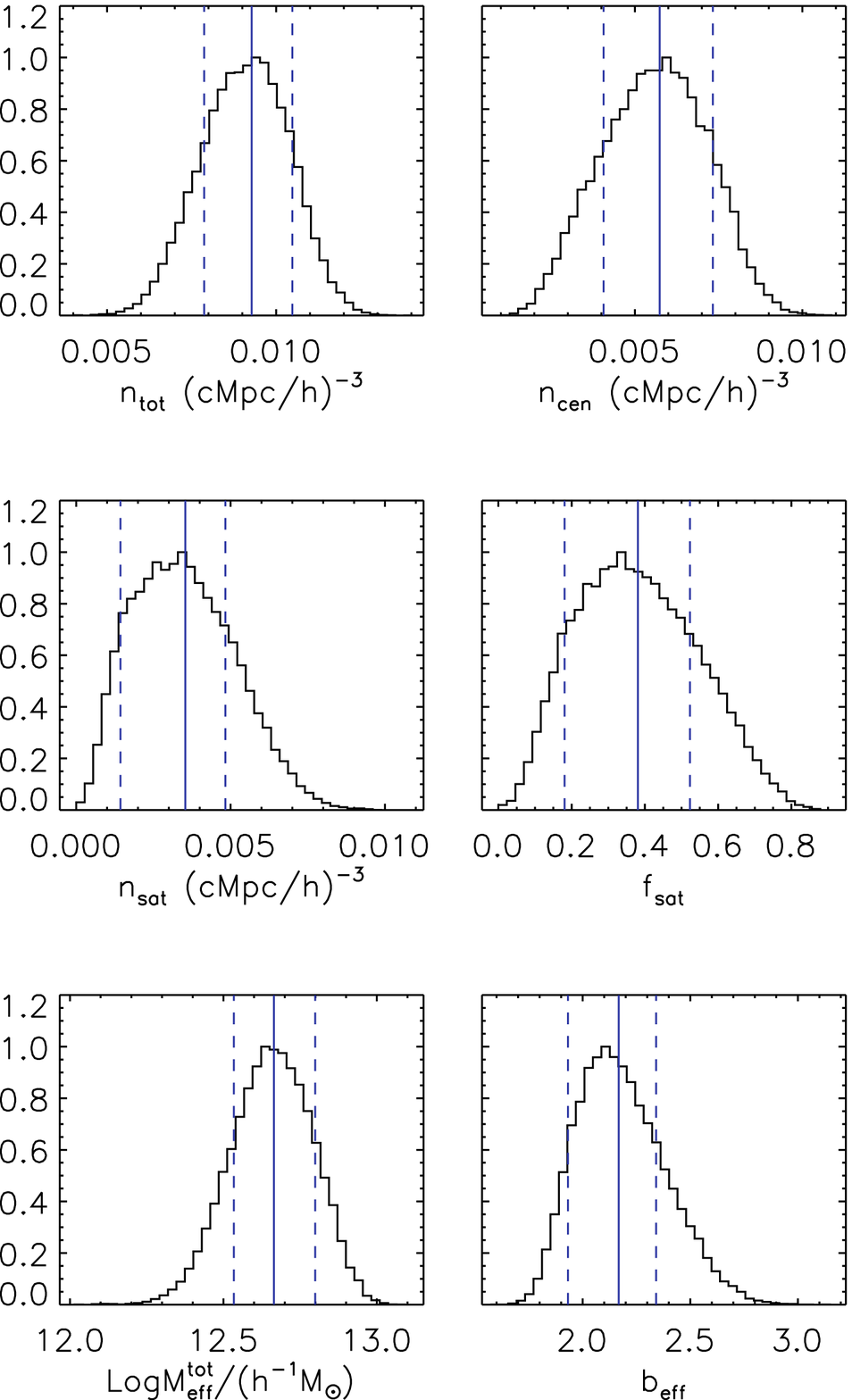} 
   \caption{Posterior distribution of the inferred physical quantities.  From the top to bottom, the average total number density ($n_\mathrm{tot}$), the average number density of central ($n_\mathrm{cen}$) and satellite galaxies ($n_\mathrm{sat}$), the satellite fraction ($f_\mathrm{sat}$), the effective halo mass ($M_\mathrm{eff}$), and the effective large scale bias ($b_\mathrm{eff}$) are shown.  Vertical solid and dashed lines indicate the best-fit value giving a minimum $\chi^2$ and the 68 percent confidence interval, respectively.}
   \label{fig:hod_phys}
\end{figure}

The three HOD parameters can be effectively constrained using our observables. The posterior distribution of the HOD and the parameters are shown in Figures \ref{fig:hod_n} and \ref{fig:hod_params}, respectively. In addition, we have also calculated various physical quantities using Equations (\ref{eq:ntot}) and (\ref{eq:beff}--\ref{eq:Meff}).  The best-fit parameters and the inferred quantities are summarized in Table \ref{tb:params}.  The posterior probability distributions are shown in Figure \ref{fig:hod_phys}.  It can been seen that $M_\mathrm{min}$ is degenerate with $M_1^\prime$.  This negative correlation is mainly caused by the prior constraint on the total abundance of galaxies because a smaller $M_\mathrm{min}$ leads to a more abundant number of centrals and thus $M_1^\prime$ needs to increase to reduce the number of satellites and vice versa.  It also seems that the data prefers a lower value for the fraction of fake detections $f_\mathrm{fake}$ compared to the prior (shown by dotted line in Figure \ref{fig:hod_params}).  We note that the results for $f_\mathrm{fake}=0$ are in agreement with the fiducial analysis within 1$\sigma$ (the rightmost column in Table \ref{tb:params}).  For the remainder of the paper, we use the results from the fiducial prior on $f_{\rm fake}$.

\begin{figure}[tbph] 
   \centering
   \includegraphics[width=3.5in]{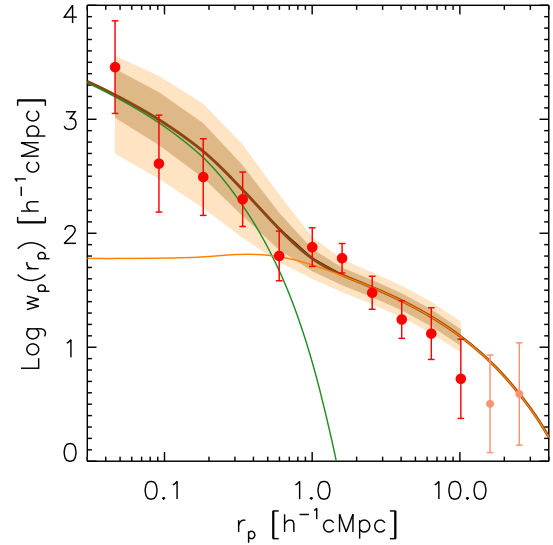} 
   \caption{Observed projected correlation function (filled circles) and the model from the best-fitting HOD (thick solid line).  Green and orange thin solid curves show the one-halo and two-halo terms, respectively.  Dark and light shaded regions indicate the 68 and 95 percent confidence intervals. The data points are corrected for the effect of fake detections with the best-fit $f_\mathrm{fake}=0.099$.}
   \label{fig:hod_wp}
\end{figure}

In Figure \ref{fig:hod_wp}, we show the observed $w_\mathrm{p} (r_\mathrm{p})$ and the model computed from the best-fitting HOD parameters.  The observed $w_\mathrm{p}(r_\mathrm{p})$ is corrected for the effect of fake detections with the best-fit value of $f_\mathrm{fake}=0.099$.  The data points are well fit with the model with a clear signature of the one-halo term.  We find that the transition from the one-halo-dominated regime at small scales to the two-halo-dominated regime at large scales happens at $r_\mathrm{p} \approx 0.5~h^{-1}~\mathrm{cMpc}$.  Since the two-halo term dominates at scales larger than the typical virial radius of largest halos, this scale is a useful probe of the virial size of halos that host the galaxies.  For SDSS galaxies ($z\lesssim0.1$), such a transition happens around $r_\mathrm{p}=1\textrm{--}2~h^{-1}~\mathrm{cMpc}$ \citep{2004ApJ...608...16Z}.  In contrast, \citet{2007ApJ...667..760Z} measured the one-to-two halo transition at $r_\mathrm{p}\sim0.4\textrm{--}0.6~h^{-1}~\mathrm{cMpc}$ at $z\sim1$, which is in good agreement with our result.  In addition, \cite{2015A&A...581A..56G} found an excess of the number of satellite galaxies up to $\sim300~\mathrm{physical~kpc}$ away from the halo centre at a similar redshifts ($z\sim1.8$) through image stacking.  This putative halo radius based on an independent technique is fully consistent with our measurement.  These results show that the transition scale decreases for high redshift galaxies and this trend likely reflects the fact that massive clusters have less formed yet at earlier epochs.   

\capstartfalse
\begin{deluxetable*}{cccc}
\tablecaption{Priors and constraints of the HOD parameters\label{tb:HODparams}}
\tablehead{\colhead{Parameters}&\colhead{Prior\tablenotemark{a}}&\colhead{Best-fit}&\colhead{Best-fit ($f_\mathrm{fake}$=0)}}
\startdata
 $\chi^2/\nu$ & -- & 1.23 & 1.01 \\
$\log M_\mathrm{min}/(h^{-1}M_\odot)$ & $U(9,15)$  & $11.71_{-0.12}^{+0.11}$ & $11.65_{-0.10}^{+0.09}$ \\
 $\sigma^2_{\log M}$ & $\ge0, G(0.24,0.03)$ for $\sigma_{\log M}$& $0.057_{-0.021}^{+0.020}$ & $0.057_{-0.024}^{+0.022}$ \\
 $\log M_1^\prime/(h^{-1}M_\odot)$ & $U(9,15) $ & $12.28_{-0.20}^{+0.18}$  & $12.40_{-0.23}^{+0.22}$  \\
$f_\mathrm{fake}$ & $\ge0, G(0.14,0.06)$ &  $0.099_{-0.092}^{+0.053}$ & --- \\
\hline
\hline
 Inferred quantities & Prior & Best-fit  & Best-fit ($f_\mathrm{fake}=0$)\\
 \hline
 $ n_\mathrm{tot}/(h^{-1} \mathrm{cMpc})^3 $\tablenotemark{b} & $>0, G(9.29\,0.93)$ & $9.27_{-1.40}^{+1.21}\times 10^{-3}$  & $9.22_{-1.56}^{+1.49}\times 10^{-3}$  \\
 $ n_\mathrm{cen}/(h^{-1} \mathrm{cMpc})^3 $ & -- & $5.74_{-1.67}^{+1.59}\times 10^{-3}$ & $6.70_{-1.68}^{+1.61}\times 10^{-3}$ \\
 $ n_\mathrm{sat}/(h^{-1} \mathrm{cMpc})^3 $ & -- & $3.53_{-2.10}^{+1.30}\times 10^{-3}$ & $2.52_{-1.94}^{+0.94}\times 10^{-3}$\\
 $ f_\mathrm{sat}$ & -- & $ 0.38_{-0.20}^{+0.14}$ & $0.27_{-0.19}^{+0.10}$ \\
 $ M_\mathrm{eff}/(h^{-1}M_\odot)$ & -- & $4.62_{-1.63}^{+1.14} \times 10^{12}$ & $3.78_{-1.45}^{+0.95} \times 10^{12}$\\
 $ b_\mathrm{eff} $ & -- & $2.17_{-0.24}^{+0.17}$ & $2.04_{-0.22}^{+0.14}$ \\
 $ \log M_1 / M_\mathrm{min}$ & -- & $0.66_{-0.27}^{+0.30}$ & $0.83_{-0.26}^{+0.31}$
\enddata
\tablenotetext{a}{Prior probability distribution of each parameter.  $U(x_1, x_2)$ is a non-informative prior with a interval [$x_1$, $x_2$].  $G(x_\mathrm{0},\sigma)$ is a Gaussian function with a mean $x_0$ and standard deviation $\sigma$.}
\tablenotetext{b}{Observed number density of the $M_\ast$-selected sample.  This is independently constrained from the photometric catalog.}
\end{deluxetable*}
\capstarttrue

\section{Discussion}
In this section, we discuss the properties of halos that contain star-forming galaxies at $z\sim1.6$ based on the physical parameters inferred from the HOD modeling of the observed correlation function, and then present a new constraint on the stellar-to-halo mass relation.  At the end, we demonstrate the capabilities of a future survey with a next-generation multi-object spectrograph.
\label{sec:discussion}

\subsection{Halo mass and large-scale bias} \label{sec:halomass}

\begin{figure*}[tbph] 
   \centering
   \includegraphics[width=5.5in]{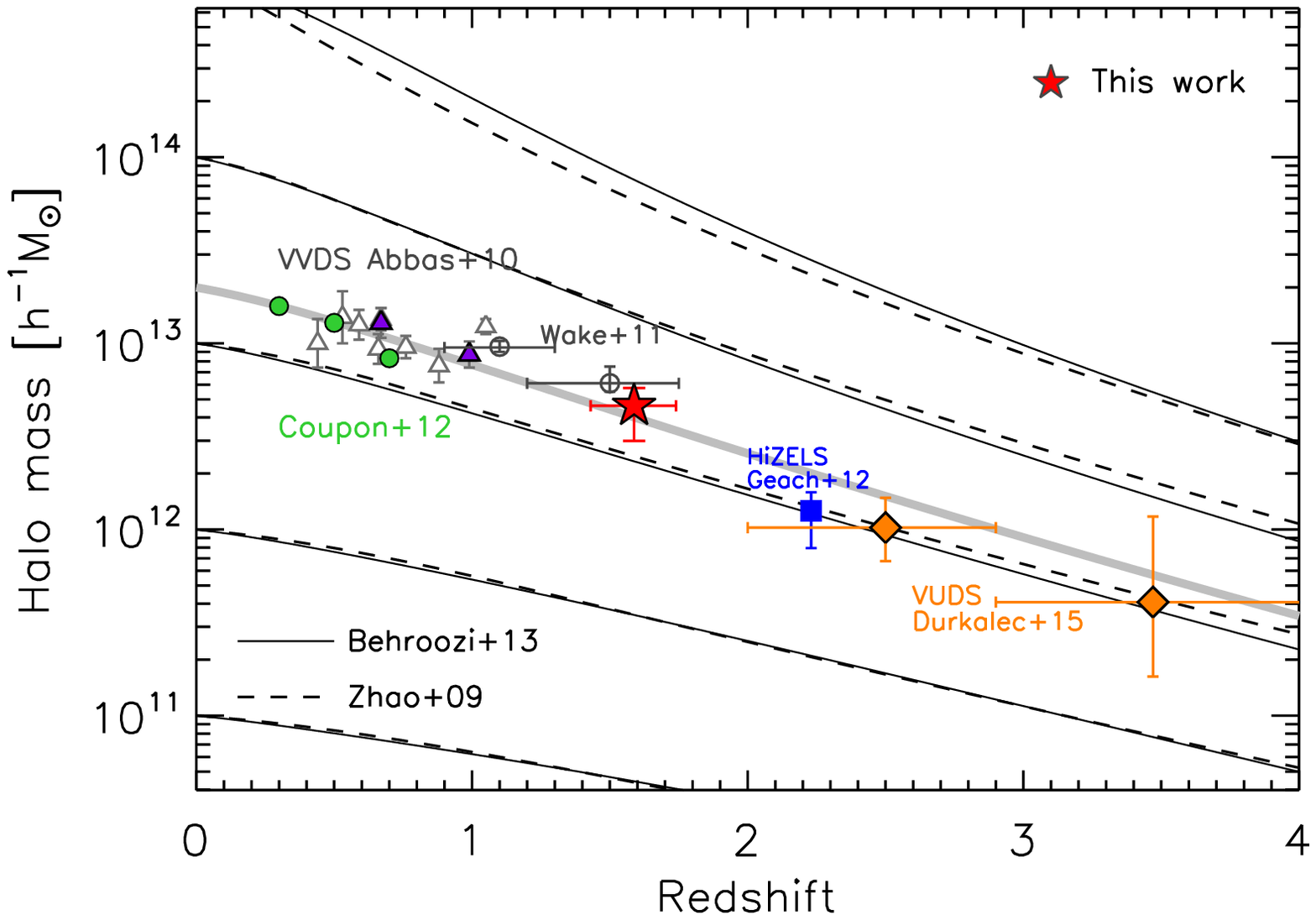} 
   \caption{The average host halo mass as a function of redshift.  A solid star indicates the effective halo mass $M_\mathrm{eff}$ estimated for our sample using HOD modeling.  Other symbols indicate results from the literature (filled green circles (error bars are comparable to the size of the symbol) -- \citealt{2012A&A...542A...5C}; empty triangle -- \citealt{2010MNRAS.406.1306A}; filled purple triangles -- $M_B<-19.5$ samples in \citealt{2010MNRAS.406.1306A}; empty circles -- \citealt{2011ApJ...728...46W}; filled blue square -- \citealt{2012MNRAS.426..679G}; filled orange diamonds -- \citealt{2015A&A...583A.128D}).  Solid and dashed curves show the mass assembly histories of halos for different present-day masses, derived by \citet{2013ApJ...770...57B} (solid lines) and \cite{2009ApJ...707..354Z} (dashed lines), respectively.  A thick gray line highlights the history for the present-day group-scale halos with $M_\mathrm{h} (z=0) =2\times 10^{13}~h^{-1}~M_\odot$, which well represents the global trend of data points shown here.}
      \label{fig:massev}
\end{figure*}

We find the effective halo mass to be $M_\mathrm{eff}=4.62_{-1.63}^{+1.14} \times 10^{12}~h^{-1}~M_\odot$ (Equation \ref{eq:Meff}) and the effective large-scale galaxy bias to be $b_\mathrm{eff}=2.17_{-0.24}^{+0.17}$ (Equation \ref{eq:beff}).  In Figure \ref{fig:massev}, we present $M_\mathrm{eff}$ in comparison to the average growth histories of halos with different present-day masses, as derived by \citet{2009ApJ...707..354Z} and \citet{2013ApJ...770...57B}.  We find that $M_\mathrm{eff}$ of our sample lies on the mass assembly history of halos having a present-day mass $M_\mathrm{h} (z=0)\approx2\times 10^{13}~h^{-1}M_\odot$ (thick gray curve).  This is equivalent to the typical mass of group-scale halos.  We also plot effective halo masses, derived through HOD modeling of galaxy clustering at different redshifts, from the literature.  At lower redshifts, we present measurements from the CFHT Legacy Survey at $z\sim0.3, 0.5$ and 0.7  \citep{2012A&A...542A...5C}.  These samples are selected based on the absolute $g$-band magnitude ($M_g-5\log h<-19.8$) and the galaxy number densities are $n_\mathrm{gal}\approx8\times10^{-3} h^{3}\mathrm{cMpc^{-3}}$, similar to our $M_\ast$-selected sample.  In addition, the results from \citet{2010MNRAS.406.1306A} at $0.2\lesssim z\lesssim1.3$ based on the VIMOS-VLT Deep Survey (VVDS; \citealt{2005A&A...439..845L}) are shown.  We highlight two subsamples in \citet{2010MNRAS.406.1306A} with $M_B<-19.5$ at $z=0.67$ and 0.99 (filled triangles) that have a number density similar to our $M_\ast$-selected sample.  We also present the results for samples with $M_\ast>10^{10}~M_\odot$ $z\sim1.1$ and $\sim1.5$ from the NEWFIRM Medium Band Survey \citep{2011ApJ...728...46W}.  These samples at lower redshifts are in excellent agreement with the same halo mass assembly history as our data.  At higher redshifts, we show results from HiZELS ($z\sim2.2$; \citealt{2012MNRAS.426..679G}) and VUDS ($z\sim2.5$ and $\sim3.5$; \citealt{2015A&A...583A.128D}), which are in broad agreement as well, favoring a slightly lower mass ($M_\mathrm{h}\approx10^{13}~h^{-1}M_\odot$ at $z=0$).  In summary, these results indicate a good agreement between the predictions of the halo mass assembly history and observations over a wide range of cosmic history since $z\sim4$ to the present.   From another perspective, it is supported that these samples at different redshifts essentially represent similar galaxy populations at different epochs.

\begin{figure}[tbph] 
   \centering
   \includegraphics[width=3.5in]{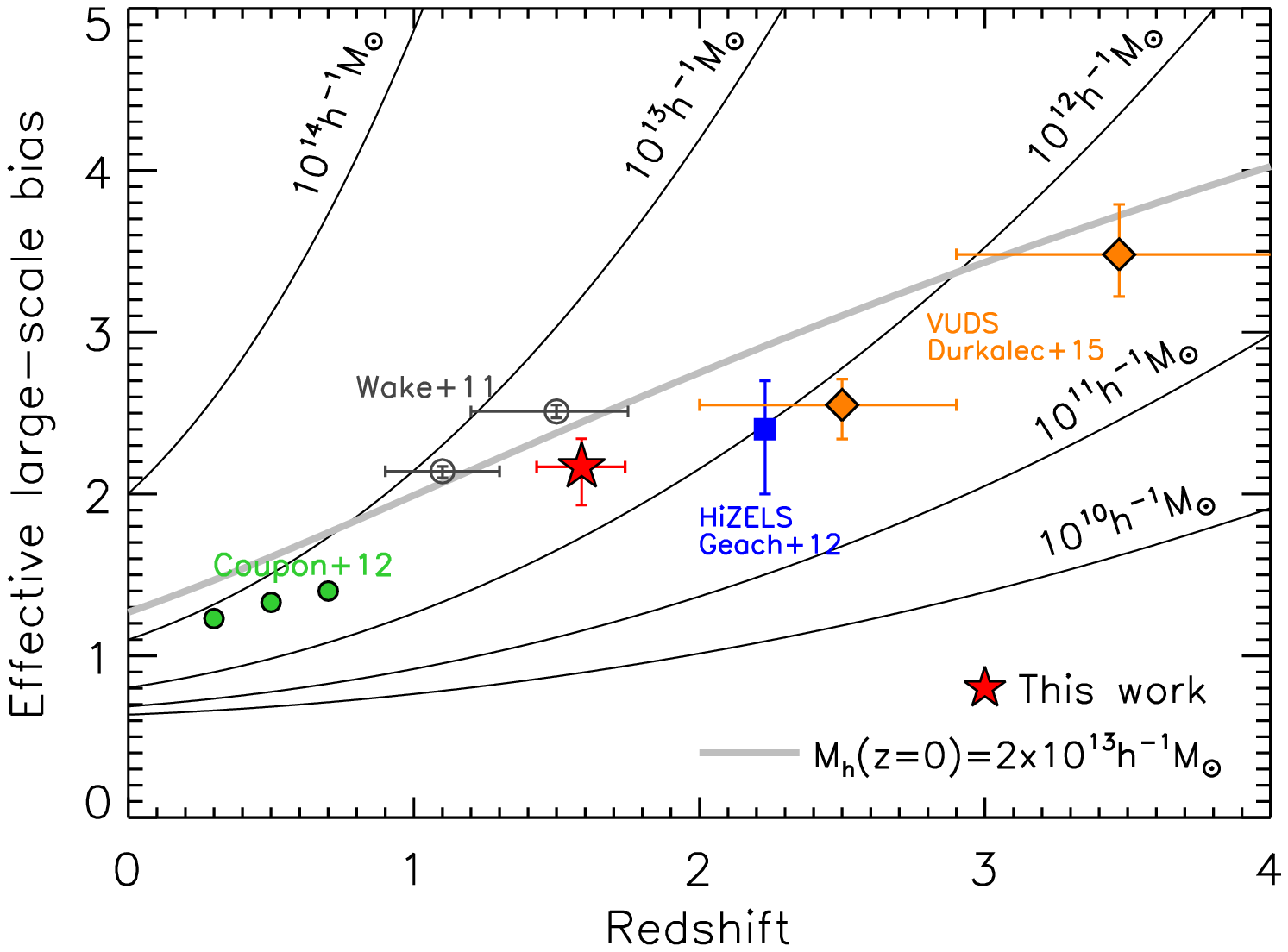} 
   \caption{The galaxy bias as a function of redshift.  The effective large-scale bias based on HOD analysis for our sample is shown by a red star.  Other symbols indicate results at different redshifts in the literature (filled green circles -- \citealt{2012A&A...542A...5C}; empty circles -- \citealt{2011ApJ...728...46W}; solid blue square -- \citealt{2012MNRAS.426..679G}; solid orange diamonds -- \citealt{2015A&A...583A.128D}).  Solid curves indicate the relation among the bias, halo mass and redshift, as labelled.  The gray thick line indicates the evolutionary track of the bias of halos of a present-day mass $M_\mathrm{h}(z=0)=2\times10^{13}~h^{-1}~M_\odot$.}
   \label{fig:bias}
\end{figure}

In Figure \ref{fig:bias}, we show the effective large-scale galaxy bias in comparison with measurements from the literature that are shown in Figure \ref{fig:massev}.  Our data in broad agreement with measurements at $1\lesssim z\lesssim2$ by \citet{2012MNRAS.426..679G} and \citet{2011ApJ...728...46W}.  Based on measurements over a wide range of redshift, it seems that the galaxy bias decreases with cosmic time.  This trend agrees with the general expectation from the standard scenario of hierarchical structure formation \citep[e.g.,][]{1996MNRAS.282..347M,1999MNRAS.307..529K}.  For reference, we show the halo bias for fixed halo masses ($\log M_\mathrm{h} / (h^{-1}M_\odot)=10,~11,~12,~13$, and 14; thin solid lines), which present a rapid decline of the bias.  As mentioned above, halos of a present-day mass $M_\mathrm{h}(z=0)=2\times10^{13}~h^{-1}~M_\odot$ are expected to be the descendants of halos containing star-forming galaxies in our sample at $z\sim1.6$ (thick gray line in Figure \ref{fig:massev}).  We further calculate the halo bias for the evolving halo mass as a function of redshift.  The global trend of the effective bias is well represented by this evolutionary track (thick gray line in Figure \ref{fig:bias}).

\subsection{Stellar mass-to-halo mass ratio} \label{sec:shmr}

The stellar mass-to-halo mass ($M_\ast/M_\mathrm{h}$) ratio encodes the efficiency of converting baryons into stars relative to the total amount of dark matter accreting onto halos.  \citet{2003MNRAS.339.1057Y} proposed a functional form with double power-law components to express the average halo mass-to-galactic luminosity ratios as a function of halo mass, motivated by the fact that the observed luminosity function is steeper (shallower) than the halo mass function at the high (low) mass end \citep[see also][]{2004MNRAS.353..189V,2013MNRAS.430..767C}.  With improvement of stellar mass measurements, this formalism has been applied to describe the stellar-to-halo mass relation (SHMR; $M_\ast/M_\mathrm{h}$ vs. $M_\mathrm{h}$) \citep[e.g.,][]{2007ApJ...667..760Z,2008ApJ...676..248Y,2009ApJ...695..900Y}.  
Past studies based on large data sets (e.g., SDSS) have shown that the $M_\ast/M_\mathrm{h}$ ratio reaches a peak around a halo mass of $M_\mathrm{h}\sim10^{12}~M_\odot$ at $z\sim0\text{--}1$ \citep[e.g.,][]{2010ApJ...710..903M,2013MNRAS.428.3121M,2010ApJ...717..379B,2010MNRAS.404.1111G,2011MNRAS.410..210M,2012ApJ...744..159L,2015MNRAS.449.1352C}.  Beyond $z>1$, there have been limited efforts to measure the SHMR based on HOD modeling of galaxy clustering \citep{2011ApJ...728...46W,2012MNRAS.426..679G,2015A&A...583A.128D,2015A&A...576L...7D,2015MNRAS.446..169M,2016ApJ...821..123H}.  Other studies use an alternative technique (i.e., abundance matching; \citealt{2004ApJ...609...35K}) to predict the SHMRs \citep{2013MNRAS.428.3121M,2013ApJ...770...57B}.  It populates halos in a $N$-body simulation with galaxies while assuming an SHMR, and the model is adjusted to match the inferred galaxy abundance to observations.

 \begin{figure*}[tbph] 
   \centering
   \includegraphics[width=5.5in]{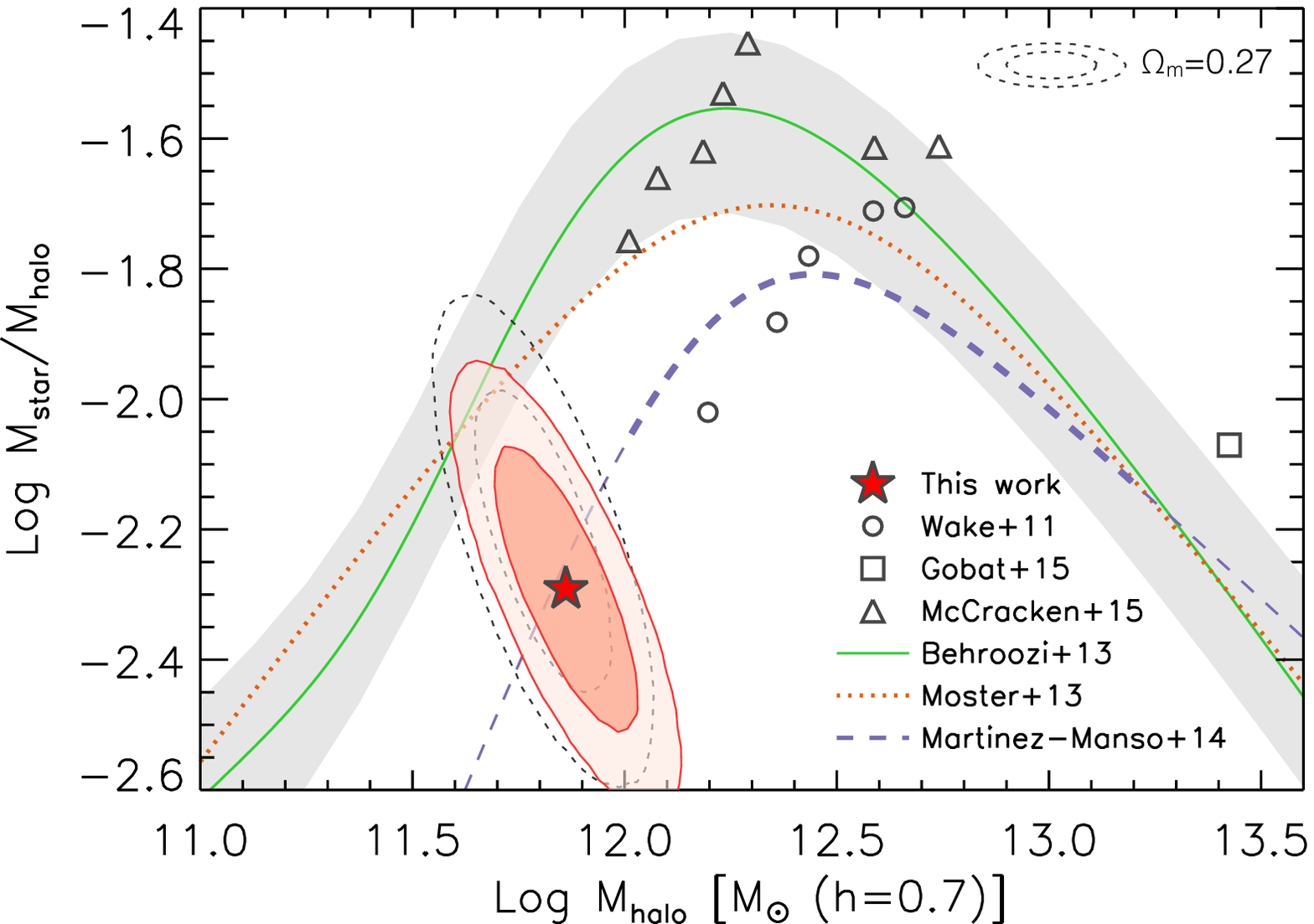} 
   \caption{Stellar mass-to-halo mass ratio ($M_\ast/M_\mathrm{h}$) as a function of halo mass (SHMR).  The star symbols indicates the ratio $M_\ast^\mathrm{lim}/M_\mathrm{min}$ with the 68\% and 95\% confidence intervals shown by red contours.  Dotted contours indicate the constraint with an alternative cosmology ($\Omega_\mathrm{m}=0.27$, see text).  A dashed line indicates the best-fit relation derived by \citet{2015MNRAS.446..169M} at $12<\log M_\mathrm{h} < 13.2$ (thick dashed line) and its extrapolation (thin dashed line).  Open circles show the measurements by \citet{2011ApJ...728...46W} for their different stellar mass threshold samples at $1.2<z<1.77$.  Triangles and a square indicate results for samples at $z\sim1.8$ \citep{2015MNRAS.449..901M,2015A&A...581A..56G}.  The SHMRs based on the abundance matching technique are shown as labelled \citep{2013ApJ...770...57B,2013MNRAS.428.3121M}.  A shaded region indicates the systematic uncertainties of the \citet{2013ApJ...770...57B} relation.}
   \label{fig:shmr}
\end{figure*}

For our sample of star-forming galaxies, we find $\log M_\ast/M_\mathrm{h}=-2.3\pm0.2$ at $M_\mathrm{h}=M_\mathrm{min}=10^{11.86}~M_\odot$ (computed with $h=0.7$), which is determined as the ratio of the threshold stellar mass $M_\ast^\mathrm{lim}$ to $M_\mathrm{min}$.  Figure \ref{fig:shmr} presents the observed $M_\ast/M_\mathrm{h}$ ratio as a function of $M_\mathrm{h}$.  We emphasize that our result probes the low-mass side of the SHMR at this epoch based on a spectroscopic sample, evidently confirming the rapid decline at $M_\mathrm{h}<10^{12}~M_\odot$.  We compare our result with past measurements at similar redshifts from the literature, which are based on HOD modeling of galaxy clustering using photometric redshifts and/or color-selected galaxy samples.  \citet{2015MNRAS.446..169M} measure the $M_\ast/M_\mathrm{h}$ ratios over $12\lesssim \log M_\mathrm{h}/M_\odot \lesssim 13.2$ by using a sample of $z\sim1.5$ galaxies selected from the {\it Spitzer}/IRAC 3.6 and 4.5 $\mathrm{\mu m}$ photometry.  Their result shows an evident peak at $M_\mathrm{h}=10^{12.4}~M_\odot$.  
Around this peak halo mass, \citet{2011ApJ...728...46W} and \citet{2015MNRAS.449..901M} measure $M_\ast/M_\mathrm{h}$ at $z\sim1.5$, which are systematically higher than our result.  We note that the $M_\ast/M_\mathrm{h}$ ratios in \citet{2011ApJ...728...46W} are multiplied by a factor of 1.5 to account for the fact that the \citet{2005MNRAS.362..799M} stellar population synthesis model induces stellar masses that are systematically lower than those derived by the \citet{2003MNRAS.344.1000B} model (on average by $\sim60\%$; \citealt{2006ApJ...652...85M}).  The measurements by \citet{2011ApJ...728...46W} are in good agreement with the result of \citet{2015MNRAS.446..169M}, whereas the measurements by \citet{2015MNRAS.449..901M} are systematically greater by a factor of $\sim2$ at fixed $M_\mathrm{h}$.  
Another study \citep{2015A&A...581A..56G} compares the average stellar mass of a sample of massive star-forming galaxies at $z\sim1.8$ ($\sim1.3\times10^{11}~M_\odot$) and the halo mass ($\sim2.4\times10^{13}~M_\odot$) induced from the average X-ray luminosity through stacking analysis.  Their independent measurement clearly indicates, with others, the decline of SHMR at high masses.  It may be worth noting that our measurement is in good agreement with the extrapolation of the SHMR derived by \citet{2015MNRAS.446..169M} (thin dashed line in Figure \ref{fig:shmr}).

Furthermore, we show in Figure \ref{fig:shmr} the SHMRs at $z=1.6$ that are derived by an abundance matching technique \citep{2013ApJ...770...57B,2013MNRAS.428.3121M}.  The \citet{2013ApJ...770...57B} SHMR presents a peak value that is higher (by $\sim0.2~\mathrm{dex}$) than as in \citet{2015MNRAS.446..169M}, while being more consistent with \citet{2015MNRAS.449..901M}.  In comparison among these studies, the three measurements based on the HOD modeling (our result, \citet{2015MNRAS.446..169M}, and \citet{2011ApJ...728...46W}) likely favor an SHMR having the low-mass part that is systematically lower than that predicted by abundance matching.  However, we note that our result is consistent with those at a $2\sigma$ confidence level, and becomes closer to the \citet{2013ApJ...770...57B} SHMR (nearly $1\sigma$) when being recalculated with the same cosmology as the reference ($\Omega_\mathrm{m}=0.27$; dotted contours in Figure \ref{fig:shmr}).  For quantitative derivation and comparison of SHMRs, one needs to carefully treat the difference between adopted cosmologies (e.g., $\Omega_\mathrm{m}, \sigma_8$) from one study to another.   We refer the reader to \citet{2013ApJ...777L..26M} for further discussions of such systematic effects on the galaxy clustering.  

\subsection{Baryon conversion efficiency} \label{sec:bar_conv}

Given an average growth history of halos and an observation of SFRs of galaxies, we can calculate the efficiency of baryon conversion $\epsilon_\mathrm{baryon}$, i.e., the fraction of the mass of baryons converted into stars per unit time to the total accretion rate of baryons falling into halos.  For this purpose, the SFRs of our spec-$z$ galaxy sample are measured from the observed H$\alpha$ luminosities by using Equation (\ref{eq:sfr}).  To derive intrinsic SFRs, we correct the observed H$\alpha$ flux for aperture loss \citep{2015ApJS..220...12S} and dust extinction assuming a \citet{2000ApJ...533..682C} reddening curve.  The level of extinction $E_\mathrm{star}(B-V)$ is estimated based on the $B_z-j$ color \citep{2007ApJ...670..156D}, and converted to the attenuation towards nebular emission lines assuming a relation $E_\mathrm{neb}(B-V)=E_\mathrm{star}(B-V)/0.66$ \citep[see][]{2013ApJ...777L...8K}.  We find the average SFR to be $\approx15~M_\odot\mathrm{yr^{-1}}$ at $M_\ast^\mathrm{lim}=10^{9.57}~M_\odot$ with a standard deviation of $0.3~\mathrm{dex}$, including errors on the flux measurement.  However, this average may be slightly biased toward a higher value as compared to the $M_\ast$-selected sample because of the imposed limit on the predicted H$\alpha$ flux and the detection bias (see Section \ref{sec:selection}).  According to SED-based SFRs shown in Figure \ref{fig:sample}, the median SFR in the spec-$z$ sample is elevated by approximately $0.1~\mathrm{dex}$ (a factor of 1.3) as compared to the entire main sequence population ($K_\mathrm{S}\le 23.5$) at the threshold stellar mass.  We do not take this bias into account since it is well below the scatter in SFR and thus does not affect our estimate of the baryon conversion efficiency with the given statistics.

We calculate the mass accretion rates by using the halo mass accretion history derived by \cite{2013ApJ...770...57B}.  With the baryon fraction of accreting matter fixed to the cosmic baryon fraction ($\Omega_b/\Omega_m=0.17$), the baryon accretion rate is found to be $\dot{M}_\mathrm{baryon}=43 ~M_\odot~\mathrm{yr^{-1}}$ for halos of $M_\mathrm{halo}=M_\mathrm{min}=10^{11.86}~M_\odot$.  We note that the value of $\dot{M}_\mathrm{baryon}$ varies almost proportionally with halo mass.  As a result, we find that the main sequence galaxies with $M_\ast\sim10^{9.57}M_\odot$ at $z\sim 1.6$ convert $\epsilon_\mathrm{baryon}=\mathrm{SFR}/\dot{M}_\mathrm{baryon}\sim 35 \%$ of baryons accreting onto halos into new stars.  This estimate is consistent with the average conversion efficiency derived by \citet{2013ApJ...770...57B}.  We further give a rough estimation on the scatter in $\epsilon_\mathrm{baryon}$ with an assumption that the scatter in SFR is caused by independent fluctuations in $\epsilon_\mathrm{baryon}$ and in $\dot{M}_\mathrm{baryon}$, which is now considered to be proportional to halo mass.  Because the scatter in SFR is approximately $0.3~\mathrm{dex}$ and the scatter in halo mass is expected to be $0.2~\mathrm{dex}$ (see Appendix \ref{sec:lim_sigmalogM}) at fixed stellar mass ($=M_\ast^\mathrm{lim}$), the scatter in $\epsilon_\mathrm{baryon}$ is to be $\sim0.2~\mathrm{dex}$ (a factor of 1.6) or less.

At $M_\mathrm{h}=M_\mathrm{min}$, the mass ratio $M_\ast/M_\mathrm{h}$ is likely to be proportional to $M_\mathrm{h}^{1.5}$, i.e., $M_\ast\propto M_\mathrm{h}^{2.5}$ \citep{2015MNRAS.446..169M,2013ApJ...770...57B}.  If the baryon conversion occurs at a constant rate for all halo masses, the stellar mass grows proportionally to the halo mass as the baryon accretion rate is expected to be almost proportional to halo mass, i.e., $\dot{M}_\mathrm{baryon}\propto M_\mathrm{h}$.  Therefore, the stronger dependence of $M_\ast$ to $M_\mathrm{h}$ indicates that the integrated past baryon conversion depends on the halo mass.  We recall that the slope of the star-forming main sequence ($M_\ast\textrm{--}\mathrm{SFR}$) is $\sim0.7\textrm{--}0.8$ \citep[e.g.,][]{2013ApJ...777L...8K}.  Given these observational facts, we find that the baryon conversion efficiency increases moderately with increasing stellar mass, as $\epsilon_\mathrm{baryon}\propto M_\ast^{0.3\textrm{--}0.4}$, and more strongly with halo mass as $\epsilon_\mathrm{baryon} \propto M_\mathrm{h}^{0.75\textrm{--}1.0}$.  A similar dependence has also been found at a higher redshift (see Figure 13 of \citealt{2016ApJ...821..123H}).  We note that these relations pertain to the stellar mass between $10^{9.57}~M_\odot$ and $M_\ast\sim10^{10.5}~M_\odot$, or halo mass from $M_\mathrm{min}=10^{11.86}~M_\odot$ to $M_\mathrm{h}\sim10^{12.2}~M_\odot$.  Since the slope of the SHMR decreases with increasing (both stellar and halo) mass (while eventually its slope becomes negative), the conversion efficiency has a peak value ($\epsilon_\mathrm{baryon} \sim 0.6$) that begins to decrease towards the highest masses (see Figure 11 of \citealt{2013ApJ...770...57B}).

\subsection{Satellite galaxies} \label{sec:satellites}

The detection of the one-halo term in the correlation function enables us to estimate the fraction of galaxies that are satellites ($f_\mathrm{sat}$).  This fraction is dependent on $M_\mathrm{min}$, $M_1$, and the halo mass function.  For our sample, we find a satellite fraction $f_\mathrm{sat}=0.38_{-0.20}^{+0.14}$ from Equation (\ref{eq:fsat}).  This is similar to the value of $f_\mathrm{sat} \approx 0.3$ for local galaxies \citep{2011ApJ...736...59Z}.  At higher redshifts, \citet{2011ApJ...728...46W} measure $f_\mathrm{sat}\approx0.28$ for stellar mass limit samples ($>10^{10}M_\odot$) at $z\sim1.5$.  \citet{2015MNRAS.446..169M} also report a similar value ($f_\mathrm{sat}\sim0.25$) at $z\sim1.5$.  Our result is consistent with these values as well.  While our data do not constrain the redshift evolution of $f_\mathrm{sat}$ with the given statistical error, \citet{2012A&A...542A...5C} suggested that the satellite fraction moderately increases with cosmic time since $z\sim1$, which can be straightforwardly interpreted as a consequence of average growth of over-dense regions.  

The ratio $M_1/M_\mathrm{min}$ determines the ``shoulder'' in $\left< N | M_\mathrm{h}\right>$ \citep{2006ApJ...647..201C}, or the gap between halo masses at which halos acquire a central galaxy and host an additional satellite galaxy (refereed to as the ``hosting gap'' by \citealt{2011ApJ...736...59Z}).  \citet{2011ApJ...736...59Z} found $M_1/M_\mathrm{min}\approx 17$ for local galaxies and that the ratio decreases with increasing threshold luminosity of a given sample.  Here we find the ratio to be $M_1/M_\mathrm{min} = 4.6_{-2.1}^{+4.5}$ for our FMOS sample, which is smaller than the local value and qualitatively consistent with the trend as reported by \citet{2006ApJ...647..201C}.  Our result is in good agreement with previous findings at intermediate redshifts. For example, \citet{2011ApJ...728...46W} find $M_1/M_\mathrm{min}\approx 6$ for samples with stellar mass above $\sim 10^{10}~M_\odot$ at $1\lesssim z \lesssim 2$.  Recently, \citet{2015MNRAS.446..169M} also found similar values $M_1/M_\mathrm{min}\approx 7$.  \citet{2006ApJ...647..201C} find a decrease in $M_1/M_\mathrm{min}$ with increasing redshift, which may reflect the increasing fraction of relatively low-mass halos that host satellites in addition to a central galaxy.  The authors argue that pairs of galaxies within the same halo with mass close to $M_\mathrm{min}$ predominantly contributes to the one-halo term of the correlation function at higher redshifts.  

\subsection{Prediction for the future Subaru/PFS survey} \label{sec:PFS}

Finally, we give predictions for the performance of the clustering measurements expected for the future Subaru galaxy survey using the Prime Focus Spectrograph (PFS; \citealt{2016SPIE.9908E..1MT}).  This is a next-generation multi-object spectrograph that follows FMOS, having 6 times larger field-of-view (1.3~degree in diameter), many more fibers ($N_\mathrm{fiber}=2400$) than FMOS, and an unprecedented wavelength coverage ($0.38\textrm{--}1.26~\mathrm{\mu m}$).  The preliminary PFS survey strategy is described in \citet{2014PASJ...66R...1T}, while the design has been modified in part to date.  The PFS galaxy evolution survey aims to have a dedicated program to observe roughly half a million color-selected galaxies at $1<z<2$ to a limiting magnitude of $J_\mathrm{AB}\approx23.3$ (or more deeper) over $\sim15~\mathrm{deg^2}$.  Galaxies will be selected from the Hyper Suprime-Cam Subaru Strategic Program (HSC-SSP) that provides deep and wide imaging with $grizY$  photometry.

Here, we aim to evaluate the statistical errors and how strongly the parameters can be constrained with a PFS-like sample, while assuming our best-fit values from the HOD modeling as fiducial.  The signal-to-noise ratio of an observed correlation function is approximately proportional to the square root of the galaxy pair counts in bins of the separation.  For an ideal survey with a contiguous survey volume and uniform sampling, the pair counts are approximately proportional to the survey volume and to the square of galaxy number density.  If the survey field is separated into distinct smaller subregions, the number of galaxy pairs, especially with large spatial separations, increases more slowly than expected.  The PFS survey will cover multiple distinct areas each having $\sim6\textrm{--}7~\mathrm{deg}^2$.  Such separate regions will not significantly affect the scales that we cover in this work ($\lesssim 20~\mathrm{cMpc}$).  Therefore, we simply scale the statistical errors on the observed correlation function to match the expected survey volume and number density for the PFS survey.

For this purpose, we adopt the same redshift range as this work, i.e., $1.43\le z\le 1.74$, resulting in a comoving volume of $1.75\times10^7 (h^{-1}~\mathrm{cMpc})^3$ for a nominal survey area of $15~\mathrm{deg^2}$, which is 20 times as large as our FMOS survey.  We assume a mass complete sample above $10^{10}~M_\sun$ with a uniform sampling rate of $75\%$.  Based on the COSMOS catalog \citep{2013A&A...556A..55I}, the sample size should be $N\sim6.5\times10^4$ with the galaxy number density of $4\times10^{-3} (h^{-1}~\mathrm{cMpc})^{-3}$, which is 7 times as high as our FMOS-spec-$z$ sample (see Table \ref{tb:samples}).  As a very rough estimate, the errors on the correlation function will decrease by a factor of 30 for the full sample with the same binning in $r_\mathrm{p}$, while, if we divide the full sample equally into 10 subsamples, the errors become smaller by a factor of 3 as compared to our data.  The values of $\log w_\mathrm{p} (r_\mathrm{p})$ are then adjusted by adding statistical noise corresponding to the scaled errors.  Again, we do not intend here to make any predictions for the shape and/or absolute values of correlation function, but just to demonstrate the improvement in statistical accuracy. 
 
\begin{figure*}[tbph] 
   \centering
   \includegraphics[width=6.5in]{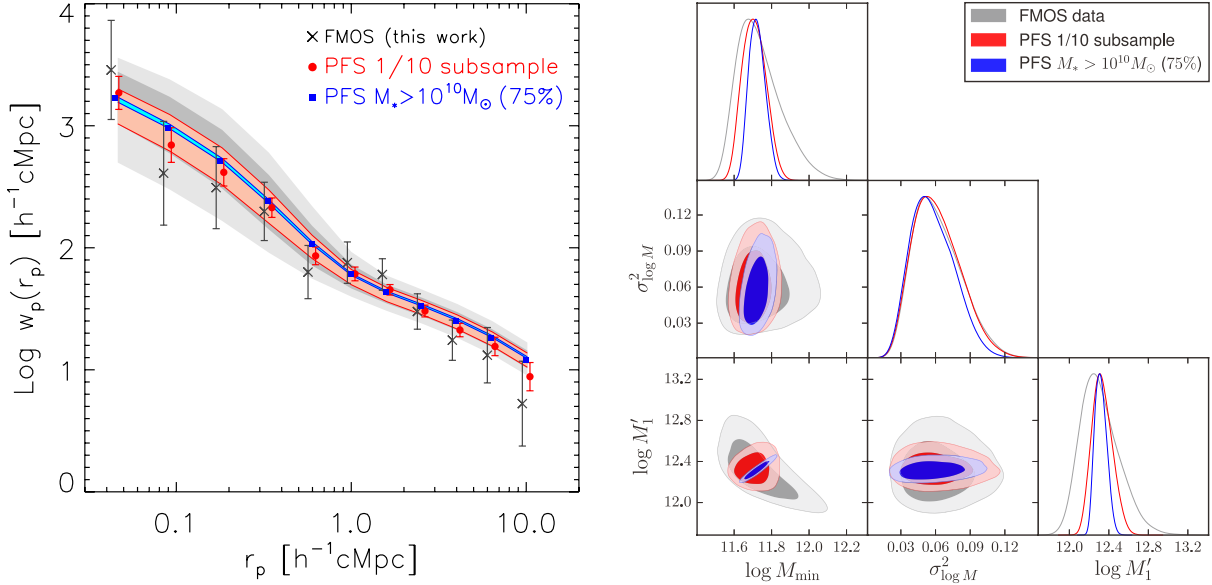} 
   \caption{Left: correlation functions for our data and hypothetic PFS samples as labeled.  Dark and light gray regions show the 68 and 95 percent confidence intervals of the HOD model for our data, and light red and cyan regions indicate the 95th percentiles for the PFS samples.  Right: the constraints of the HOD parameters for each sample.}
   \label{fig:pfs}
\end{figure*}

For these two cases (the full sample with a 75\% sampling rate and the 1/10 binned subsample), we show in Figure \ref{fig:pfs} the expected measurements and parameter constraints in comparison with our results.  For simplicity, we assume the pair counting with the same ($r_\mathrm{p}$, $\pi$) grid and employ the same priors as our work with the FMOS data.  As evident, the statistical uncertainties are remarkably improved for the hypothetic samples for a future survey, even if the samples are limited to a narrow range of redshift.  For the full sample, it is obvious that the systematic effects that we have discussed in this paper, i.e., the impact of fiber allocation and/or inhomogeneous line-of-sight detection rate, will significantly influence the measurements, and thus the physical interpretations.  In the right panel of Figure \ref{fig:pfs}, it is shown that the constraints on $M_\mathrm{min}$ and $M_\mathrm{1}^\prime$ are significantly improved, while those on $\sigma^2_{\log M}$ are almost identical to our FMOS result.  In fact, this parameter is essentially determined by the prior.  This means that even very statistically-accurate clustering measurements cannot constrain all parameters strongly.  Therefore, alternative methods are still necessary to fully understand the galaxy-halo relation.  In conclusion, the future multi-object spectroscopic surveys (e.g., PFS, MOONS, DESI) assure significant improvement in the statistics with respect to clustering measurements, while systematic effects must be excluded more carefully and precise models are imperative to interpret such upcoming observational data.

\section{Summary}
\label{sec:summary}

We have investigated the clustering properties of star-forming galaxies at $1.43\le z\le 1.74$ using the data set from the FMOS-COSMOS survey.  With 516 galaxies having an H$\alpha$ detection, we measured the projected two-point correlation function and investigated the properties of dark matter halos through the HOD modeling of galaxy clustering.  Our main results are as follows:

\begin{enumerate}
	\item The observed correlation function indicates a significant clustering at $0.04\lesssim r_\mathrm{p}/(h^{-1}~\mathrm{cMpc})\lesssim10$.  By modeling with a power-law function, we find a correlation length of $r_0 = 5.21^{+0.70}_{-0.67}~h^{-1}~\mathrm{cMpc}$, which is consistent with preceding studies using galaxy samples with stellar masses similar to our study. 
	
	\item We model the observed correlation function using an HOD model (Equations \ref{eq:cen}--\ref{eq:sat}).  The HOD parameters are effectively constrained with our current data and a significant one-halo term is confirmed with the transition scale of $r_\mathrm{p}\simeq 0.5~h^{-1}~\mathrm{cMpc}$, where the one-halo and two-halo terms are equivalent.  
	
	\item We derive an effective large-scale bias, $b_\mathrm{eff}=2.17^{+0.17}_{-0.24}$, and find that star-forming galaxies with $M_\ast\ge 10^{9.57}~M_\odot$ reside on average in halos with mass $M_\mathrm{eff}=4.62^{+1.14}_{-1.63}\times10^{12}~h^{-1}M_\odot$, which is consistent with other measurements from HOD modeling.  With predictions of the mass assembly histories, we find that these halos will have grown into group-scale halos ($\sim 2\times10^{13}~h^{-1}~M_\odot$) at the present epoch.
	
	\item Our work investigates the low-mass part of the stellar-to-halo mass relation at this redshift, with a new constraint, $\log M_\ast / M_\mathrm{h} = -2.3\pm0.2$ at $M_\mathrm{h}=10^{11.86}~M_\odot$, while confirming the decline in the stellar-to-halo mass ratio at $M_\mathrm{h}<10^{12}~M_\odot$.  However, there are discrepancies between SHMRs based on different observations and/or analyses.  In particular, SHMRs based on HOD modeling, including our result, seem to have systematically lower values at lower halo masses than those based on abundance matching.

	\item We find the efficiency of converting baryons into stars relative to the total amount of baryons accreting onto halos to be $\sim35\%$ with a maximum scatter of a factor of $\sim1.6$, and that the baryon conversion efficiency $\epsilon_\mathrm{baryon}$ depends on stellar and halo mass as $\epsilon\propto M_\ast^{0.3\textrm{--}0.4}$ and $\epsilon\propto M_\mathrm{h}^{0.75\textrm{--}1.0}$, respectively, around the threshold mass up to the peak mass of the SHMR ($\log M_\ast / M_\odot \approx 9.57 \textrm{--}10.5$, $\log M_\mathrm{h} / M_\odot \approx 11.86\textrm{--}12.2$).

	\item We find that the satellite fraction to be $f_\mathrm{sat}=0.38^{+0.14}_{-0.20}$, which is consistent with other measurements at both local and higher redshifts.  In addition, the $M_1 / M_\mathrm{min}$ ratio is found to be $\sim 5$, which is lower than as typically seen for low-$z$ galaxies ($\simeq 17$ at $z\lesssim0.1$) while consistent with other studies at higher redshifts ($z>1$), which show $M_1 / M_\mathrm{min}<10$.  These results suggests that galaxy pairs in the same, relatively low-mass halos significantly contribute to the one-halo term of the correlation function.
	 
\end{enumerate}

In addition, we established correction schemes for fiber allocation and inhomogeneous detection due to sky contamination, and demonstrated the effectiveness of these methods even for small scales (i.e., the one-halo regime).  Such techniques will be effective to measure the intrinsic galaxy clustering in the future spectroscopic surveys with the next-generation multi-object spectrographs such as Subaru/PFS.  Future surveys will provide a sample of $\sim5\times10^5$ galaxies at $z>1$, and allow to measure galaxy clustering as a function of galactic properties at much elevated statistical accuracy.  Therefore, the systematic effects must be excluded carefully and precise models are required to draw information maximally from such upcoming observational data.

\acknowledgements
This work is based on data collected at the Subaru telescope, which is operated by the National Astronomical Observatory of Japan.  We greatly thank the staff of the Subaru telescope, especially K. Aoki, for supporting the observations.  We greatly thank T. Ishiyama and the $\nu^2$GC team, and the Bolshoi simulation collaboration group for making their halo/subhalo catalogs public.  We appreciate P. Behroozi for providing their data.  J.D.S is supported by JSPS KAKENHI grand Number 26400221 and the World Premier International Research Center Initiative (WPI), MEXT, Japan.  D.K. was supported through the Grant-in-Aid for JSPS Fellows (No. 26-3216).

\appendix
\section{Correction for critical biases in the observed correlation function}
\label{sec:mock_corr}

\subsection{Mock samples from the $\nu^2$GC simulation}
\label{sec:mock_n2gc}

We construct mock samples to evaluate our correction scheme for the effects of fiber allocation and sky contamination using the new numerical Galaxy Catalog ($\nu^2$GC) cosmological simulation \citep{2015PASJ...67...61I}.  We use the middle size simulation ($\nu^2$GC-M) that was conducted using $4096^3$ particles with a mass of $2.20\times 10^8~h^{-1} M_\odot$ in a comoving cube having a side length of $560~h^{-1}~\mathrm{cMpc}$.  We employ a halo/subhalo catalog at a scale factor $a=0.384871$ ($z=1.598$), which is close to the median redshift of our spectroscopic sample.  Halos and subhalos are identified using the Rockstar algorithm \citep{2013ApJ...762..109B}, resolving subhalos with $M_\mathrm{h} \sim 10^{10}~h^{-1}M_\odot$.  The lengths of our survey volume in the plane of the sky and along the line-of-sight are $\sim 60~h^{-1}~\mathrm{cMpc}$ and $\sim400~h^{-1}~\mathrm{cMpc}$ over $1.43\le z\le 1.74$, respectively.  Hence, we can divide the simulation box into 64 sub-boxes on the $x$-$y$ plane each having a volume of $70 \times 70 \times 560~(h^{-1}\mathrm{cMpc})^3$ that can enclose the survey volume.  In doing so, we implicitly ignore the redshift evolution between this redshift range.  Then, we convert the comoving coordinates ($x,y,z$) into angular coordinates and redshift, associating the middle point in a box to the center of the survey volume.  We take the peculiar velocity along $z$-axis of individual (sub)halos into account to derive the true redshift (as opposed to the cosmological redshift).

For each subbox, we select halos and subhalos from the catalog to create a sample that has a number density similar to the FMOS-parent sample ($n=3.07\times10^{-3}~h^3\mathrm{cMpc}^{-3}$) and a correlation function whose amplitude is also similar to the observed one (see Figure \ref{fig:wp_mock}).  We extract halos having a maximum circular velocity $V_\mathrm{max}$ greater than $170~\mathrm{km~s^{-1}}$ and subhalos having $V_\mathrm{max}>160~\mathrm{km~s^{-1}}$.  It is known that the highest $V_\mathrm{max}$ that a halo can attain over its past history (often called $V_\mathrm{peak}$) is a better proxy of stellar mass, rather than $V_\mathrm{max}$ at the epoch of interest, especially in order to better reproduce galaxy clustering \citep{2013ApJ...771...30R}.  This is reflective of the fact that subhalos lose their mass and $V_\mathrm{max}$ when they accrete onto their parent halos.  However, such information is not yet available for this simulation.  To roughly account for the mass loss of subhalos, we here use a lower threshold of $V_\mathrm{max}$ for subhalos than for parent halos.  However, the precise selection does not matter because we are just interested in the relative change due to the observational biases, but not in any absolute quantities predicted by the simulation. 

To ensure that the redshift distribution of the mock samples matches that of the FMOS-parent sample, we assign {\it photometric} redshifts with additional random fluctuations of $\mathrm{rms} (\Delta z)=0.062$ to the true redshift of each halo.  Then we extract objects from the same range of the photometric redshift as we applied for the real data ($1.46\le z_\mathrm{phot}\le 1.72$).  From them, we randomly select $\sim2300$ objects to match the number of the FMOS-parent sample, and refer these as the {\it Mock-parent} samples.  From these samples, we construct two types of mock samples that take into account the inhomogeneous detection along the line-of-sight (mainly due to the atmospheric lines) and fiber allocation effects.  

Then we randomly select the same number of objects as present in our FMOS-spec-$z$ sample (516) from a subset of the Mock-parent sample restricted to having a true redshift $1.43\le z \le1.74$.  During this selection, we account for the weight as a function of redshift, as shown in Figure \ref{fig:weight_z}, to reflect the non-uniform detectability of the H$\alpha$ emission line along the line of sight.  These {\it Mock-zweight} samples are used in the subsequent section.  

Next we prepare mocks to assess the impact of fiber allocation algorithm.  For each Mock-parent sample, we mimic the selection function induced by the FMOS fiber allocation software to construct the {\it Mock-fiber-target} samples.  We also account for fibers which broke down, some of these were regularly out-of-order and the others depending on the observing runs.  These samples have approximately the same number of galaxies ($\sim1200$) as the FMOS-fiber-target sample, representing the observed (fiber-allocated) galaxies.  Then we create the {\it Mock-spec-$z$} samples by randomly extracting 516 objects with a true redshift between $1.43\le z\le 1.74$ from each Mock-fiber-target catalog.  These final samples correspond to the FMOS-spec-$z$ sample that is used for clustering measurement.  We note that the fiber allocation is not considered for the Mock-zweight samples, and the inhomogeneous detection rate is not taken into account for the Mock-fiber-target/spec-$z$ samples to examine each of these effects individually.

\capstartfalse
\begin{deluxetable*}{lcl}
\tablecaption{Mock samples from the $\nu^2$GC simulation \label{tb:mocks}}
\tablehead{\colhead{Sample name}& $N$ &\colhead{Conditions\tablenotemark{a}}}
\startdata
Mock-parent & $\sim2300$ & $1.46\le z_\mathrm{phot}\le 1.72$\tablenotemark{b}\\
Mock-zweight & 516 &  $1.46\le z_\mathrm{phot}\le 1.72$, $1.43\le z_\mathrm{spec}\le 1.74$, and non-uniform detectability\tablenotemark{d} \\
Mock-fiber-target & $\sim1200$ &  $1.46\le z_\mathrm{phot}\le 1.72$  and fiber allocation\\
Mock-spec$z$ & 516 &  $1.46\le z_\mathrm{phot}\le 1.72$, fiber allocation, and $1.43\le z_\mathrm{spec}\le 1.74$\\
\enddata
\tablenotetext{a}{The thresholds on $V_\mathrm{max}$ and the realistic shape of the FMOS survey area are applied for all the mock samples listed here.}
\tablenotetext{b}{Photometric redshift of a (sub)halo is generated by addition a random error on the true redshift.}
\tablenotetext{c}{Spectroscopic redshift of a (sub)halo includes the effect of peculiar motion.}
\tablenotetext{d}{The inhomogeneous weight function shown in Figure \ref{fig:weight_z} is applied.}
\end{deluxetable*}
\capstarttrue

\subsection{Inhomogeneous line-of-sight detectability}

The inhomogeneous detection of the H$\alpha$ emission line along the redshift direction due to the OH lines/mask and the instrumental characteristics can induce an artificial clustering signal along the direction parallel to the line of sight.  As a result, the amplitude of the correlation function is expected to be artificially enhanced if corrections are not applied.  In our analysis, we used a random sample where the redshifts were sampled from the intrinsic distribution of the FMOS-parent sample, but with a probability weighted by the non-uniform detectability as a function of redshift (Figure \ref{fig:weight_z}).  

\begin{figure}[tbph]
   \centering
   \includegraphics[width=3.5in]{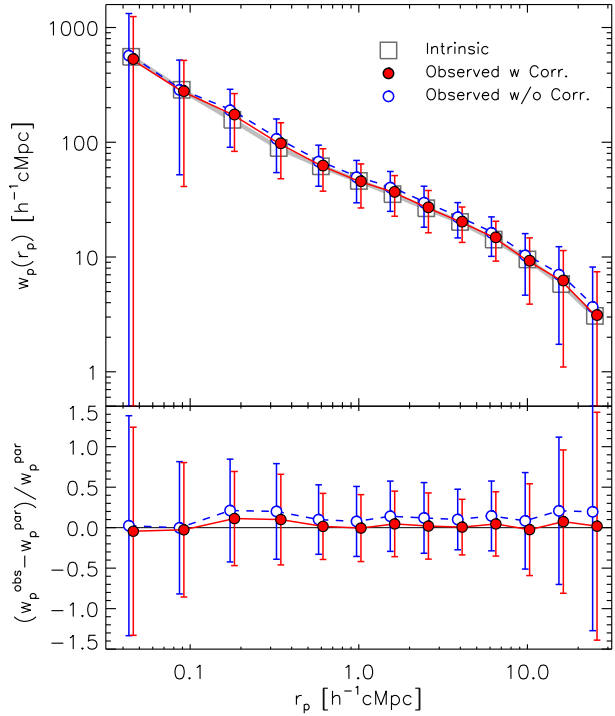} 
   \caption{{\it Upper panel}: Projected two-point correlation functions from the mock samples.  The average {\it observed} $w_\mathrm{p} (r_\mathrm{p})$ is calculated for the Mock-zweight samples with (filled red circles) and without (blue empty circles) the correction in the random sample for the non-uniform detection rate along the line-of-sight.  The $x$-axis values of these data points are slightly shifted for the purpose of display.  Large squares indicate the intrinsic correlation function computed for the Mock-parent-$z_\mathrm{spec}$ samples.  {\it Lower panel}: Difference between the intrinsic and the observed $w_\mathrm{p} (r_\mathrm{p})$ with or without the correction.  Symbols are the same as the upper panel.}
   \label{fig:wp_mock_zweight}
\end{figure}

We demonstrate the effects and the reliability of our method.  We calculate the correlation functions of a series of Mock-zweight samples, in which the non-uniform detectability is taken into account by selecting 516 objects with the weighted probability (Appendix \ref{sec:mock_n2gc}).  To assess the effect on our calculation of the correlation function using Equation (\ref{eq:LS}), we inspect differences in the clustering using random samples with and without the non-uniform detection rate corrections.  In Figure \ref{fig:wp_mock_zweight}, the average correlation function of 64 mock samples is shown corresponding to the use of these random catalogs, and compared to the intrinsic correlation function of the sample.  It can be seen that the use of the modified random samples accurately recovers the intrinsic one (filled red circles).  In contrast, the values of $w_\mathrm{p}(r_\mathrm{p})$ are slightly, but systematically enhanced over the range of $r_\mathrm{p}>0.1~h^{-1}\mathrm{Mpc}$ if we do not consider the non-uniform weight in the random sample (open blue circles).  This enhancement of the clustering amplitude results in an overestimate of the correlation length by $\Delta r_0\sim0.2\textrm{--}0.3~h^{-1}~\mathrm{cMpc}$.  Although this systematic error is well below the uncertainty on the measurement from our current data, it could be critical in future studies based on large samples, which provide much higher statistical accuracy than our study (see Section \ref{sec:PFS}).  This exercise evidently indicates that the effects of non-uniform detectability along the redshift direction can be mitigated by applying an appropriate weight function for the random sample.  

\subsection{Fiber allocation}

The selection of targets is dependent on the software that allocates fibers to objects \citep{2008SPIE.7018E..94A}.  As a result, there can be artificial biases in the on-sky distribution of objects since not all galaxies in the input catalog are observed.  After allocating a pair of fibers to one galaxy, the opportunity for its neighboring galaxies to be observed at the same time decreases due to the lack of fibers and to avoid fiber entanglement.  In contrast, galaxy pairs having separations of several times the fiber separation are easy to simultaneously assign as targets.  This results in a bias in the pair counts, hence the observed correlation function.  In addition, the sampling rate varies across the survey area because of the different number of exposures among the four footprints and the existence of the small areas where the footprints overlap.  Moreover, some broken fibers result in holes, where the sampling rate is reduced.  In order to correct for these biases, we apply a weight for each galaxy pair as a function of their angular separation \citep[see e.g.,][]{2011MNRAS.412..825D,2015A&A...583A.128D}).  The weight is defined by Equation (\ref{eq:weight}) using the ratio of the angular correlation function of the input catalog and that of the sample of galaxies for which fibers were allocated.  

\begin{figure}[tbph]
   \centering
   \includegraphics[width=3.5in]{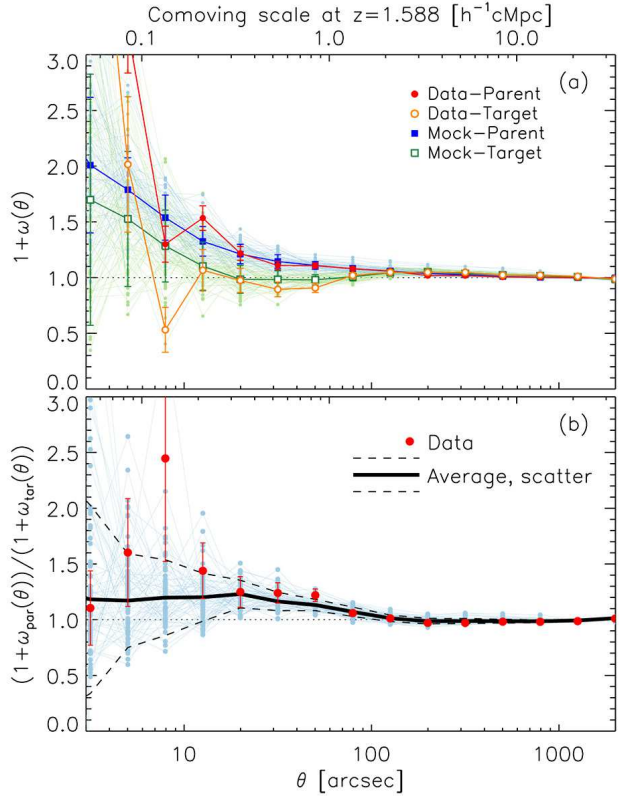} 
   \caption{Panel (a): angular correlation function $\omega \left( \theta \right)$ for the data and the mock samples.  The corresponding comoving separation at $z=1.588$ is shown in the upper $x$-axis.  Filled red and open orange circles show $1+\omega (\theta)$ for the FMOS-parent and FMOS-fiber-target samples, respectively.  Filled blue and open green squares show the average values of the 64 Mock-parent and Mock-fiber-target samples, respectively, with the individual measurements (light color dots).  There is a suppression at scales $\theta\lesssim100~\mathrm{arcsec}$ due to the fiber allocation.  Panel (b): ratios of $1+\omega(\theta)$ between the parent and fiber-target samples for the data (red circles) and the individual mocks (light blue dots).  The average of the mocks, which is used as the weight function for the analysis, and the scatter for the mock samples are indicated by a thick solid and thin dashed lines, respectively.}
   \label{fig:wang_mock}
\end{figure}

In Figure \ref{fig:wang_mock}, we demonstrate the effects of fiber allocation.  The angular correlation functions $\omega (\theta)$ of the real data, FMOS-parent and FMOS-fiber-target samples, are shown in panel (a).  It is evident that the angular clustering amplitude of the FMOS-fiber-target sample is suppressed compared to the FMOS-parent sample at scales $\theta \lesssim 100~\mathrm{arcsec}$.  This angular scale is similar to the fiber separation, corresponding to $r_\mathrm{p}\lesssim 2~h^{-1}~\mathrm{cMpc}$ at $z\sim 1.6$.  We also show the angular correlation functions for 64 sets of Mock-parent and Mock-fiber-target samples.  Although the scatter is large, the suppression at small scales is evident in their average values, in agreement with the real data.  This decrease in the angular clustering amplitude is expected to results from the fiber allocation.  Figure \ref{fig:wang_mock}b shows the ratios of $1+\omega (\theta)$ between the parent and fiber-target samples for both the data (red circles) and the mocks (light blue dots).  For our analyses, we use the average ratios calculated from the mock samples (thick solid line) as the weight function (Equation \ref{eq:weight}).  The weight function is $>1$ at angular scales of $\theta \lesssim 100$ arcsec and reaches $\sim 1.2$ at $\theta <20$ arcsec, while it is slightly smaller than unity at scales at several hundreds arcsec.  On large scale, it reflects the ease of observing pairs with a separation several times larger than the fiber separation simultaneously.

\begin{figure}[tbph]
   \centering
   \includegraphics[width=3.5in]{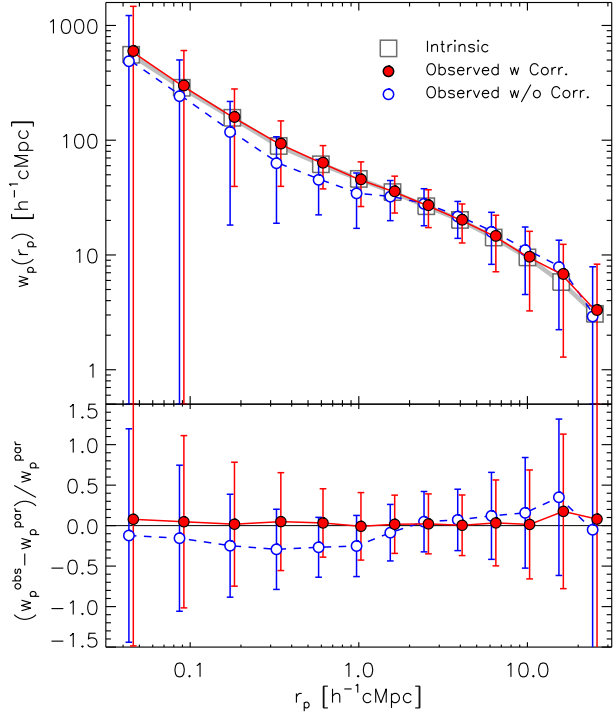} 
   \caption{The symbols are the same as in Figure \ref{fig:wp_mock_zweight}, but the average {\it observed} $w_{p}(r_\mathrm{p})$ are calculated for the Mock-spec-$z$ samples with and without the correction for the effects of fiber allocation.}
   \label{fig:wp_mock}
\end{figure}

We examine the effectiveness of our correction scheme.  In Figure \ref{fig:wp_mock}, we compare the average correlation functions $w_\mathrm{p} (r_\mathrm{p})$ of 64 Mock-spec-$z$ samples with and without the correction to the intrinsic correlation function.  The average amplitude of the Mock-spec-$z$ samples without the correction is reduced at scales $r_\mathrm{p}\lesssim 2~h^{-1}~\mathrm{cMpc}$, while it is slightly enhanced at $r_\mathrm{p}\gtrsim 3~h^{-1}~\mathrm{cMpc}$, as clearly seen in the lower panel.  In contrast, the application of the correction scheme recovers the true $w_\mathrm{p} (r_\mathrm{p})$ over the entire scale of interest in a relatively unbiased manner.  These biases are not negligible even for studies using a relatively small sample, as in our study, and necessarily will has a more significant impact for studies with larger samples.  We conclude that the proposed correction scheme based on the angular correlation functions works appropriately in our analysis (and should work for future multi-object surveys) even for small scales down to $r_\mathrm{p}\sim0.1~h^{-1}~\mathrm{cMpc}$.

\section{The effects of sample selection}
\label{sec:mock_am}

We assess the effect of the stellar mass completeness on the constraints of the HOD model.  For this purpose, we construct mock samples using the Bolshoi simulation \citep{2011ApJ...740..102K} to take the advantage of its better mass resolution and the availability of merger histories of halos, instead of the $\nu^2$GC simulation used above.  The simulation consists of $2048^3$ particles with a mass of $1.35\times 10^8~h^{-1}M_\odot$ in a comoving cube having a side of $250~h^{-1}~\mathrm{cMpc}$.  The volume of the simulation box is 19 times as large as our survey volume, while the side length is shorter than the radial extend of the survey ($\sim380~h^{-1}~\mathrm{cMpc}$).  We utilize the public catalog at $a=0.38435$ ($z=1.602$), in which halos are identified down to a mass of $10^{10}~h^{-1}M_\odot$ based on the Rockstar halo finding algorithm \citep{2013ApJ...762..109B}.  For the following exercise, we use the same cosmology as the Bolshoi simulation $(\Omega_m, \Omega_\Lambda) = (0.27, 0.73)$.

For abundance matching, we use the peak maximum circular velocity ($V_\mathrm{peak}$) provided in the public catalog as the indicator of the stellar mass of a galaxy hosted by a (sub)halo.  Assuming a tight correlation between $V_\mathrm{peak}$ and stellar mass with no scatter, we chose the lower limit $V_\mathrm{peak}^\mathrm{lim}$ by matching the (sub)halo number density $n\,(>V_\mathrm{peak})$ to the number density of the $M_\ast$-selected galaxy sample (Table \ref{tb:samples}).  Note that the number density is recomputed, as $n=8.47\times10^{-3} (h^{-1} \mathrm{cMpc})^{-3}$ based on the above cosmology.  In total, 132408 (sub)halos with $V_\mathrm{peak}>V_\mathrm{peak}^\mathrm{lim}\approx165~\mathrm{km~s^{-1}}$ are selected.  For each (sub)halo, a stellar mass is assigned, while keeping the relation  $n\,(>V_\mathrm{peak})=n\,(>M_\ast)$, in order to reproduce the stellar mass function of the $M_\ast$-selected sample (see Figure \ref{fig:histm}).   We denote this mock catalog as {\it Mock-$V_\mathrm{peak}$-limited} sample.  From this sample, we construct the {\it Mock-$M_\ast$-incomplete} sample, in which the stellar mass incompleteness is taken into account, by randomly select (sub)halos with a probability as a function of the assigned stellar masses, such that the resulting stellar mass function equals that of the FMOS-parent sample (Figure \ref{fig:histm}).  In total, 43777 (sub)halos are selected.

\begin{figure}[tbph]
   \centering
   \includegraphics[width=3.5in]{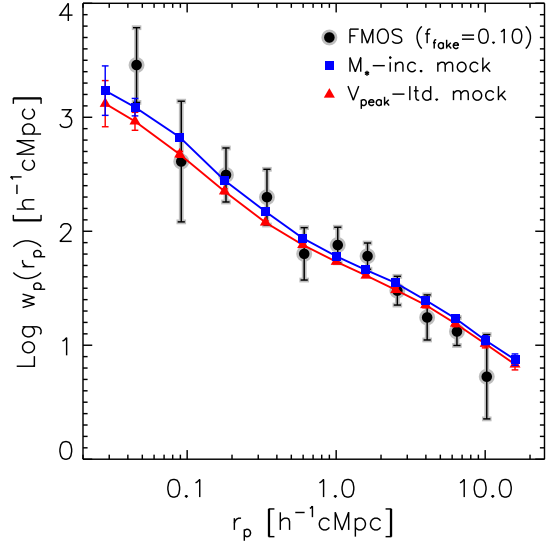} 
   \caption{Projected two-point correlation functions for the FMOS-spec-$z$ sample (circles), the Mock-$V_\mathrm{peak}$-limited sample (red triangles), and the Mock-$M_\ast$-incomplete sample (blue squares).}
   \label{fig:mock_am}
\end{figure}

\begin{figure}[tbph]
   \centering
   \includegraphics[width=3.5in]{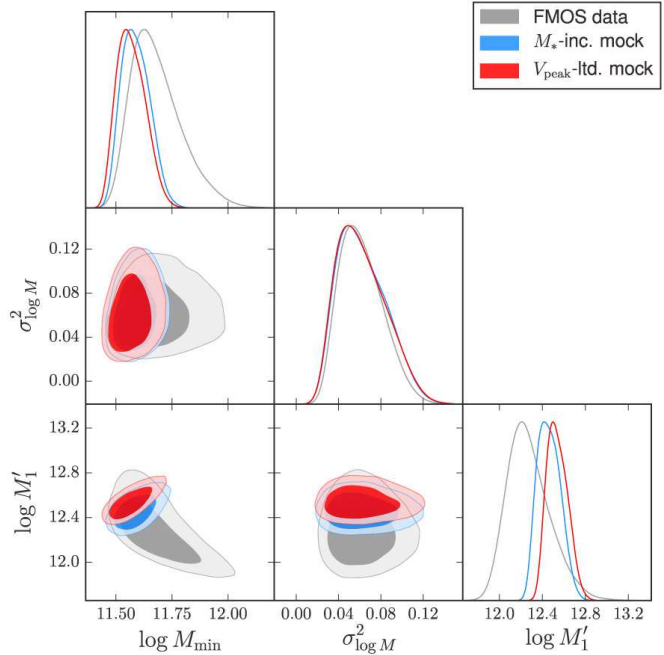} 
   \caption{Constraints of the HOD parameters ($M_\mathrm{min}$, $\sigma^2_{\log M}$, $M_1^\prime$) for the Mock-$V_\mathrm{peak}$-limited sample (red) and the Mock-$M_\ast$-incomplete sample (blue) in comparison with the FMOS results (gray).}
   \label{fig:par_mock_am}
\end{figure}

We measure the projected correlation function of these selected halos.  Figure \ref{fig:mock_am} compares the correlation functions for the Mock-$V_\mathrm{peak}$-limited (red triangles), Mock-$M_\ast$-incomplete (blue squares), and the FMOS-spec-$z$ (black circles) samples.  We correct the correlation functions of the mock samples for the effect of the finite size of the simulation box \citep[see][for details]{2013MNRAS.430..725V}.  We also note that the correlation function for the real data is multiplied by $1/(1-f_\mathrm{fake})^2$ where $f_\mathrm{fake}=0.099$, which is the expected value from the HOD fitting (Table \ref{tb:HODparams}).  It can be seen that the clustering amplitude of the Mock-$M_\ast$-incomplete sample (blue squares) is slightly enhanced relative to the Mock-$V_\mathrm{peak}$-limited sample (red triangles).  This is naturally expected from the fact that the $M_\ast$-incomplete sample is on average biased towards massive halos, which are more strongly clustered.  We fit the HOD model (Equations \ref{eq:cen} and \ref{eq:sat}) to the correlation functions of these two mock samples.  Here we scale the covariance matrix of the mock measurements by a factor of 19, which is the ratio of the volumes of the simulation box and the survey, to match the level of the statistics to the real data.  The parameter constraints are shown in Figure \ref{fig:par_mock_am} separately for the $V_\mathrm{peak}$-limited and $M_\ast$-incompleteness samples in comparison to the results for the real FMOS sample.  We find that all parameters are consistent within their 1~$\sigma$ confidence level, although the peak positions of the posterior distributions are slightly different among data and mocks.  In particular, the constraints are almost identical for the two mock samples.  Theses results indicate that the stellar mass incompleteness in our FMOS sample should not have a significant impact on our conclusions, given the current level of statistical errors.

\section{Limitation on $\sigma_{\log M}$}
\label{sec:lim_sigmalogM}

In this section, we present the prior that we assign to the HOD parameter, $\sigma_{\log M}$ (see Section \ref{sec:prior}).   We use the SHMR derived by \cite{2013ApJ...770...57B}, who parametrized its evolution and intrinsic scatter over $0<z<8$ and constrained it using a variety of observational constraints.  Although our constraint on the stellar-to-halo mass ratio deviates from the \cite{2013ApJ...770...57B} SHMR (see Section \ref{sec:shmr}), this discrepancy is not significant in the following calculations because only the scatter and the slope of the SHMR matter for the prior that we use for $\sigma_{\log M}$.

The scatter in stellar masses ($\log M_\ast$) of galaxies that inhabit halos of a given mass is constrained to be $\sigma_{\log M_\ast}^\mathrm{int}=0.23\pm0.04$ at $z=1.588$.  In addition, we incorporate another scatter ($\sigma^\mathrm{sample}_{\log M_\ast}$) which accounts for sample selection.  We select galaxies based on the {\it best} estimate of the stellar mass from the SED, which should differ from the {\it true} value due the uncertainties in the photometric redshift, flux measurement, and the population synthesis model.  Therefore, a threshold in the estimated stellar mass does not correspond to a threshold in the true stellar mass, and one can expect low stellar mass objects to scatter into our sample.  The width of this scatter will correspond to the typical error in the stellar mass estimation.  We compute for each galaxy the probability that its true stellar mass satisfies $M_\ast\ge M_\ast^\mathrm{lim}$, and derive the probability-weighted stellar mass distribution of our sample such that it has a smoothed lower limit at $M_\ast=M_\ast^\mathrm{lim}\equiv10^{9.57}~M_\odot$.  We find $\sigma^\mathrm{sample}_{\log M_\ast}=0.09$ by deconvolving the distribution into a sharp-cutoff function and a gaussian function with a standard deviation $\sigma^\mathrm{sample}_{\log M_\ast}$.  The total scatter in $\log M_\ast$ of galaxies within halos of the threshold mass is thus given as
\begin{equation}
\sigma_{\log M_\ast} = \sqrt{(\sigma_{\log M_\ast}^\mathrm{int})^2 + (\sigma_{\log M_\ast}^\mathrm{sample})^2} = 0.24\pm0.04.
\end{equation}

We then calculate the probability of finding a galaxy of $M_\ast=M_\ast^\mathrm{lim}$ in a halo as a function of halo mass:
\begin{eqnarray}
P \left(\log M_\ast^\mathrm{lim} | M_\mathrm{h}\right) = \nonumber \\ 
\frac{1}{\sqrt{2}\pi \sigma_{\log M_\ast}} \exp \left( -\frac{\left( \log M_\ast^\mathrm{lim} - \log M_\ast (M_\mathrm{h})\right)^2}{2 \sigma_{\log M_\ast}^2}\right),
\end{eqnarray}
where $M_\ast (M_\mathrm{h})$ is the SHMR parametrized by \citet{2013ApJ...770...57B}.  The expected mean and mean square of halo mass to have a galaxy of $M_\ast=M_\ast^\mathrm{lim}$ are given by 
\begin{eqnarray}
\left< ( \log M)^m \right> = \nonumber \\
 \frac{\int P \left( \log M_\ast^\mathrm{lim} | M_\mathrm{h}\right) \left(\log M_\mathrm{h}\right)^m n \left(M_\mathrm{h}\right) \drm M_\mathrm{h}}{\int P \left( \log M_\ast^\mathrm{lim} | M_\mathrm{h}\right) n \left(M_\mathrm{h}\right) \drm M_\mathrm{h}}.
\end{eqnarray} 
where $n (M_\mathrm{h})$ is a halo mass function at a given redshift, and $m=1 (2)$ corresponds to the mean (mean square).  We then find the deviation in halo mass at our threshold stellar mass:
\begin{eqnarray}
\sqrt{\mathrm{Var} \left(\log M_\mathrm{h}\right)} = \nonumber \\
\sqrt{\left< (\log M_\mathrm{h})^2 \right> - \left< \log M_\mathrm{h}\right>^2} \approx 0.17\pm0.02.
\end{eqnarray}
The HOD parameter $\sigma_{\log M}$ is then given by $\sigma_{\log M} = \sqrt{2} \times \sqrt{\mathrm{Var} (\log M_\mathrm{h})}=0.24\pm0.3$.  This is the prior information we adopt for the parameter in our HOD modeling.  We have checked that the inclusion of the redshift evolution of the \citet{2013ApJ...770...57B} SHMR over $1.43\le z\le 1.74$ does not change these values.  The derived value of $\sigma_\mathrm{\log M}$ is consistent with an estimate in previous studies \citep[e.g.,][]{2011MNRAS.410..210M}, albeit at local redshifts.

\end{document}